\documentclass{article}

\usepackage{PRIMEarxiv}

\usepackage[utf8]{inputenc} 
\usepackage[T1]{fontenc}    
\usepackage{hyperref}       
\usepackage{url}            
\usepackage{booktabs}       
\usepackage{amsfonts}       
\usepackage{nicefrac}       
\usepackage{microtype}      
\usepackage{lipsum}
\usepackage{fancyhdr}       
\usepackage{graphicx}       

\usepackage{multirow}%
\usepackage{amsmath,amssymb,amsfonts}%
\usepackage{amsthm}%
\usepackage{mathrsfs}%
\usepackage[title]{appendix}%
\usepackage{xcolor}%
\usepackage{textcomp}%
\usepackage{manyfoot}%
\usepackage{booktabs}%
\usepackage{algorithm}%
\usepackage{algorithmicx}%
\usepackage{listings}%
\usepackage[numbers]{natbib}
\usepackage{pdfpages}

\title{Data Driven Deep Learning for Correcting Global Climate Model Projections of Sea Surface Temperature and Dynamic Sea Level in the Bay of Bengal}

\title{Data Driven Deep Learning for Correcting Global Climate Model Projections of Sea Surface Temperature and Dynamic Sea Level in the Bay of Bengal}

\author{
Abhishek Pasula\\
Department of Computational and Data Sciences,\\ Indian Institute of Science,\\
Bengaluru, Karnataka,\\
\texttt{abhishekp1@iisc.ac.in}\\
\And
Deepak N. Subramani*\\
Department of Computational and Data Sciences, \\
Divecha Center for Climate Change, \\
Indian Institute of Science,\\
Bengaluru, Karnataka,\\
\texttt{deepakns@iisc.ac.in}\\
}
\begin{document}
\maketitle

\begin{abstract}
Climate change alters ocean conditions, notably temperature and sea level. In the Bay of Bengal, these changes influence monsoon precipitation and marine productivity, critical to the Indian economy. In Phase 6 of the Coupled Model Intercomparison Project (CMIP6), Global Climate Models (GCMs) use different shared socioeconomic pathways (SSPs) to obtain future climate projections. However, significant discrepancies are observed between these models and the reanalysis data in the Bay of Bengal for 2015-2024. Specifically, the root mean square error (RMSE) between the climate model output and the Ocean Reanalysis System (ORAS5) is 1.2$^{\circ}$C for the sea surface temperature (SST) and 1.1 m for the dynamic sea level (DSL). We introduce a new data-driven deep learning model to correct for this bias. The deep neural model for each variable is trained using pairs of climatology-removed monthly climate projections as input and the corresponding month's ORAS5 as output. This model is trained with historical data (1950 to 2014), validated with future projection data from 2015 to 2020, and tested with future projections from 2021 to 2023. Compared to the conventional EquiDistant Cumulative Distribution Function (EDCDF) statistical method for bias correction in climate models, our approach decreases RMSE by 0.15 $^{\circ}$C for SST and 0.3 m for DSL. The trained model subsequently corrects the projections for 2024-2100. A detailed analysis of the monthly, seasonal, and decadal means and variability is performed to underscore the implications of the novel dynamics uncovered in our corrected projections.
\end{abstract}

\keywords{CMIP6, Deep Learning, Artificial Intelligence, Bias correction, Bay of Bengal}



\maketitle

\section{Introduction}
    \label{intro}    
    Anthropogenic climate change is a significant concern for the health of the planet. Warming of the ocean affects the energy budget of the Earth system, leading to an increase in extreme events \citep{chaudhury2000impact, goswami2006increasing, rajesh2016role}. Changes in the ocean alter the biogeochemical balance affecting marine life \citep{gruber2011warming, bopp2013multiple}. Developing economies such as India face the brunt of a changing climate. During the last five decades, the Indian Ocean has been warming throughout the basin \citep{swapna2014indian}, significantly increasing the frequency and magnitude of cyclones and other extreme rain events during the Indian summer monsoon.  
    
    Among all ocean variables, the sea surface temperature (SST) is widely recognized as a crucial indicator of climate change. Changes in global and regional SST patterns can affect zonal and meridional circulations, which can alter precipitation and temperature patterns throughout the world \citep{bjerknes1969atmospheric, ashfaq2011influence, srinivas2018simulation}. The Dynamic Sea Level (DSL) is also an important ocean variable to assess the impact of climate change in space and time \citep{lyu2020regional, chen2023performance}. 

\begin{figure}[ht]
            \centering
            \includegraphics[width=\textwidth]{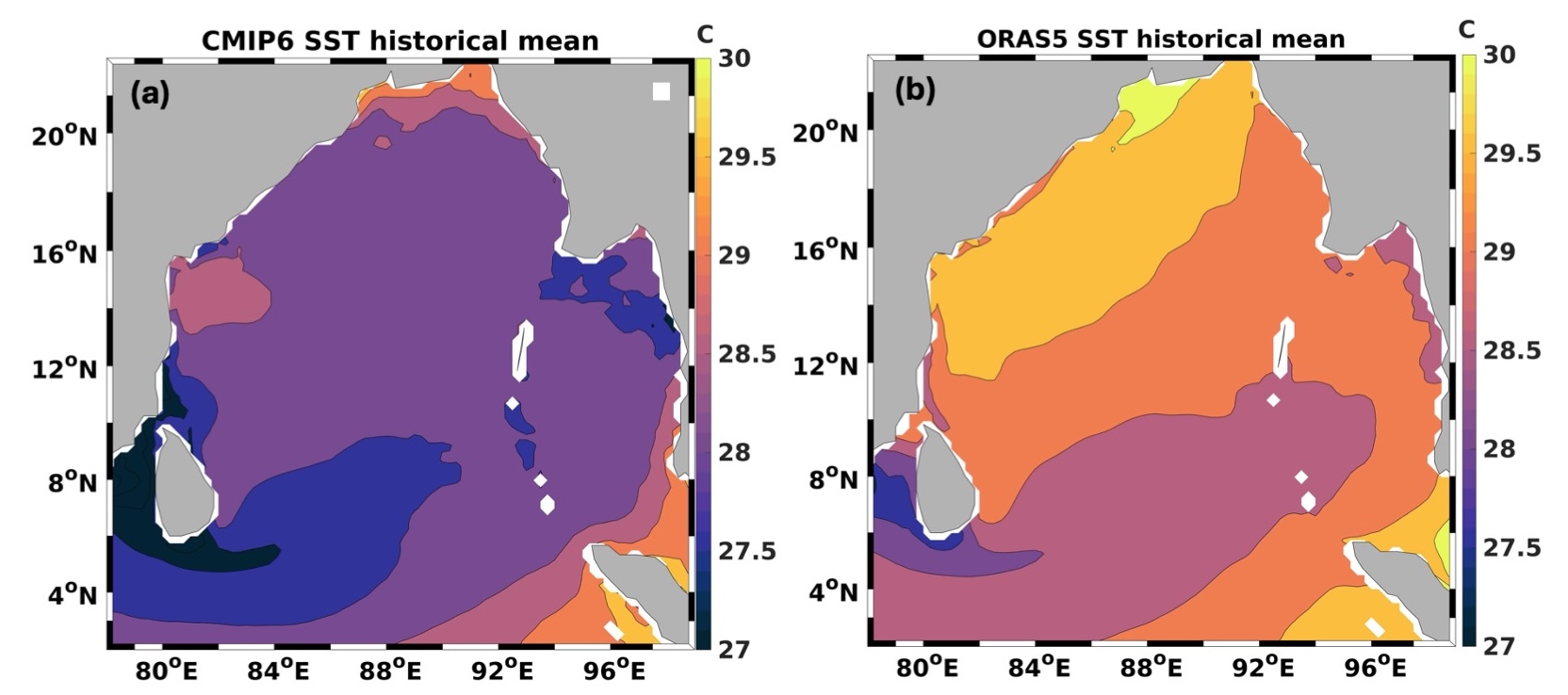}
            \caption{June climatology SST for the Bay of Bengal from 1958 to 2014 using (a) CNRM-CM6 historical (CMIP6) and (b) ORAS5 reanalysis. }\label{fig: intro}
        \end{figure}

    Forecasting future climate scenarios based on climate models is essential to understand anthropogenic causes and inform science-based climate policy. Global climate models (GCMs) are complex multiscale nonlinear numerical models used to obtain such climate projections. GCMs simulate the Earth's climate system by considering factors such as atmospheric composition, ocean circulation, and land surface processes. The World Climate Research Program (WCRP) under the Intergovernmental Panel on Climate Change (IPCC) periodically approves a family of climate models that contain past, present, and future climate change scenarios for use by scientific and policy communities. The climate models developed under this program provide projections of probabilistic scenarios of future change that address a range of outcomes, from sustainable pathways to unsustainable growth. The Phase 5 and 6 models of the Coupled Model Intercomparison Project (CMIP) are the most recent results of this project \citep{reichstein2019deep, gruber2011warming}. These models provide essential information on the future atmosphere and ocean states. Crucially, the CMIP6 family of GCMs employs Shared Socioeconomic Pathways (SSP) scenarios, which represent a range of potential global social developments in the future, without taking into account climate change or measures for mitigation or adaptation \citep{su2021insight}. The main demerit of GCMs is their potential for error, which can have significant implications for decision-making and policy making. In the period for which both projections and reanalysis are available, it is possible to qualitatively and quantitatively analyze this error. As an example, Fig.~\ref{fig: intro} shows the climatology of June SST (1958-2014 mean) in the Bay of Bengal ($2^{o}$ N to $22.5^{o}$ N; $78^{o}$E to $99^{o}$E) from CNRM-CM6 (one of the CMIP6 Global Climate Models) and an ocean reanalysis product ORAS5. The historical mean of CMIP6 (CNRM-CM6 historical) differs significantly in the SST pattern compared to ORAS5, with the extent of the surface temperature signature of the southwest monsoon current being different in both. Thus, a crucial need in climate science is the availability of data-driven bias-corrected climate projections.

    This paper aims to develop and utilize a deep learning model to correct the bias in GCM SST and DSL projections in the Bay of Bengal. It also seeks to document new dynamical insights from the corrected projections for each season. 

    The remainder of the paper is organized as follows. Section~\ref{sec: prior} describes the previous progress in bias correction. In Section~\ref{sec: method}, we define the problem statement, the data sets used, and describe the development of the UNet bias correction model. Section~\ref{sec: analysis} presents the evaluation of the bias correction models for the SST and DSL in the test years, focusing on the overall error and the month-by-month analysis of the projections. Section~\ref{sec:analysisfuture} presents analyses of the annual cycle and variability in the corrected future projections, juxtaposing it with the raw projections. Key dynamical insights with respect to important features that get affected are documented. Finally, Section~\ref{sec: conclusion} offers conclusions and future research directions.

\subsection{Prior work in bias correction of climate projections}\label{sec: prior}
        Several error or bias correction methods have been proposed in the literature to correct GCM projections. These techniques are divided into interpolation methods, dynamical downscaling methods, and machine learning methods. Within interpolation, delta change methods \citep{graham2007assessing} and quantile-based mapping methods \citep{piani2010statistical} have been proposed. These two differ according to whether the correction takes into account the probability distribution of the climate model data \citep{maraun2016bias}. The delta change point methods are simple and usually involve calculating biases in the mean, variance, or other moments of the climate model data during the historical period relative to observations. These biases are then removed directly or proportionally from future climate projections to correct for any biases \citep{schmidli2006downscaling, graham2007assessing}. However, such methods often neglect the stochastic characteristics of climate variables \citep{maraun2016bias}. Quantile mapping-based strategies overcome the limitations of the delta-change method by separately determining the Cumulative Distribution Function (CDF) for both observed and model data in the historical period and applying a transfer function (TF) to modify the simulation CDFs in other periods, thereby reducing bias. However, they are limited to addressing errors in the simulated frequency distributions in each cell of the grid individually \citep{li2010bias}. Recent developments in bias correction methods have been aimed at improving spatial dependencies \citep{li2010bias, cannon2018multivariate}. However, these techniques have issues such as instability and overfitting \citep{franccois2020multivariate}. 

        The Equi-Distant Cumulative Distribution Function (EDCDF) is a version of the quantile mapping method that is the de facto standard for the correction of climate model bias \cite{li2010bias}. The EDCDF method corrects the model projections by comparing them with observations in the historical period and using a closed-form expression that involves the CDF of both \citep{li2010bias}. In the present paper, we compare our proposed deep learning model for bias correction with the EDCDF method for benchmarking. 

        Recent studies indicate that machine learning proves highly effective in meteorological forecasting and in addressing data gaps in climatic variables, with notable advances documented \citep{reichstein2019deep}. Deep neural networks originally developed for computer vision tasks such as image processing have been successfully adapted for numerous applications in Earth system science, including weather prediction \citep{bi2023accurate, rasp2021data, zhang2023skilful}. Recently, deep learning techniques have been used to mitigate biases in Tmin, Tmax \citep{wang2022deep}, precipitation \citep{fulton2023bias, huang2023investigating, huang2024unsupervised}, and wind energy \citep{zhang2021future}.

\section{Model Development } \label{sec: method}   
    \subsection{Problem Definition}
    Consider a 2D geographic domain $\Omega \subset \mathcal{R}^2$ with coordinates $(x,y)$ and a time horizon with a temporal coordinate $t \in [0,T)$. Let $U(x,y,t)$ denote the climate projection from a GCM and $U^\dagger(x,y,t)$ be the reanalysis obtained after observations are available. There exists a bias-correcting mathematical operator $U^\dagger = \mathcal{G}(U)$ that maps $U\rightarrow U^\dagger$. We propose learning a parametrized neural network $\mathcal{N}_\theta$ to approximate $\mathcal{G}$ using historical data of climate projections and reanalysis. By assuming that the reanalysis is the ground truth state of sea surface temperature and dynamic sea level, the learned neural operator is a neural bias correction system.

    Different architectures and learning paradigms could be used to train $\mathcal{N}_\theta$. In a concurrent work, we have explored multiple architectures and conducted an extensive ablation study to arrive at the UNet architecture used in the present paper \cite{pasula2025globalclimatemodelbias}. Next, we discuss the data sets and describe the architecture and training procedure.

    \subsection{Data Sets}
    The Bay of Bengal, extending from 2$^o$N to 22.5$^o$N and 78$^o$E to 99$^o$E, is the domain in focus for our model development and analysis.
    
    Among the different global climate models (GCMs) in the CNRM-CM6 suite, the Center National de Recherches Meteorologiques and Cerfacs' CNRM-CM6-1-HR AOGCM climate model is selected for the present study. This choice is based on the following factors. The ocean temperature and salinity bias is the lowest reported for this model \citep{voldoire2019evaluation}. The projected ocean heat uptake is reported to be closely aligned with the observed data \citep{kuhlbrodt2023historical}. CNRM-CM6 models have been shown to better represent the mean state and variability of the climate system, showing moderate skill in local climate simulation compared to other GCMs \citep{fan2022assessment}. On the atmospheric side, this model has improved Indian summer monsoon simulations due to updated convective parameterization, cloud fraction, and better representation of the western ghats, the orography of northeast India, and the Madden-Julian oscillation (MJO) \citep{gusain2020added,chen2022mjo}.

    CNRM-CM6-1-HR AOGCM includes physical processes of land, ocean, atmosphere, and sea ice, together with their dynamical interactions \citep{voldoire2019evaluation}. It was built using the SURFEX land component version 8.0, the NEMO ocean component version 3.6, and the ARPEGE-Climate atmosphere component version 6.3 \citep{madec2017nemo,voldoire2019evaluation}. The model has a 25 km ocean grid resolution and a 50 km atmosphere grid resolution, resolving regional ocean currents, as well as atmospheric general circulation patterns, convection, microphysics, and turbulence processes \citep{voldoire2019evaluation}. The upper water column is resolved with a layer thickness of 1 m from the surface to 200~m depth, and the atmospheric boundary layer is resolved with 15 levels below 1500 m \citep{voldoire2019evaluation}. Historical simulations were available from 1850 to 2014. CNRM-CM6-1-HR provides four SSP scenarios, namely SSP1-2.6, SSP2-4.5, SSP3-7.0, and SSP5-8.5, where each SSP considers different pathways for energy, land use, and emissions from 2015 to 2100.

    To correct the CNRM-CM6 projections, we require reanalysis data. We selected ORAS5, a global Ocean Reanalysis System (ORAS5) developed by ECMWF, as the ocean ground truth to train our CNRM-CM6 correction model. ORAS5 uses the NEMO ocean model version 3.4.1 \citep{madec2017nemo} with a horizontal resolution of 25 km and assimilates SST, subsurface temperature, salinity profiles, and satellite sea level and sea ice concentration data using 3D-Var with a 5-day assimilation cycle. It also includes a constraint on sea surface salinity by nudging to climatology \citep{zuo2019ecmwf}.

        The modeling time period and the ocean model are the same for the historical CNRM-CM6-1-HR and ORAS5. The difference between them is in the initial and forcing conditions. CNRM-CM6-1-HR considers observations about greenhouse gas concentrations, global gridded land use forcing, solar forcing, stratosphere aerosol data set, Atmospheric Model Intercomparison Project (AMIP) SST and sea ice concentration, ozone chemistry, and aerosol forcing. ORAS5 uses initial conditions from ERA-40 \citep{uppala2005era}, ERA-Interim forcing fields \citep{dee2011era}, different ocean observations assimilated in an operational ensemble reanalysis \citep{zuo2019ecmwf}, and nudging the salinity of the sea surface to climatology. Furthermore, CNRM-CM6 and ERA5 use different atmosphere models. ERA5 uses the Integrated Forecast System (IFS) model coupled with a land surface model (HTESSEL), and CNRM-CM6 uses the atmosphere-ocean general circulation model (AOGCM) with different initial conditions. ORAS5 includes freshwater discharge, while CNRM uses its own model, Total Runoff Integrating Pathways, to simulate river discharge \citep{seferian2019evaluation}.

        A summary of the data source and its temporal and spatial resolution is listed in Table~\ref{tab: data_sources}.

                \begin{table}[]
                \centering
                    \caption{Data Sources}
                    \label{tab: data_sources}
                    \begin{tabular}{|c|c|c|c|}
                    \hline
                    \textbf{Climate Dataset}                                                         & \textbf{Time frame}     &  \begin{tabular}[c]{@{}c@{}}\textbf{Temporal}\\ \textbf{Resolution}\end{tabular}&  \begin{tabular}[c]{@{}c@{}}\textbf{Spatial}\\ \textbf{Resolution}\end{tabular} \\ \hline
                    CNRM-CM6 historical run                                                     & 1958-2014      & monthly             & $\sim 0.25^{\circ} \times 0.25^{\circ}$         \\ \hline
                    \begin{tabular}[c]{@{}c@{}}CNRM-CM6 Projections\\ (SSP1,2,3,5)\end{tabular} & 2015-2100      & monthly             &  $\sim 0.25^{\circ} \times 0.25^{\circ}$          \\ \hline
                    ORAS5 Reanalysis                                                        & 1958-2024 & monthly             & $\sim 0.25^{\circ} \times 0.25^{\circ}$         \\ \hline
                    \end{tabular}
                \end{table}
    
    \subsection{UNet Model for SST and DSL Bias Correction} 
\begin{figure}[ht]
    \centering
    \includegraphics[width=\linewidth]{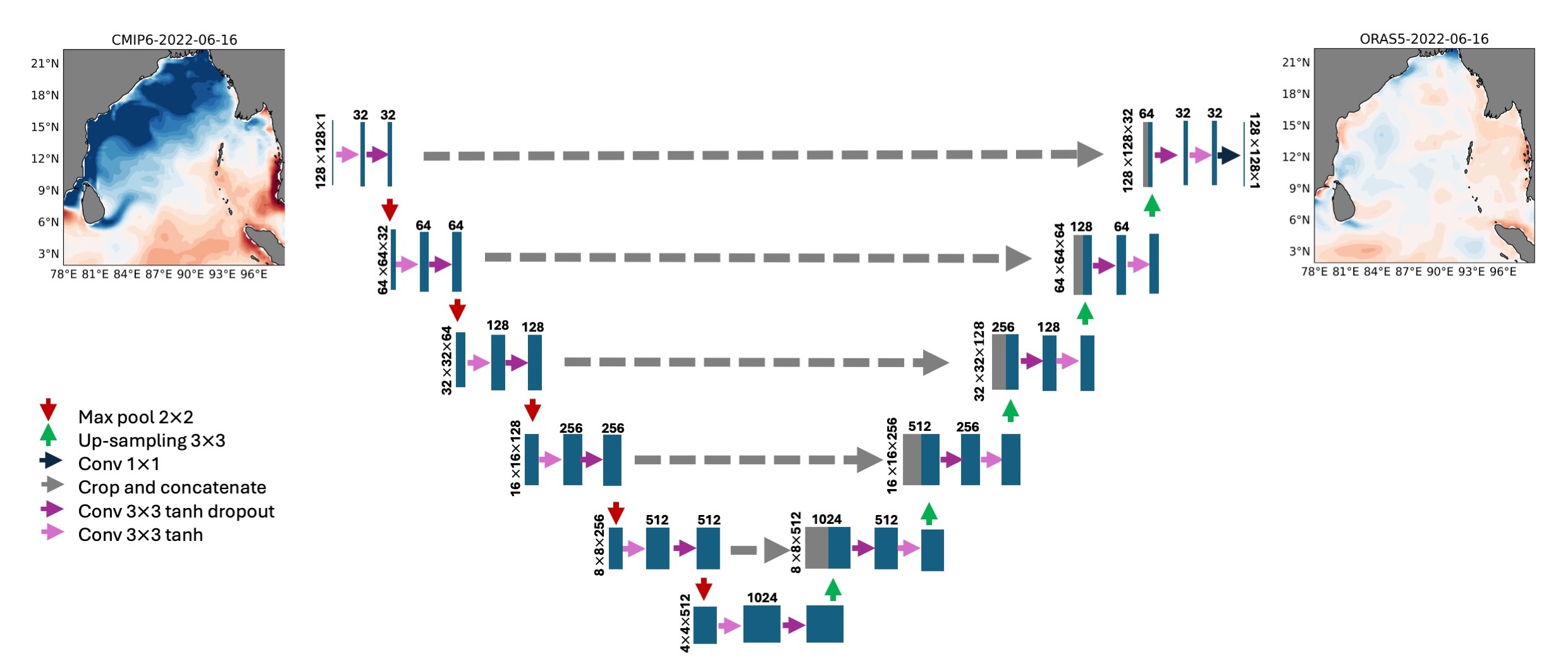}
    \caption{Components of the neural network named UNet developed for correcting the bias in CNRM-CM6 projections of SST and DSL in the Bay of Bengal.}\label{fig:unetdes}
    \end{figure}

    The UNet architecture (Fig.~\ref{fig:unetdes}) has a classical fully convolutional encoder-decoder structure consisting of a downsampling encoder arm followed by an upsampling decoder arm, giving it the shape of the alphabet $U$ and hence the name. A symmetric structure is utilized with an equal number of downsampling blocks and upsampling blocks. 
        
    The encoder starts with an input layer that accepts $128 \times 128$ dimensional images with one channel. Then, it applies five consecutive downsampling blocks, each doubling the number of filters (32, 64, 128, 256, 512) while reducing spatial dimensions. Each downsampling block comprises two $3\times 3$ convolutions with tanh activation, followed by 2x2 max pooling and dropout (30\%). The bottleneck uses two 3x3 convolutions with 1024 filters. The encoder path progressively reduces the spatial resolution while increasing the feature depth, capturing hierarchical representations of the GCM. The decoder mirrors the encoder with five upsampling blocks through $ 3\times3$ transposed convolutions with stride 2. Skip connections are used to concatenate the corresponding encoder features with the upsampled decoder features, preserving fine-grained details. These connections avoid vanishing gradient problems at the encoder layers and also capture multiscale features. This process allows the model to generate corrected GCM projections that preserve fine-scale spatial information. The final layer is a $1\times 1$ convolution that produces the image. Tanh activation is used. We provide data processing and UNet neural network for correcting biases in CNRM-CM6 SST and DSL, with our code available on GitHub\footnote{https://github.com/AbhiPasula/CNRM-CM6-SST-and-DSL-Bias-Correction.git}.

    Crucially, modifying the neural operator to learn the map between climatology-removed $U^\dagger$ and $U$ helped to achieve the best neural model. Two UNet models are developed separately, one for SST and the other for DSL.

    \subsection{Training Procedure}
                  
            For the study area in the Bay of Bengal (2$^o$N to 22.5$^o$N and 78$^o$E to 99$^o$E), the grid is of size $85\times 85$ with the resolution of CNRM-CM6 and ORAS5 (0.25$^o$ resolution). We have historical data from 1958 to 2014 and projections from 2015 to 2100. For training and validation, we used data from 1958 to 2020 (972 months), and these shuffled data are divided into training data (768 months) and validation data (204 months). For the test data set, we used data from 2021 to June 2024 (42 months). The monthly climatology is calculated using ORAS5 data and is removed from both the input and target data. This helps the model to learn the corrections in the projected anomaly. In our ablation study, we find that the anomaly correction yields the best results. All input monthly data from CNRM-CM6 and ORAS5 were resized to $128 \times 128$ for use in our UNet model. During training, for all SSP projections, the corresponding ORAS5 reanalysis is considered the ground truth.

            The validation data set comprises 20\% of historical and projection data from 1958 to 2020. Hyperparameters are tuned via cross-validation on these data. The hyperparameters that gave the highest performance in cross-validation are ultimately chosen for the final model.
            
\section{Analysis of SST and DSL projections in the test years} \label{sec: analysis}
    The trained UNet model for each variable is used to correct the projections from 2021 to 2100. For the 2021 to 2024 period, reanalysis data is also available. Hence, we first use this period to evaluate the performance of our UNet model in terms of RMSE and PCC, comparing the performance with the statistical EDCDF method (Sect.~\ref{sec: monthly_ana}). Next, we examine the monthly projections in the test year 2022 to study how our UNet correction faithfully reproduces the dynamics in the region for SST and DSL. 

    \subsection{Evaluation of the corrected projections in the test period (2021-2024)} \label{sec: monthly_ana}
    \begin{figure}[ht]
            \centering
            \includegraphics[width=\linewidth]{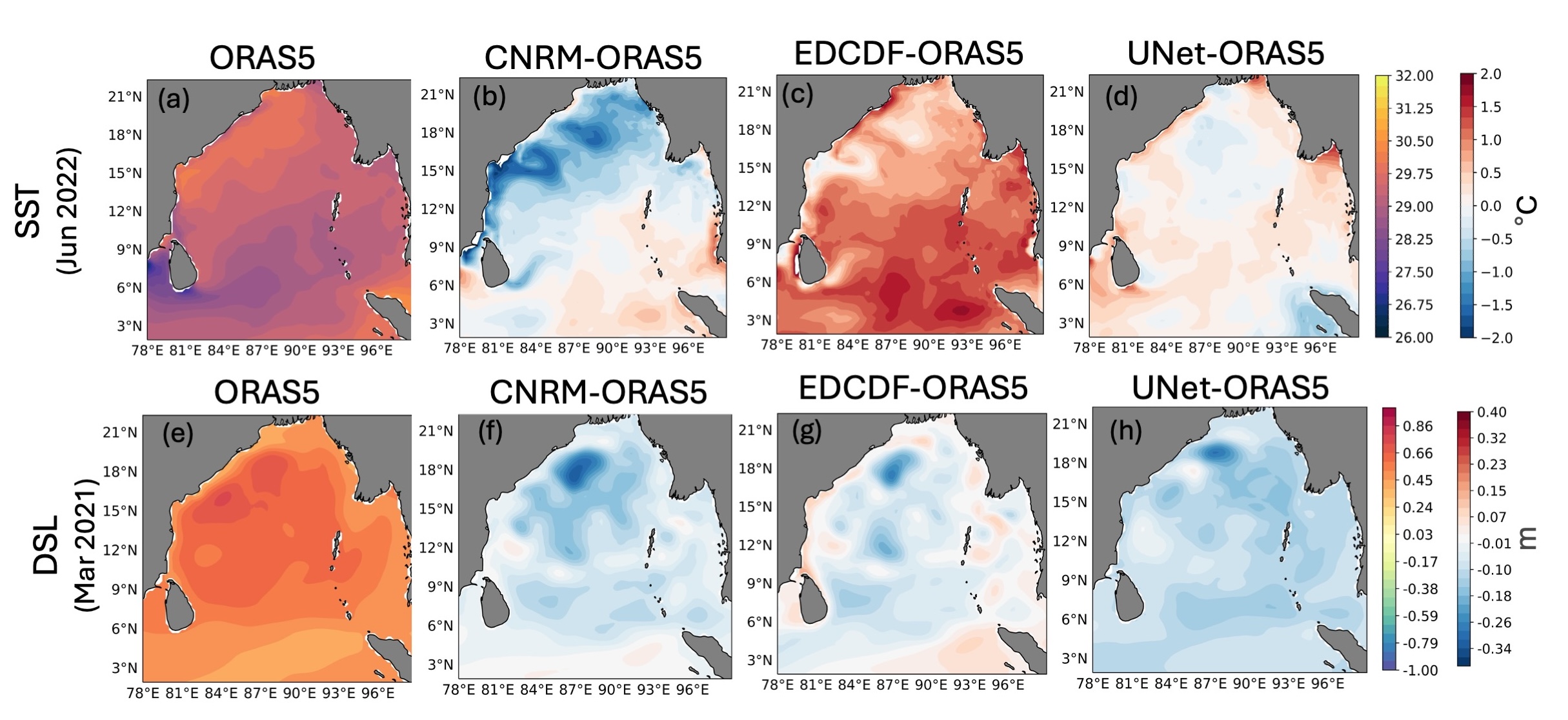}
            \caption{Comparison of Sea Surface Temperature (SST) and Dynamic Sea Level (DSL) during the test period in the Bay of Bengal. (a-d) SST for June 2022: ORAS5 observations and bias comparison with CNRM-ORAS5, EDCDF-ORAS5, and UNet-ORAS5. (e-h) DSL for March 2021: ORAS5 observations and bias comparison with CNRM-ORAS5, EDCDF-ORAS5, and UNet-ORAS5.}
            \label{fig: monthly_snapshot}
    \end{figure}     

    We start by assessing the error in the corrected projections for the 2021–2024 test years. Table~\ref{tab: metrics} presents the root mean square error (RMSE) and the pattern correlation coefficient (PCC) for the raw and corrected projections of CNRM-CM6 against the ORAS5 reanalysis for SST and DSL. The EDCDF correction method (Sect.~\ref{sec: prior}) serves as a comparison baseline, with metrics listed accordingly. RMSE and PCC are determined monthly and then averaged over the test years. The corrected projections from our UNet model exhibit reduced RMSE and enhanced PCC for both variables relative to the EDCDF method. Across all SSPs, RMSE reduces by 9-38\%, and PCC increases by 0.02-0.12.
        
\begin{table}
\centering
\begin{tabular}{lllllll}
\multicolumn{7}{c}{SST}                       
\\ \hline  \hline
\multicolumn{1}{|c|}{} & \multicolumn{2}{c|}{CNRM-CM6}                                & \multicolumn{2}{c|}{EDCDF}                                & \multicolumn{2}{c|}{UNet}   \\ \hline
\multicolumn{1}{|l|}{SSP}   & \multicolumn{1}{l|}{RMSE} & \multicolumn{1}{l|}{PCC} & \multicolumn{1}{l|}{RMSE} & \multicolumn{1}{l|}{PCC} & \multicolumn{1}{l|}{RMSE} & \multicolumn{1}{l|}{PCC}
\\ \hline
\multicolumn{1}{|l|}{1}   & \multicolumn{1}{l|}{1.2264} & \multicolumn{1}{l|}{0.5747} & \multicolumn{1}{l|}{0.6871} & \multicolumn{1}{l|}{0.6812} & \multicolumn{1}{l|}{0.5018} & \multicolumn{1}{l|}{0.7768} \\ \hline

\multicolumn{1}{|l|}{2}   & \multicolumn{1}{l|}{1.3480} & \multicolumn{1}{l|}{0.5721} & \multicolumn{1}{l|}{0.6352} & \multicolumn{1}{l|}{0.6750} & \multicolumn{1}{l|}{0.5091} & \multicolumn{1}{l|}{0.7613} \\ \hline

\multicolumn{1}{|l|}{3}   & \multicolumn{1}{l|}{1.3494} & \multicolumn{1}{l|}{0.5743} & \multicolumn{1}{l|}{0.6255} & \multicolumn{1}{l|}{0.6809} & \multicolumn{1}{l|}{0.5092} & \multicolumn{1}{l|}{0.7615} \\ \hline

\multicolumn{1}{|l|}{5}   & \multicolumn{1}{l|}{1.2368} & \multicolumn{1}{l|}{0.7285} & \multicolumn{1}{l|}{0.7898} & \multicolumn{1}{l|}{0.6515} & \multicolumn{1}{l|}{0.4920} & \multicolumn{1}{l|}{0.7666} \\ \hline

\multicolumn{7}{c}{DSL}                              \\ \hline  \hline
\multicolumn{1}{|c|}{} & \multicolumn{2}{c|}{CNRM-CM6}                                & \multicolumn{2}{c|}{EDCDF}                                & \multicolumn{2}{c|}{UNet}   \\ \hline
\multicolumn{1}{|l|}{SSP}   & \multicolumn{1}{l|}{RMSE} & \multicolumn{1}{l|}{PCC}  & \multicolumn{1}{l|}{RMSE} & \multicolumn{1}{l|}{PCC} & \multicolumn{1}{l|}{RMSE} & \multicolumn{1}{l|}{PCC}
\\ \hline
\multicolumn{1}{|l|}{1}   & \multicolumn{1}{l|}{1.1886} & \multicolumn{1}{l|}{0.4623} & \multicolumn{1}{l|}{0.83} & \multicolumn{1}{l|}{0.6613} & \multicolumn{1}{l|}{0.5967} & \multicolumn{1}{l|}{0.7535} \\ \hline

\multicolumn{1}{|l|}{2}   & \multicolumn{1}{l|}{1.148} & \multicolumn{1}{l|}{0.4697} & \multicolumn{1}{l|}{0.8421} & \multicolumn{1}{l|}{0.6633} & \multicolumn{1}{l|}{0.6599} & \multicolumn{1}{l|}{0.7358} \\ \hline

\multicolumn{1}{|l|}{3}   & \multicolumn{1}{l|}{1.2516} & \multicolumn{1}{l|}{0.4327} & \multicolumn{1}{l|}{0.909} & \multicolumn{1}{l|}{0.6404} & \multicolumn{1}{l|}{0.6586} & \multicolumn{1}{l|}{0.7490} \\ \hline

\multicolumn{1}{|l|}{5}   & \multicolumn{1}{l|}{1.144} & \multicolumn{1}{l|}{0.5} & \multicolumn{1}{l|}{0.75} & \multicolumn{1}{l|}{0.701} & \multicolumn{1}{l|}{0.66} & \multicolumn{1}{l|}{0.7401} \\ \hline
\end{tabular}

\caption{Performance Metrics: Root Mean Square Error (RMSE) and Pattern Correlation Coefficient (PCC) of raw CNRM-CM6, EDCDF corrected, and UNet corrected projections, evaluated against ORAS5 reanalysis for test years.}
\label{tab: metrics}
\end{table}
        
        For SST, UNet corrected projections achieve the lowest RMSE values ($<0.51^o$C) in all scenarios, while EDCDF shows higher values ($>0.78^o$C), indicating UNet's superior bias correction ability.  For DSL, the UNet corrected projections have lower RMSE ($<0.66$~m) compared to EDCDF ($>0.75$~m). The pattern correlation coefficient captures how well the projections represent the spatial patterns compared to the ORAS5 reanalysis. For SST projections under different SSP scenarios, the raw CNRM-CM6 projection has a PCC ranging from 0.5747 to 0.5721, EDCDF corrected projections have a PCC from 0.6812 to 0.6515, and UNet corrected projections have the highest PCC ($>0.72$). For DSL, the UNet corrected projections have a PCC of around 0.74, whereas EDCDF achieves a PCC of 0.64 to 0.7. Thus, on average, our UNet correction model produces projections more aligned with ORAS5. 
        
        To emphasize the dynamics captured in the corrected projections, Fig.~\ref{fig: monthly_snapshot} provides randomly selected examples of the monthly projections of the test years. Column 1 presents the SST for June 2022 and the DSL for March 2021. The subsequent three columns show the differences between uncorrected CNRM-CM6 and ORAS5 (CNRM-ORAS5), EDCFD-corrected CNRM-CM6 and ORAS5 (EDCDF-ORAS5), and UNet-corrected CNRM-CM6 and ORAS5 (UNet-ORAS5).

        In June 2022, ORAS5 shows a sea surface temperature pattern with elevated temperatures in the northern bay relative to the central bay. Moreover, it distinctly captures a thermal feature extending from the south of Sri Lanka to the Andaman Islands, accurately representing the known surface dynamics of June in the Bay of Bengal \citep{vecchi2002monsoon}. CNRM-CM6 does not depict this pattern well (showing a negative error). CNRM-ORAS5 projection (SSP2-4.5; Fig.~\ref{fig: monthly_snapshot}(b)) is generally cooler than ORAS5 (Fig.~\ref{fig: monthly_snapshot}(a)). The EDCDF-ORAS5 (Fig.~\ref{fig: monthly_snapshot}(c)) shows that the EDCDF projections are warmer than ORAS5. We see that the EDCDF method excessively corrected the SST, with temperatures of 32$^{\circ}$C in the central and northern areas. To its credit, the EDCDF accurately identified several warm anomalies along the western Bay of Bengal, but unexpectedly did not capture the temperature feature south of Sri Lanka accurately (showing a higher error). In contrast, our UNet model excellently reproduces the structural characteristics visible in ORAS5 by rectifying the bias in the CNRM-CM6 projection (Fig.~\ref{fig: monthly_snapshot}(d)), particularly evident in coastal and deepwater temperature errors $\approx\pm 0.5^{\circ}$C. Unlike EDCDF, UNet avoids overcorrection problems.

        The DSL for March 2021 has positive values as seen in the ORAS5 data (Fig.~\ref{fig: monthly_snapshot}(e)). CNRM-ORAS5 projections (Fig.~\ref{fig: monthly_snapshot}(f)) show higher bias, and the EDCDF corrected DSL somewhat matches the ORAS5 pattern but still with errors (Fig.~\ref{fig: monthly_snapshot}(g)), particularly in the central and eastern areas. The UNet-corrected projections exhibit a more tempered increase in the DSL values, with UNet-ORAS5 (Fig.~\ref{fig: monthly_snapshot}(h)) indicating lower error margins compared to EDCDF.

        Overall, UNet outperforms EDCDF by providing more consistent and accurate corrections that result in a better match with the reanalysis.

    \subsection{Comparison of SST projections in 2022} 

\begin{figure}[htbp]
            \centering
            \includegraphics[width=1\linewidth]{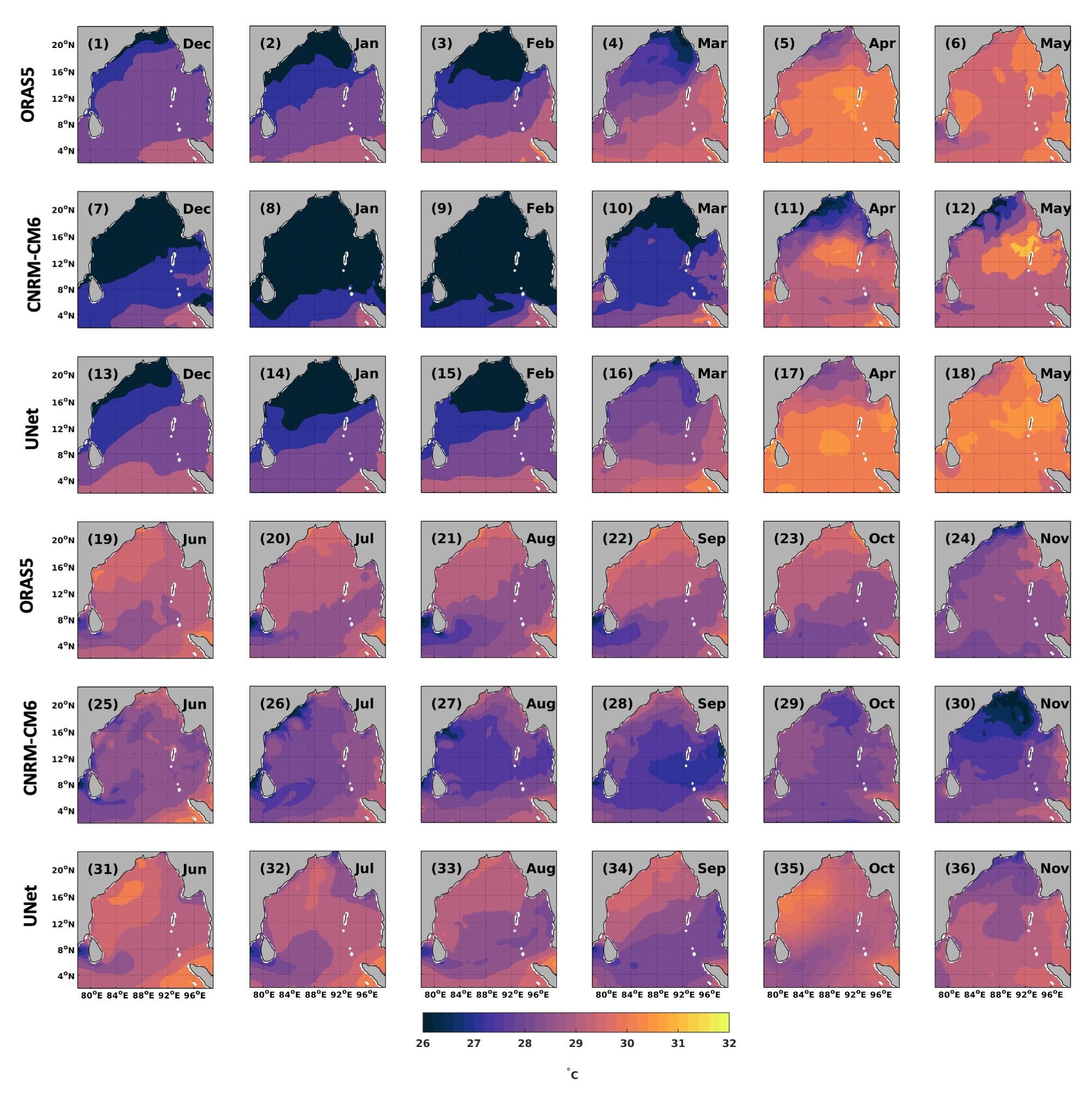}
            \caption{Monthly sea surface temperature (SST) in the BoB during 2022, comparing raw CNRM-CM6 model output (CNRM-CM6), UNet-corrected SST (UNet), and ORAS5 reanalysis data.}\label{fig: sst_monthly_2022}        
        \end{figure}

We select the year 2022 from the test period to analyze the projections month-by-month.
Figure~\ref{fig: sst_monthly_2022} displays the monthly SST for 2022 from the reanalysis (ORAS5), raw projections of CNRM-CM6, and UNet corrected projections in the Bay of Bengal for SSP2-4.5. Similar analyses of the corrected projections of SSP1-2.6, SSP3-7.0, and SSP5-8.5 are also shown in the supplementary information section S2.1.

\paragraph{Winter} December marks the establishment of winter conditions with ORAS5 showing temperatures of 28-29$^{\circ}$C in the northern bay and 29-30$^{\circ}$C in the southern regions. The UNet-corrected SST projections demonstrate strong agreement with these observations. In contrast, the raw CNRM-CM6 projection has a cold bias, with values of 26-27$^{\circ}$C in the northern region and 28-30$^{\circ}$C in the southern region, as shown in Fig.~\ref{fig: sst_monthly_2022}(1, 7, 13). January shows a clear temperature gradient, with ORAS5 showing 26-28$^{\circ}$C in the northern BoB and 29-30$^{\circ}$C in the southern regions (Fig.~\ref{fig: sst_monthly_2022}(2, 8, 14)). This period is characterized by the winter monsoon current (WMC), which flows eastward south of Sri Lanka and India, creating a distinct temperature band $>29.5^{\circ}$C along its path in both ORAS5 and UNet SST. February shows a similar transitional warming pattern as shown in Fig.~\ref{fig: sst_monthly_2022}(3, 9, 15). Throughout winter, UNet corrections demonstrate strong agreement with ORAS5 observations.


\paragraph{Pre-monsoon} The pre-monsoon warming phase starts in March (Fig.~\ref{fig: sst_monthly_2022}(4, 10, 16)), seen in ORAS5 temperatures ranging from 28-29$^{\circ}$C in the northern bay to 29-30$^{\circ}$C in the southern regions. Warming begins in the eastern bay, with temperatures exceeding 29$^{\circ}$C. April (Fig.~\ref{fig: sst_monthly_2022}(5, 11, 17)) experiences intense warming in the basin, with ORAS5 showing 29-30$^{\circ}$C in the northern regions and 30-31$^{\circ}$C in the southern and central regions. Warm circulation in spring develops fully during this month, with temperatures exceeding 31$^{\circ}$C in the central and eastern bays, which is crucial for the onset of the monsoon. May (Fig.~\ref{fig: sst_monthly_2022}(6, 12, 18)) represents peak pre-monsoon conditions, and ORAS5 shows temperatures reaching 30-31$^{\circ}$C in the northern, central, and eastern bays. UNet-corrected SST closely matches these ORAS5 patterns, while CNRM-CM6 consistently underestimates SST (Fig.~\ref{fig: sst_monthly_2022}).

\paragraph{Monsoon} June (Fig.~\ref{fig: sst_monthly_2022}(19, 25, 31)) marks the onset of the monsoon, where ORAS5 SST shows temperatures of 29-30$^{\circ}$C in the southern bay and 30-31$^{\circ}$C in the northern regions. The Summer Monsoon Current (SMC) is established south of Sri Lanka, while the Western Boundary Current flows northwards as a warm coastal current along the Indian coast. July (Fig.~\ref{fig: sst_monthly_2022}(20, 26, 32)) and August (Fig~\ref{fig: sst_monthly_2022}(21, 27, 33)) maintain temperatures of 29-30$^{\circ}$C throughout most of the bay, with the southeastern regions reaching 29$^{\circ}$C. The surface SMC temperature patterns, such as the distinct thermal signature south of Sri Lanka, are well represented in the UNet-corrected SST, with values close to ORAS5 reanalysis. In contrast, the CNRM-CM6 projection underestimates this feature in July and fails to capture it in August. September (Fig.~\ref{fig: sst_monthly_2022}(22, 28, 34)) is the start of monsoon withdrawal, showing temperatures from 30-31$^{\circ}$C in the northern bay to 29-30$^{\circ}$C in the southern region.

\paragraph{Post-monsoon} October (Fig.~\ref{fig: sst_monthly_2022}(23, 29, 35)) shows post-monsoon warming patterns, with ORAS5 showing temperatures of 30-31$^{\circ}$C in the northern bay and $<30^{\circ}$C in the southern region. The East India Coastal Current (EICC) begins its seasonal reversal to southward flow. November (Fig.~\ref{fig: sst_monthly_2022}(24, 30, 36)) marks the transition to winter cooling, with temperatures ranging from 28-29$^{\circ}$C in the northern bay to 29-30$^{\circ}$C in the southern region. The EICC fully reverses to southward flow ($27^{\circ}$C along the Indian north coast), while the Andaman Sea region maintains its warm circulation above 29$^{\circ}$C. 

Throughout 2022, UNet-corrected projections show better agreement with ORAS5 reanalysis, while CNRM-CM6 consistently underestimates temperatures and fails to capture the important features of the region.

          \begin{figure}[htbp]
            \centering
            \includegraphics[width=1\linewidth]{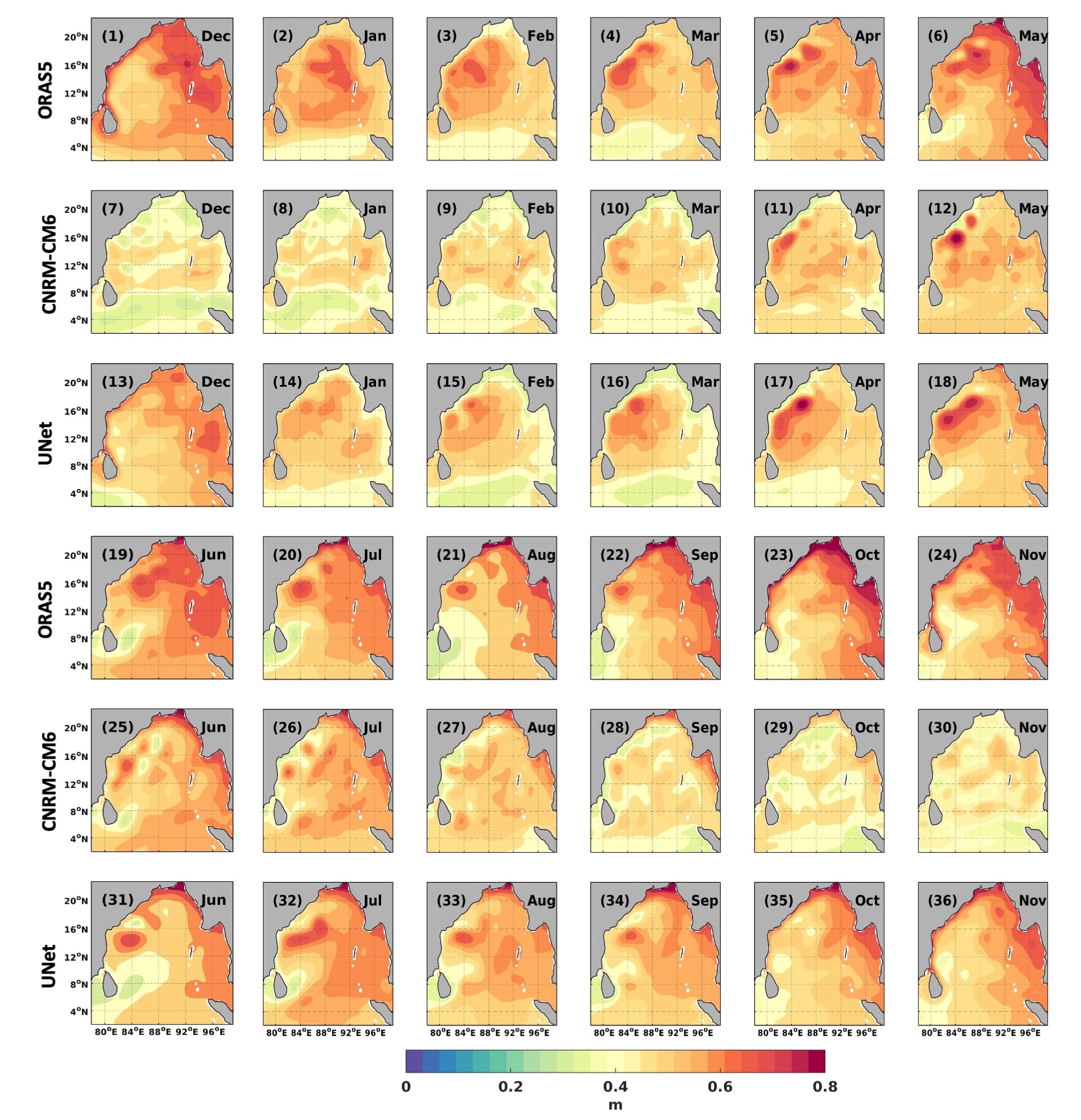}
            \caption{Monthly dynamic sea level (DSL) in the BoB during 2022, comparing raw CNRM-CM6 model output (CNRM-CM6), UNet-corrected DSL (UNet), and ORAS5 reanalysis data.}\label{fig: dsl_2022_monthly}        
        \end{figure}

\subsection{Comparison of DSL projections in 2022}
Figure~\ref{fig: dsl_2022_monthly} displays the monthly DSL for 2022 from the reanalysis (ORAS5), raw projections of CNRM-CM6 (SSP2-4.5), and UNet-corrected projections (USSP2-4.5) in the Bay of Bengal. Same as our SST analysis, we report only SSP2-4.5 in the main paper, and provide similar analyses of the corrected projections of SSP1-2.6, SSP3-7.0, and SSP5-8.5 in the supplementary information section S2.1.
        
\paragraph{Winter} The winter season (December-February) in the Bay of Bengal has DSL values of 0.4-0.6~m in the central bay, a distinct maximum (0.5-0.6~m) centered around 15-18$^{\circ}$N, 85-90$^{\circ}$E, with a secondary peak along the eastern boundary near the Andaman Sea, as seen in the ORAS5 reanalysis (Fig.~\ref{fig: dsl_2022_monthly}(1, 2, 3)). UNet-corrected DSL accurately captures this dual maximum structure and the associated spatial patterns as seen in Fig.~\ref{fig: dsl_2022_monthly}(13, 14, 15), while CNRM-CM6 significantly underestimates these features, showing only weak values (0.3-0.5~m) as seen in Fig.~\ref{fig: dsl_2022_monthly}(7, 8, 9).

\paragraph{Pre-monsoon} The pre-monsoon (March-May) season is characterized by a reorganization of the DSL patterns compared to winter. The ORAS5 reanalysis shows that in March (Fig.~\ref{fig: dsl_2022_monthly}(4)), the DSL values are 0.5-0.6~m in the central bay around 15-18$^{\circ}$N, 85-90$^{\circ}$E. In the next two months (Fig.~\ref{fig: dsl_2022_monthly}(5, 6)), there is progressive intensification with DSL reaching maximum values (0.7-0.8~m) in the central and eastern regions. This development coincides with an increase in spring eddy activity \cite{chen2018origins}. Peak pre-monsoon conditions occur in May with maximum DSL values observed throughout the year. Multiple mesoscale eddies are evident through localized DSL variations, particularly in the central and western regions. Throughout the season, enhanced boundary current activity is reflected in higher DSL values (0.6-0.7~m) along the eastern boundary. The western boundary of BoB consistently maintains elevated DSL values (0.6-0.7~m), indicating strong boundary current activity (WBC flow) along the Indian coast. UNet-corrected DSL successfully captures these complex features, including the warm circulation structure and eddy patterns (Fig.~\ref{fig: dsl_2022_monthly}(16, 17, 18)). In contrast, the raw CNRM-CM6 projections significantly underestimate intensity and project a different spatial structure (Fig.~\ref{fig: dsl_2022_monthly}(10, 11, 12)). 

\paragraph{Monsoon} The monsoon season (June-September) exhibits complex DSL patterns influenced by strong monsoon winds and ocean currents (WBC and the Summer Monsoon Current). The ORAS5 reanalysis shows a well-defined SMC signature (Sri Lankan Dome) near Sri Lanka and the WBC along the coast of India, as shown in Fig.~\ref{fig: dsl_2022_monthly}(19, 20, 21, 22). The western boundary exhibits increased variability with alternating bands of higher (0.5-0.6~m) and lower (0.3-0.4~m) DSL values, reflecting enhanced eddy activity during the monsoon. UNet-corrected DSL successfully captures these patterns as shown in Fig.~\ref{fig: dsl_2022_monthly}(31, 32, 33, 34), including mesoscale features and boundary current influences, demonstrating that the UNet correction provides significant improvements over the raw CNRM-CM6 projections (Fig.~\ref{fig: dsl_2022_monthly}(25, 26, 27, 28)) to represent monsoon dynamics.

\paragraph{Post-monsoon} The post-monsoon period is characterized by coastal Kelvin wave propagation along the eastern boundary, resulting in higher DSL values along the boundary. As seen in the ORAS5 reanalysis (Fig.~\ref{fig: dsl_2022_monthly}(23, 24)), the EICC reversal starts, indicated by elevated levels of DSL near the Indian coast. The presence of eddies is indicated by zones of varying DSL intensities in the western and central regions. UNet-corrected DSL accurately captures these transitional features (Fig.~\ref{fig: dsl_2022_monthly}(35, 36)), whereas raw CNRM-CM6 projection continues to show systematic errors (Fig.~\ref{fig: dsl_2022_monthly}(29, 30)). 

The east-west asymmetry in DSL patterns persists annually--the eastern boundary consistently displays higher values than the western counterpart. This aspect is inadequately captured by the raw CNRM-CM6 DSL projections. In contrast, our UNet-corrected DSL accurately replicates these characteristics year-round.

\section{Analysis of future projections}\label{sec:analysisfuture}
We analyze the corrected SST and DSL projections to study how monthly climatology (annual cycle) is affected by climate change. Subsequently, we present a detailed discussion of the EOF analysis of the raw and corrected CNRM-CM6 projections, along with near-, mid-, and far-future projections of the SST and DSL in the region. This paper defines ``near-future projections (near of the century)" as the average for 2040-2059, ``mid-future projections (mid of the century)" as the average for 2060-2079, and ``far-future projections (end of the century)" as the average for 2080-2099. 
  \subsection{Annual cycle of spatially averaged SST and DSL of raw and corrected projections for different SSPs}
    \label{sec: SSTana}
        \begin{figure}[ht]
            \centering
            \includegraphics[width=\textwidth]{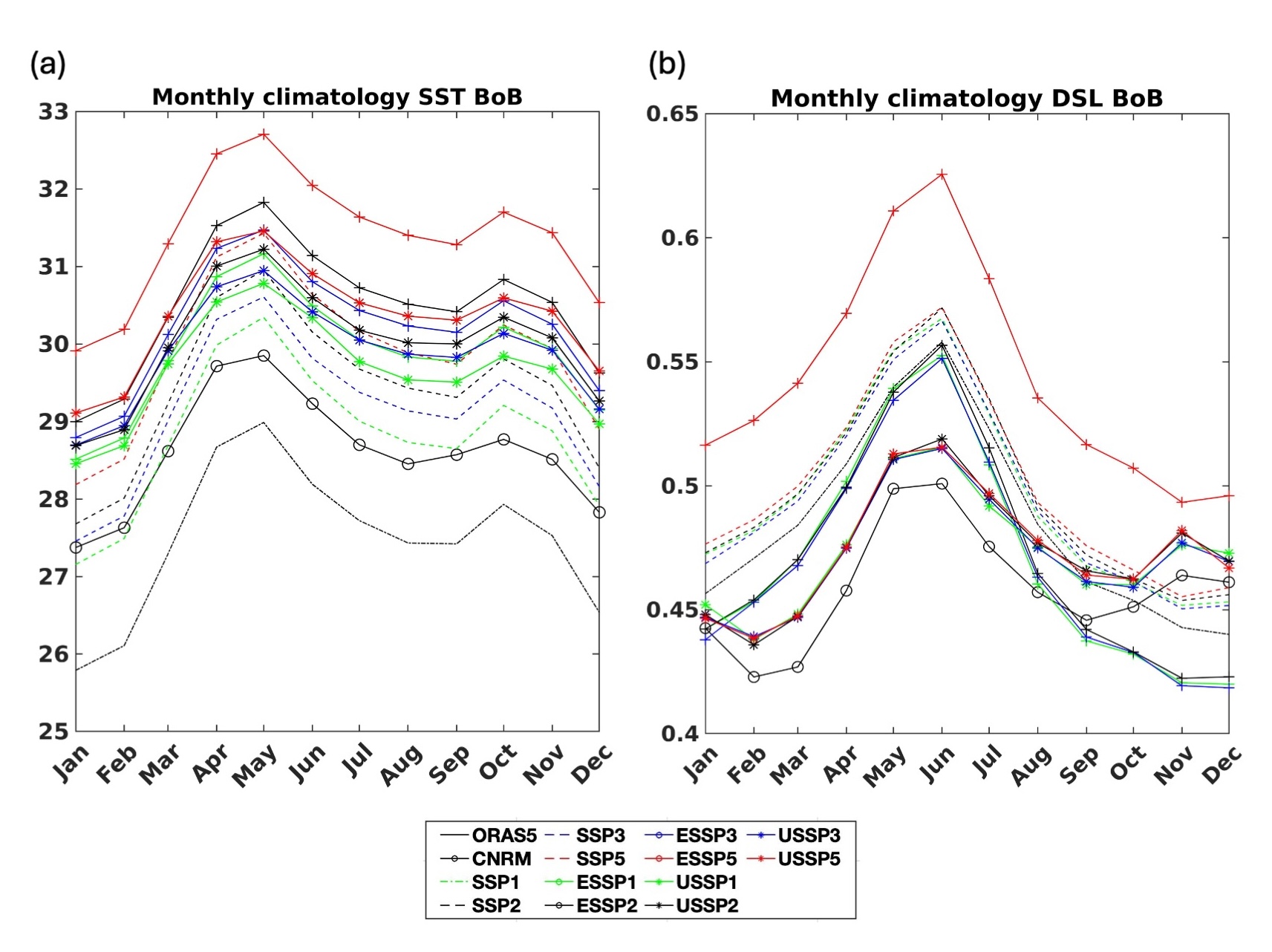}
            \caption{Monthly climatology of (a) SST, and (b) DSL by CNRM-CM6, corrected CNRM-CM6 by EDCDF, and UNet.}\label{fig: monthly_clim}        
        \end{figure}
        Figure~\ref{fig: monthly_clim}(a) presents the monthly climatology of the spatially averaged SST from 1958 to 2100 to study the differences in the projected annual cycle. The data are from reanalysis (ORAS5), historical CNRM-CM6 simulations (Hist. CNRM-CM6), raw CNRM-CM6 projections with four SSP scenarios, and CNRM-CM6 projections corrected by UNet (USSP) and EDCDF (ESSP) for each SSP scenario. The monthly climatology for the reanalysis (ORAS5) is computed from 1958 to 2023. For historical CNRM-CM6, the climatology is computed from 1958 to 2014. For the raw SSP and all corrections (ESSP and USSP), the climatology is computed from 2015 to 2100. 

        SST in the Bay of Bengal demonstrates an annual cycle with a maximum mean temperature in the Boreal Summer (around May) and a minimum mean temperature in the Boreal Winter (around January). The Bay of Bengal cools from its May high during the summer monsoon (JJAS) and then briefly rises to a post-monsoon high temperature in October before cooling again in winter. All SSP scenarios faithfully reproduce the annual cycle of the spatial mean SST, with more extreme SSP scenarios showing an overall warming pattern compared to less extreme SSP scenarios. The difference between historical CNRM-CM6 and ORAS5 is striking, indicating the need for correction. For every SSP future projection scenario, the EDCDF correction overestimates the warming, and the UNet correction places the projection between the raw projections and the EDCDF corrected projections. For SSP5, the mean May SST from the EDCDF-corrected projection is 32.8$^{\circ}$C, compared to 31.3$^{\circ}$C given by the UNet corrected projection.

        A similar analysis is also performed for spatially averaged DSL (Fig.~\ref{fig: monthly_clim}(b)). The monthly climatology of the spatially averaged DSL shows that the highest DSL is during the summer monsoon (June) and the lowest is during the boreal winter (January). The DSL values range from 0.42-0.45~m in winter (January- February), gradually increasing through spring, and reaching peak values of 0.5-0.62~m in June-July. The USSP5-8.5 scenario exhibits the highest spatially averaged DSL values throughout the year, with a pronounced peak of approximately 0.62~m in June. The CNRM-CM6 projections (dashed lines) generally show higher values compared to their corrected counterparts, particularly during the monsoon season. Although historical CNRM-CM6 captures the overall seasonal trend, it consistently shows higher values compared to ORAS5 reanalysis, indicating a bias that requires correction. Compared to the DSL projections among the raw CNRM-CM6 and corrected DSL (UNet and EDCDF), EDCDF-corrected DSL exhibits greater variability and higher values, especially in SSP5-8.5.  The intraannual variability of the EDCDF projection aligns with the CNRM-CM6 pattern. The UNet corrections significantly modify the CNRM-CM6 projections, generally reducing the DSL values and providing more realistic seasonal variations. Corrections are most pronounced during the monsoon season (June-September), where the raw CNRM-CM6 projections tend to overestimate DSL values. After the monsoon peak, all scenarios show a gradual decrease in DSL values from August to December, returning to winter minimum values.

\subsection{Interannual Variability of Seasonal SST}

\begin{figure}[htbp]
            \centering
            \includegraphics[width=0.95\linewidth]{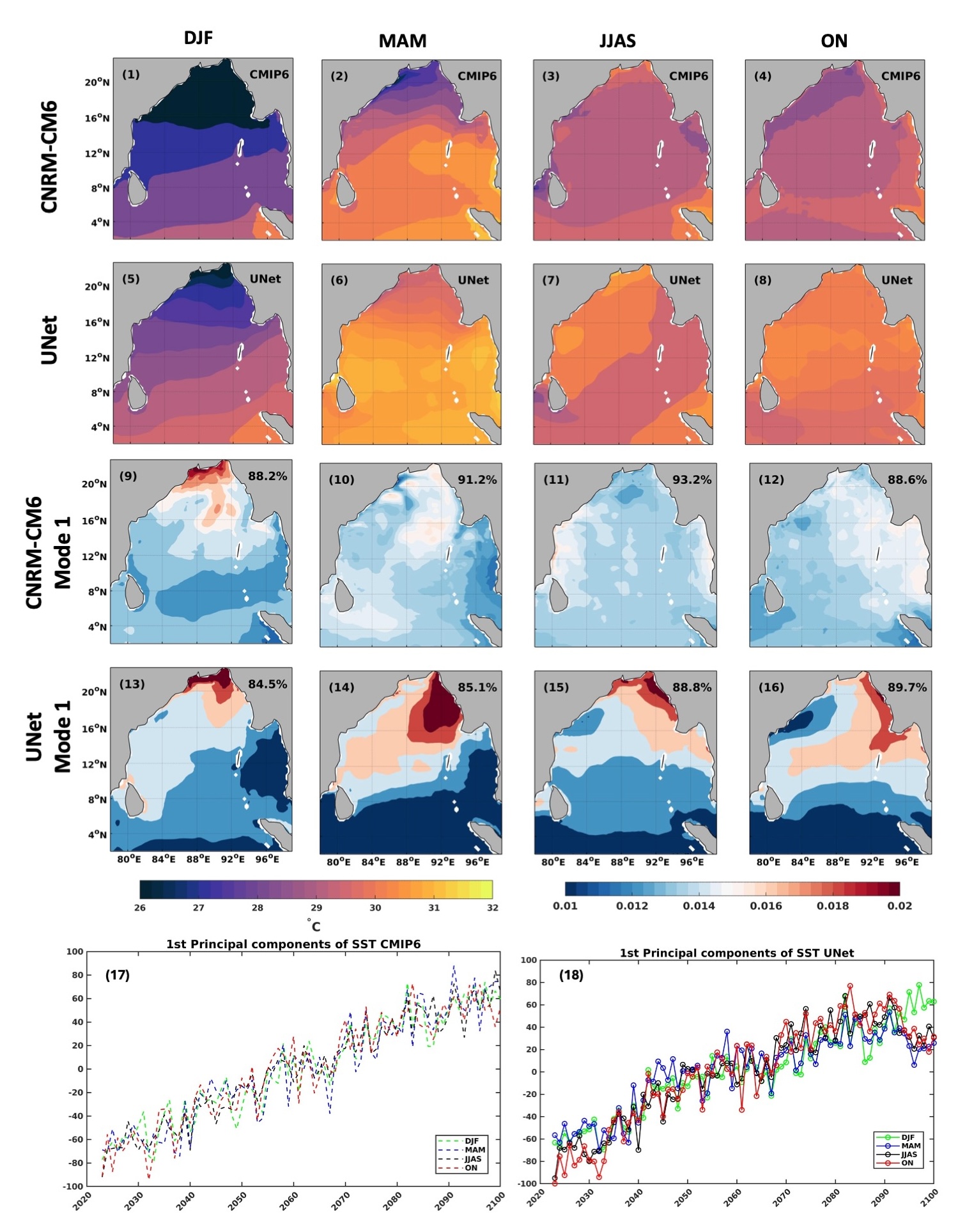}
            \caption{ Seasonal SST temporal evolution from 2023 to 2100:
            Seasonal mean SST over the Bay of Bengal comparing raw CNRM-CM6 (1-4) and UNet-corrected (5-8) projections across DJF (winter), MAM (pre-monsoon), JJAS (monsoon), and ON (post-monsoon) seasons. First EOF modes of CNRM-CM6 (9-12) with variances (DJF: 88.2\%, MAM: 91.2\%, JJAS: 93.2\%, ON: 88.6\%) and UNet-corrected projections (13-16; DJF: 84.5\%, MAM: 85.1\%, JJAS: 88.8\%, ON: 89.7\%). First Principal Component time series for CNRM-CM6 (17) and UNet corrected (18) SST .}\label{fig: sst_m1}
        \end{figure}

For analyzing the spatiotemporal patterns of seasonal SST (winter, pre-monsoon, monsoon, and post-monsoon) from 2023-2100, we performed an EOF analysis for each season using raw CNRM-CM6 and UNet-corrected CNRM-CM6 projections for SSP2-4.5, SSP3-7.0, and SSP5-8.5 scenarios. We report our results for SSP2-4.5, SSP3-7.0,  and SSP5-8.5. Fig.~\ref{fig: sst_m1} shows the mean, mode 1, and principal component 1 of the raw and corrected CNRM-CM6 projections for all seasons. Each column represents one season. The PC for all seasons is combined in panels 17 and 18. In addition, to better understand the implications of the warming, we visualize the raw and corrected mean SST for the near (2040-2059), mid (2060-2079), and far (2080-2100) future periods (Fig.~\ref{fig: sst_proj}) for all seasons. The analysis for each season follows.

\subsubsection{Winter Season}

The mean winter SST (2023 to 2100) is less than $26^{\circ}$C in the northern bay and greater than $29.5^{\circ}$C in the southern region (Fig.~\ref{fig: sst_m1}(1)). However, the UNet-corrected projection (UNet) shows higher temperatures throughout the bay, with the northern region showing temperatures $1^{\circ}C$ warmer than the raw CNRM-CM6. In particular, the spatial pattern of the mean UNet-corrected projection shows enhanced gradients, particularly in the northern bay, suggesting modifications in coastal/bay circulation patterns.

The spatial patterns of the first EOF mode reveal substantial differences between the raw and UNet-corrected SST projections.  Mode 1 of raw CNRM-CM6 explains 88.2\% of the total variance (Fig.~\ref{fig: sst_m1}(9)), while Mode 1 of UNet corrected CNRM-CM6 explains $84.5\%$ of the variance (Fig.~\ref{fig: sst_m1}(13)). In CNRM-CM6, Mode 1 shows maximum variability in the northern bay, reflecting the strong influence of the winter monsoon and freshwater influx (Fig.\ref{fig: sst_m1}(9)). However, the UNet-corrected mode 1 for SSP2-4.5 shows higher variability in the northern and western regions, suggesting more complex interactions between the winter monsoon and coastal processes (Fig.\ref{fig: sst_m1}(13)) than in the raw projections.

For DJF, both raw and corrected PC 1 show an upward trend, indicating a gradual warming of the winter SST. The temporal pattern is similar for both, highlighting that the corrections have maintained the interannual variability while enhancing the spatial pattern of mode 1. 

Similar to SSP2-4.5, for the SSP3-7.0 and SSP5-8.5 scenarios, the northern bay (16-20$^{\circ}$N) shows the strongest variability, with UNet-corrected projections showing more intense warming in this region than the raw SSPs. The central bay (12-16$^{\circ}$N) shows moderate variability in the raw projections but enhanced patterns in the UNet-corrected projections. The southern bay (4-12$^{\circ}$N) exhibits relatively stable warming patterns in raw projections, but shows increased warming and variability in UNet-corrected projections, indicating potential changes in equatorial interactions that affect WMC. 

        \begin{figure}[htbp]
            \centering
            \includegraphics[width=0.85\textwidth]{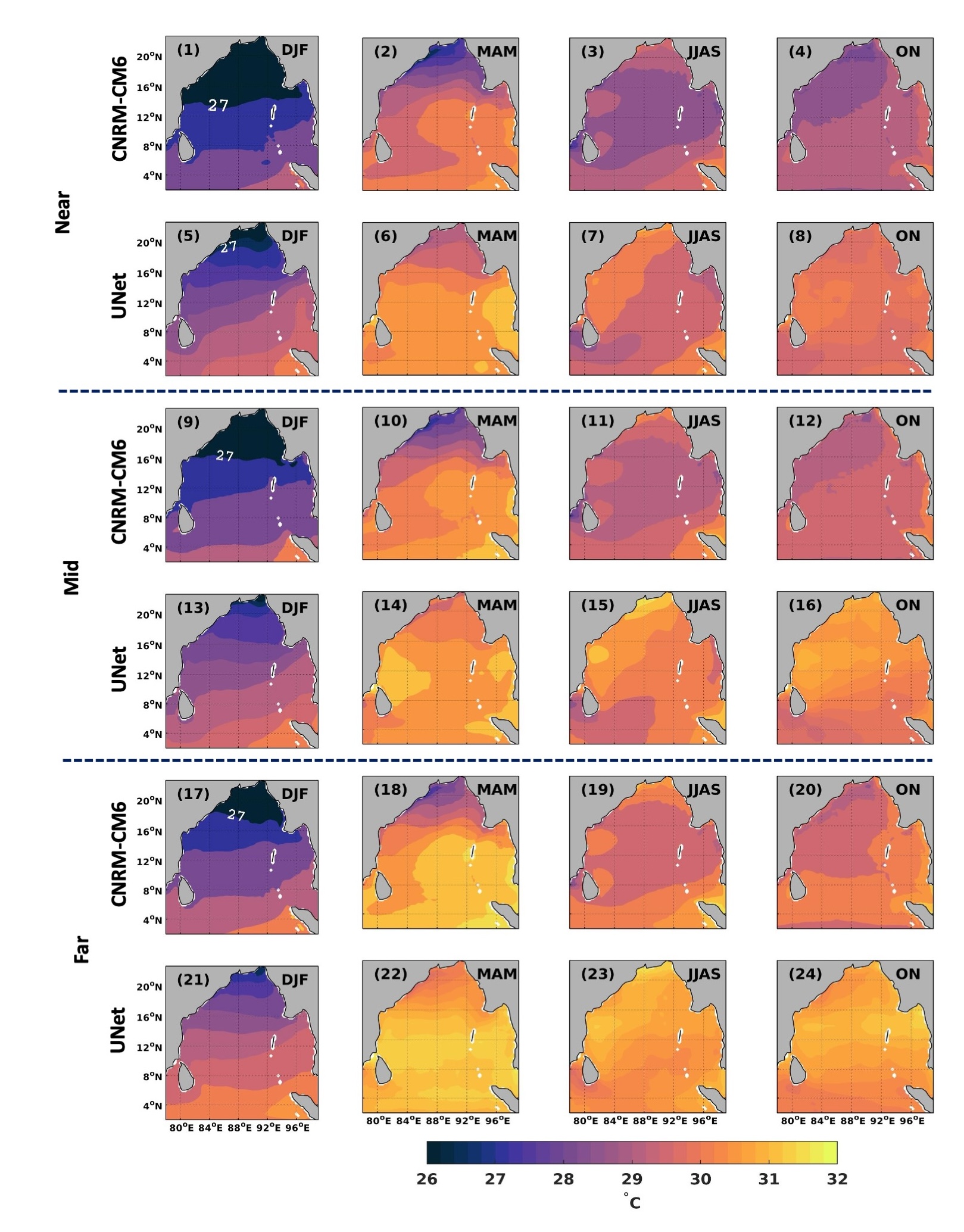}
            \caption{Seasonal SST patterns over the Bay of Bengal comparing raw CNRM-CM6 SSP2-4.5 and UNet-corrected projections for Near future of CNRM-CM6 (1-4) and UNet (5-8), Mid future of CNRM-CM6 (9-12) and UNet (13-16), and Far future of CNRM-CM6 (17-20) and UNet (21-24). Each row shows seasonal means for DJF (winter), MAM (pre-monsoon), JJAS (monsoon), and ON (post-monsoon) seasons.}\label{fig: sst_proj}        
        \end{figure}


For the near future, the raw projections show temperatures ranging from 26$^{\circ}$C in the northern bay to 28.5$^{\circ}$C in the southern region (Fig.~\ref{fig: sst_proj}(1)), reflecting the effect of winter surface cooling in the region \citep{shetye1996hydrography}. However, the UNet-corrected projections show an intense SST pattern with significantly reduced winter cooling effects. Specifically, the UNet-corrected SST is $\approx1^{\circ}$ C higher than the corresponding raw projections, particularly in the central and southern regions shown in Fig.~\ref{fig: sst_proj}(5). For the mid-future period, the raw projections show progressive warming with consistently higher SST values than for the near-future period. The corresponding UNet-corrected projections show accelerated warming compared to the raw projections, with $\approx1.5^{\circ}$C higher SST compared to the raw projections. The far future period shows more significant changes, demonstrating the most extreme winter warming with high temperatures in both the raw projections (Fig.\ref{fig: sst_proj}(17)) and UNet-corrected projections (Fig.\ref{fig: sst_proj}(21)). Specifically, the far-period corrected mean is $\approx2.5^{\circ}$ C warmer. 

\paragraph{Dynamical implication}
These projected changes have significant implications for winter dynamics in the Bay of Bengal. Specifically, changes in winter monsoon characteristics and freshwater influence in the northern bay may differ from the belief based on raw CNRM-CM6 projections. Northern Bay (16-20$^{\circ}$N) shows the strongest warming between the Near and Far periods, with UNet corrections predicting more intense warming than raw projections. The 27$^{\circ}$C isotherm acts as a significant thermal boundary in the northern bay during the winter season, typically positioned around 14$^{\circ}$N in the CNRM-CM6 projections for the three periods. Critically, in the UNet-corrected projections, this isotherm shows a gradual northward shift and completely disappears in the far future period. This increase in temperature in the northern region leads to a weakening of the north-south temperature gradient (as the northern bay warms faster than the southern regions), which could reduce the thermal gradient that traditionally drives the dynamics of the northeast monsoon \citep{prasanna2020precursors}. However, a warmer SST in the central and southern bays could enhance evaporation and moisture availability and eventually intensify rainfall events (extreme rainfall events) when other favorable conditions align during the winter monsoon.

 \subsubsection{Pre-monsoon Season}
               
        The mean pre-monsoon SST (2023 to 2100) in raw CNRM-CM6 projections (Fig.~\ref{fig: sst_m1}(2)) shows a distinctive warming pattern with temperatures ranging from $<29^{\circ}$C in the northern bay to $30-31^{\circ}$C in the southern and eastern regions. However, the UNet-corrected projections (Fig.~\ref{fig: sst_m1}(6)) reveal more intense warming patterns, with higher temperatures throughout the bay. In particular, the central and eastern regions show an enhanced warming compared to raw CNRM-CM6, suggesting a modification in the general circulation and coastal upwelling. 

        The spatial patterns of the first EOF mode demonstrate substantial differences between the raw and UNet-corrected SST projections. Mode 1 of raw CNRM-CM6 SST for MAM explains 91.2\% of the total variance (Fig.~\ref{fig: sst_m1}(10)), reflecting the typical variability of the pre-monsoon warming with warmer temperatures in the eastern bay due to the developing summer monsoon conditions. In contrast, the UNet-corrected mode 1  explains only 85.1\% of the variance (Fig.~\ref{fig: sst_m1}(14)), showing greater variability in the central and eastern regions, suggesting modified dynamics in the deep sea and coast. Specifically, there is enhanced variability in the coastal warming, leading to changes in coastal upwelling characteristics. Recent research suggests that the BoB warming during the pre-monsoon contributes to a weaker monsoon circulation, resulting in a reduction in rainfall during the Indian summer monsoon \citep{goswami2022role}. Thus, the correction paints a more worrying spatial pattern than the raw CNRM-CM6, suggesting that monsoon onset and propagation patterns over India are likely to be different from the raw CNRM-CM6 projections.

        Like DJF, PC 1 of MAM of the raw and UNet-corrected CNRM-CM6 projections show similar variability (Fig.~\ref{fig: sst_m1}(17)).

        In the near-future period, the mean raw CNRM-CM6 projection shows the SST ranging from $28-29^{\circ}C$ in the northern bay to $29-30^{\circ}C$ in the southern regions (Fig.~\ref{fig: sst_proj}(2)). The near-future UNet-corrected SST projections show temperatures approximately $1.2^{\circ}C$ higher than the raw SST projections, particularly pronounced in the eastern and central regions. The mid-future reveals progressive warming, with raw projections showing approximately $1^{\circ}$C higher SST than near-future, whereas the UNet-corrected projection for the same period shows approximately $2^{\circ}C$ higher SST near-future raw CNRM-CM6 projection. The far future period reveals the most extreme scenarios, with raw SSP2-4.5 showing temperatures approximately $1.5^{\circ}C$ higher than near SSP2-4.5. In contrast, far-future UNet-corrected projection shows temperatures approximately $2.5^{\circ}C$ higher than near SSP2-4.5.

        Similarly, SSP3-7.0 and SSP5-8.5 also show further enhanced warming during the pre-monsoon season (Supplementary Information Section S3.1).
         
        \paragraph{Dynamics implication}
        The basin-wide warming, especially in the central bay (increasing by $\approx2-3^{\circ}$C in the far-future period), could significantly impact the pre-monsoon conditions necessary for monsoon onset. The traditional north-south temperature gradient that helps establish the monsoon circulation appears modified, with a more uniform warming that potentially delays or weakens the monsoon onset process \citep{choudhury2019impact}. The increased warming in the eastern bay (reaching $31.5-32^{\circ}C$) could modify the formation and propagation of pre-monsoon depressions, while the reduced temperature gradients along the western boundary could affect the intensity of pre-monsoon rainfall along the eastern Indian coast. These changes suggest a potential shift in the pre-monsoon environment that could significantly influence the timing and intensity of the subsequent onset of the summer monsoon \citep{choudhury2019impact}.

        There are also significant implications for pre-monsoon cyclogenesis in the Bay of Bengal. Warming patterns show a concerning trend. In the near-future period, temperatures already reach $30-30.5^{\circ}$C in the central and southern bays, providing enhanced surface support for cyclone development during this season (Fig.\ref{fig: sst_proj}). Warming intensifies through the mid- and far-future periods, with the far-future showing a UNet-corrected SST exceeding $31.5^{\circ}$ C across most of the bay. This substantial warming, particularly in the central bay (12-16$^{\circ}$ N), where cyclogenesis occurs frequently \citep{neetu2019premonsoon}, could significantly affect the characteristics of the cyclones \citep{vissa2013intensity}. The increased warming along the eastern boundary ($\approx2.5-3^{\circ}$C in far-future) could also modify the moisture content and atmospheric instability patterns that typically support pre-monsoon cyclogenesis.
        
        In addition to an increase in cyclogenesis, the enhanced pre-monsoon warming predicted by the UNet-corrected SST in the far-future period could result in cyclonic systems becoming more frequent and severe \citep{ray2024role}. The eastern bay, traditionally a region of cyclogenesis during MAM \citep{neetu2019premonsoon}, shows a particularly pronounced warming ($2.5-3.5^{\circ}$C increase in the far-future), which could alter the spatial patterns of cyclone formation. The reduced temperature gradients between coastal and offshore waters seen in the far-future UNet-corrected projections could affect cyclone tracks and intensification patterns (Fig.\ref{fig: sst_proj}(22)). Furthermore, the warming of the northern bay (16-20$^{\circ}$N) could extend the spatial domain favorable to the intensification of cyclones \citep{ray2024role}, potentially allowing systems to maintain their strength further north than historically observed \citep{neetu2012influence} and the raw projections of CNRM-CM6.  
        
        Overall, the UNet corrected projections indicate a possible rise in the number and strength of pre-monsoon cyclones, with the far-future period having conditions favoring quicker intensification and possibly more damaging systems. This is especially worrisome for coastal areas around the bay, as stronger pre-monsoon cyclones could elevate risks to human communities and coastal ecosystems. Even more worrisome is that the SSP3-7.0 and SSP5-8.5 pathways show more intense warming after the correction by our UNet model indicating an increased risk than the raw projections (Supplementary Information Section S3.2).

    \subsubsection{Monsoon Season}

        The mean monsoon SST (2023 to 2100) in raw CNRM-CM6 projections show temperatures ranging from $\approx28.5^{\circ}$C in the northern bay to $\approx30^{\circ}$C in the southern region (Fig.\ref{fig: sst_m1}(3)). This mean pattern reflects monsoon-induced cooling, with relatively cooler temperatures throughout the bay due to cloud cover and rainfall during the monsoon. However, the UNet-corrected projections reveal weakened monsoon cooling, with higher temperatures throughout the bay, particularly in the northern and central regions, where temperatures are $\approx1^{\circ}$C warmer than the raw CNRM-CM6 (Fig.\ref{fig: sst_m1}(7)). Along the western coast (Indian coast), the raw projections show temperatures of $\approx29-29.5^{\circ}$C with upwelling signatures, while the UNet-corrected projections show temperatures of $\approx29.5-30^{\circ}$C, potentially affecting upwelling patterns and coastal productivity. Further warming is observed on the eastern coast, with temperatures reaching $30^{\circ}$C, creating two distinct warming zones in the western and eastern bays.

        Mode 1 of the raw CNRM-CM6 SST for JJAS explains 93.2\% of the total variance and shows the lowest variability in the central bay during the strong influence of monsoon processes (Fig.\ref{fig: sst_m1}(11)). However, the UNet-corrected mode 1 explains only 88.8\% of the variance, showing increased variability in the eastern and western regions, suggesting more complex interactions between monsoon and coastal upwelling processes (Fig.\ref{fig: sst_m1}(15)). Like DJF and MAM, PC 1 of JJAS of the raw and UNet-corrected CNRM-CM6 projections show similar variability (Fig.~\ref{fig: sst_m1}(17, 18)).

        In the near-future projections, the raw SST projections show monsoon-induced cooling patterns in the region. The mid-future projections reveal basin-wide warming, where the mid-future SSP2-4.5 shows temperatures $1^{\circ}$C higher than the near-future raw SSP2-4.5. The far-future reveals the most intense warming changes, with the far-future SSP2-4.5 showing temperatures $2^{\circ}$ C higher than the near-future raw SSP2-4.5. Spatial patterns show significantly weakened monsoon cooling effects, particularly in the northern and central bays.

        Similarly, SSP3-7.0 and SSP5-8.5 also show a further warming during the monsoon season.  
        
        \paragraph{Dynamics implication}
        These changes have profound implications for monsoon dynamics and regional circulation patterns. The warm front of the coastal BoB SST is crucial for the rainfall in central India \citep{roxy2017threefold, samanta2018impact}. This feature is absent in the raw CNRM-CM6 (Fig.\ref{fig: sst_proj}(3,7)). The distribution and magnitude of the warm coastal SST front are increasing in the UNet-corrected monsoon SST, which requires a more detailed analysis of their impact on Indian summer monsoon \citep{roxy2017threefold}. The monsoon cooling effect, traditionally strongest in the northern bay, appears substantially weakened, potentially impacting the monsoon progression as the north-south temperature gradient that typically drives the monsoon circulation \citep{sheehan2023influences} becomes more pronounced. The increase in basin-wide temperatures (by $\approx2.5-4^{\circ}$ C in the far-future projections) could modify atmospheric moisture content and convection patterns, potentially leading to more intense but spatially irregular rainfall \citep{sakthivel2024sea}(Fig.\ref{fig: sst_proj}(23)). The warming pattern in the eastern bay (reaching 31-32$^{\circ}$C in the far UNet-corrected SSP5-8.5; Supplementary Information S3.3) could affect the formation and propagation of monsoon depressions, while the reduced temperature gradients along the western boundary could affect coastal upwelling and associated biological productivity. These changes suggest a fundamental modification of the monsoon system, with potential implications for regional climate and marine ecosystems.

        The JJAS SST projections suggest significant changes in the SST pattern that affect SMC. The near-future SST projections (Fig.\ref{fig: sst_proj}(7)) show an amplified warming in the region 8$^{\circ}$N-12$^{\circ}$N, where the SMC typically manifests the strongest. This warming trend intensifies in the UNet-corrected projections for the mid-future period, with temperatures reaching $30.5-31^{\circ}$C (Fig.\ref{fig: sst_proj}(15)) in the region south of Sri Lanka (traditionally occupied by the SMC). The SST projections for the far-future period show extreme warming (31-32$^{\circ}$C in UNet-corrected SSP5-8.5 projections) throughout this region. The east-west thermal gradient that is historically shown as the SMC feature \citep{vinayachandran1998monsoon} appears significantly modified and could have the following implications. First, increased temperature gradients have an effect on SMC, which could affect the transport of heat and salt to the Bay of Bengal during the monsoon season. Second, increasing SST could modify the vertical structure of current \citep{thushara2019vertical}. This could affect nutrient transport \citep{jardine2025asymmetric} and biological productivity along the SMC path \citep{jyothibabu2015phytoplankton}. 
                
    \subsubsection{Post-monsoon Season}
    
        The mean post-monsoon SST (2023 to 2100) in raw CNRM-CM6 projections shows a transitional pattern with temperatures ranging from $\approx28.5^{\circ}$C in the northern bay to $\approx29.5^{\circ}$C in the southern region (Fig.\ref{fig: sst_m1}(4)). This pattern reflects the post-monsoon characteristics, with gradual warming in the bay following monsoon withdrawal. The UNet-corrected projections (SSP2) show higher temperatures throughout the bay, particularly in the eastern and central regions, where temperatures are $1^{\circ}$C warmer than the raw projections (Fig.\ref{fig: sst_m1}(8)). The spatial pattern shows enhanced gradients, especially in the eastern bay, suggesting modifications in post-monsoon circulation patterns.

         EOF mode 1 of raw CNRM-CM6 explains 88.6\% of the total variance and shows the highest variability in the eastern bay, reflecting the influence of retreating monsoons and the beginning of winter conditions (Fig.\ref{fig: sst_m1}(12)). The UNet-corrected mode 1 explains 89.7\% of the variance. Crucially, it shows enhanced variability in both the eastern and central regions, suggesting more complex interactions between the retreating monsoon and regional processes (Fig.\ref{fig: sst_m1}(16)). These differences in spatial patterns indicate potential modifications in the post-monsoon transition dynamics that could affect both atmospheric and oceanic processes in the region. Like DJF, MAM, and JJAS, PC 1 of JJAS of the raw and UNet-corrected CNRM-CM6 projections shows similar variability (Fig.~\ref{fig: sst_m1}(17, 18)).

         In the near-future period, the raw CNRM-CM6 SST ranges from $\approx28-29^{\circ}$ C, reflecting typical post-monsoon surface warming after monsoon withdrawal. The near-future UNet-corrected SST projections show temperatures $\approx1.5^{\circ}$ C higher than the raw SST projections, especially along the coasts. The mid-future projections indicate a gradual increase in SST within the basin, with the raw SST showing a $\approx1.2^{\circ}$C higher than the near-future raw SST. The UNet-corrected projections for the mid-future show accelerated warming, with temperatures approximately $\approx2.2^{\circ}$C above the near-future raw SST. The far-future projections reveal the most extreme temperature changes, with raw SST showing temperatures approximately $\approx2^{\circ}$C higher than near-future SST, while UNet-corrected far-future projections show temperatures $\approx3^{\circ}$C above near-future raw SST.
         
        Similarly, SSP3-7.0 and SSP5-8.5 also show a further warming of the post-monsoon SST pattern (Supplementary Information section S3.4).  
        
        \paragraph{Dynamics implication}
        These changes have profound implications for the BoB features. The timing and intensity of the EICC reversal could be affected by the modified temperature gradients, particularly in the far-future SST projections, where the coastal-offshore temperature differences are significantly reduced. Weaker temperature gradients along the western boundary (evident in the far-future) could affect the development and strength of the southward flow current during the winter season \citep{dandapat2018interannual}. 

        The basin-wide warming of $2.5-4^{\circ}$ C in the far-future projections could impact cyclogenesis patterns, as warmer SSTs provide more energy (support) for cyclone development during the post-monsoon \citep{jyoteeshkumar2021impact}. This is particularly concerning for the central and southern BoB, where cyclones usually form during this period \citep{vissa2013intensity}. Reduced temperature gradients and higher SST are attributed to coastal downwelling processes \citep{wentz2000satellite, rao2010interannual}, which can affect the availability of nutrients and biological productivity. The warming pattern on the eastern boundary (reaching $31-32^{\circ}$C in the far-future USSP5) could modify the development of eddies and thermal fronts that typically characterize this region during the post-monsoon season, potentially affecting the overall circulation pattern and heat distribution of the bay. These changes suggest a fundamental modification of post-monsoon oceanographic conditions, with potential implications for both regional climate patterns and marine ecosystem dynamics.

\subsection{Interannual Variability of Seasonal DSL} \label{sec: SSHana}
\begin{figure}[htbp]
            \centering
            \includegraphics[width=0.9\textwidth]{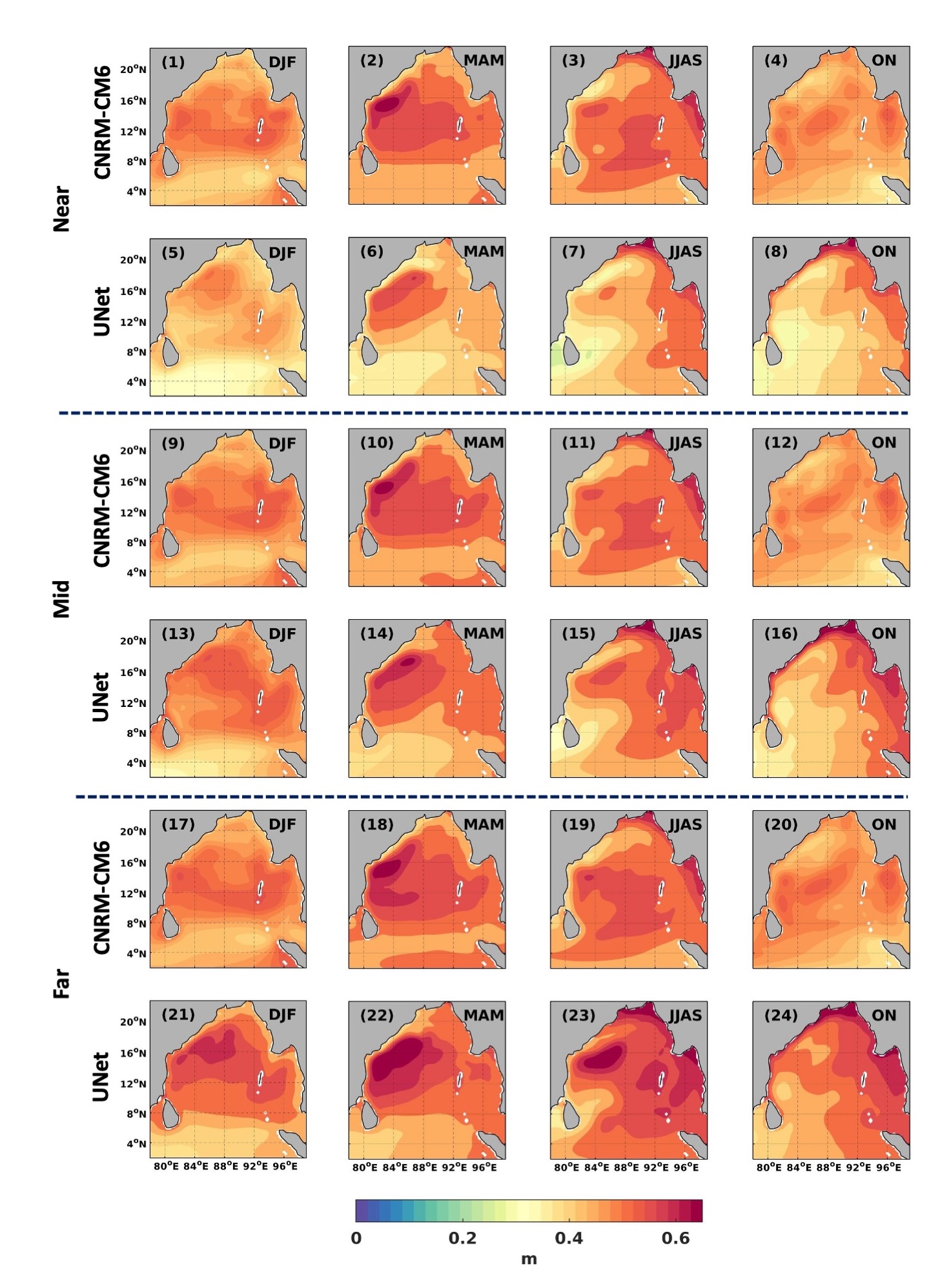}
            \caption{Seasonal DSL patterns over the Bay of Bengal comparing raw CNRM-CM6 SSP2-4.5 and UNet-corrected projections for Near future of CNRM-CM6 (1-4) and UNet (5-8), Mid future of CNRM-CM6 (9-12) and UNet (13-16), and Far future of CNRM-CM6 (17-20) and UNet (21-24). Each row shows seasonal means for DJF (winter), MAM (pre-monsoon), JJAS (monsoon), and ON (post-monsoon) seasons.}\label{fig: dsl_proj}        
        \end{figure}

For analyzing the spatiotemporal patterns of seasonal DSL (winter, pre-monsoon, monsoon, and post-monsoon) from 2023-2100, we performed an EOF analysis for each season using raw and UNet-corrected CNRM-CM6 projections for SSP2-4.5, SSP3-7.0, and SSP5-8.5 scenarios. Like SST, we report our results for SSP2-4.5 (figures in the main document) and note the parallels with SSP3-7.0 and SSP5-8.5 (figures in the supplementary material in section S4). Fig.~\ref{fig: dsl_m1_ssp2} shows the mean, mode 1, and principal component 1 of the raw and corrected CNRM-CM6 DSL projections for all seasons. Each column represents one season. The PC for all seasons is combined in panels 17 and 18. Fig.~\ref{fig: dsl_proj} shows the mean of the raw and corrected CNRM-CM6 DSL projections in the near (2040-2059), mid (2060-2079), and far (2080-2100) future periods. 

\begin{figure}[htbp]
            \centering
            \includegraphics[width=\textwidth]{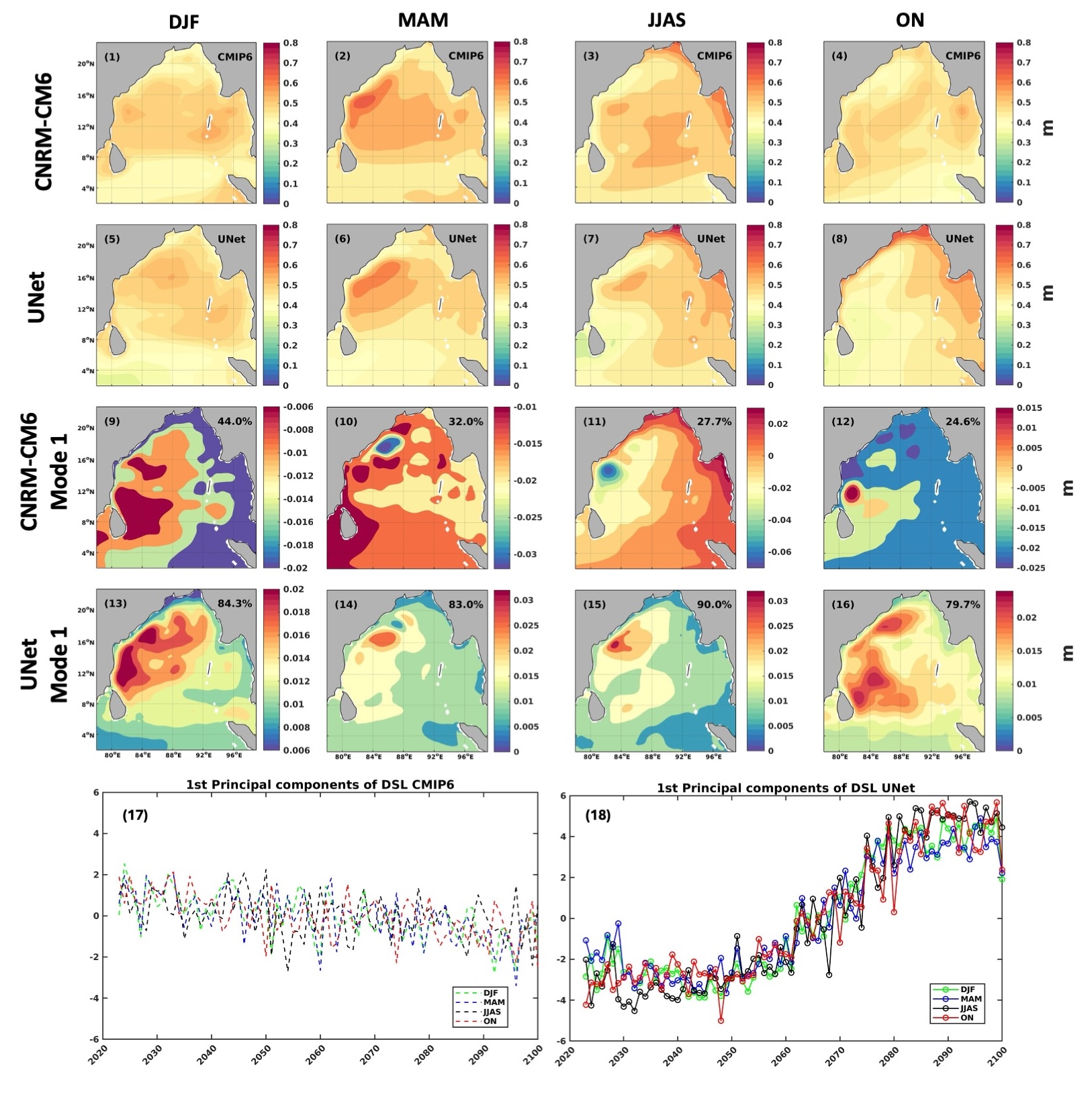}
            \caption{
            Seasonal DSL temporal evolution from 2023 to 2100:
            Seasonal mean DSL over the Bay of Bengal comparing raw CNRM-CM6 (1-4) and UNet-corrected (5-8) projections across DJF (winter), MAM (pre-monsoon), JJAS (monsoon), and ON (post-monsoon) seasons. First EOF modes of CNRM-CM6 (9-12) show higher variance explained (DJF: 44.0\%, MAM: 32.0\%, JJAS: 27.7\%, ON: 24.6\%) compared to UNet-corrected projections (13-16; DJF: 84.3\%, MAM: 83.0\%, JJAS: 90.0\%, ON: 79.7\%). First Principal Component time series for seasonal CNRM-CM6 (17) and UNet corrected (18) DSL.}\label{fig: dsl_m1_ssp2}
        \end{figure}
        
        \subsubsection{Winter Season}
        The mean winter DSL (2023 to 2100) in raw CNRM-CM6 projections shows moderate values ranging from 0.3-0.5~m with a relatively uniform distribution, exhibiting slightly higher values in the central bay (Fig.~\ref{fig: dsl_m1_ssp2}(1)). The UNet-corrected DSL projections reveal enhanced spatial patterns with more distinct features, particularly in the northern region (16$^o$N-20$^o$N) (Fig.~\ref{fig: dsl_m1_ssp2}(5)). Both raw and UNet-corrected DSL show lower values in the southern bay (4$^o$N-8$^o$N), reflecting typical winter circulation patterns. The eastern coastal region maintains moderate DSL values, influenced by the winter coastal current system.

        The first EOF mode spatial patterns demonstrate substantial differences between the raw and UNet-corrected DSL projections. In raw CNRM-CM6, mode 1 explains 44.0\% of the total variance and shows two distinct structures with lower variability in the eastern bay and high variability in the western region (Fig.~\ref{fig: dsl_m1_ssp2}(9)). After UNet correction, the explained variance increases significantly to 84.3\%, with enhanced east-west contrast showing less variability in the central bay and higher DSL variability along the Indian coast (Fig.~\ref{fig: dsl_m1_ssp2}(13)). UNet-corrected DSL patterns better capture winter circulation features, including the East India Coastal Current (EICC) and associated eddy activities, suggesting that UNet corrections improve the representation of important regional oceanographic processes during winter.

        The principal components show notable differences in the temporal evolution between the projections. Although the raw CNRM-CM6 PC exhibits moderate variability (Fig.~\ref{fig: dsl_m1_ssp2}(17)), the UNet-corrected projections show stronger variability with a clear increasing trend after 2060, suggesting an intensification of the winter monsoon circulation pattern (Fig.~\ref{fig: dsl_m1_ssp2}(18)). This temporal difference highlights the potential impact of UNet corrections on capturing long-term changes in winter circulation patterns, with implications for regional climate dynamics and coastal processes in the Bay of Bengal region.

        In raw projections of CNRM-CM6 SSP2-4.5, the near-future period (Fig.\ref{fig: dsl_m1_ssp2}(1)) shows moderate DSL values with a clear north-south gradient, ranging from 0.3-0.4~m in the northern Bay of Bengal to 0.2-0.3~m in the southern region. The mid-future (Fig.\ref{fig: dsl_m1_ssp2}(9)) shows similar spatial patterns but with slightly enhanced values, particularly in the central region. The far-future period (Fig.\ref{fig: dsl_m1_ssp2}(17)) demonstrates further intensification of these patterns, with higher DSL values in the northern and central bay, indicating a strengthening winter circulation pattern. Throughout all periods, the raw CNRM-CM6 projections maintain a consistent spatial structure with maximum values concentrated in the northwestern bay. The UNet-corrected DSL shows distinct modifications to the DSL patterns across all periods, characterized by increased DSL values in the northwest bay and decreased DSL values in the southern bay. In the near-future period (Fig.\ref{fig: dsl_m1_ssp2}(5)), the spatial structure reveals enhanced gradients with more pronounced regional structures compared to raw CNRM-CM6. The mid-future (Fig.\ref{fig: dsl_m1_ssp2}(13)) exhibits intensified DSL patterns, particularly along the western boundary, suggesting stronger winter circulation. The far-future (Fig.\ref{fig: dsl_m1_ssp2}(21)) shows the most dramatic changes, with significantly higher DSL values (0.5-0.6~m) in the northwestern bay and along the western boundary, accompanied by intensified eddy structures. The UNet corrections consistently improve the representation of winter circulation features and coastal dynamics across all time periods.

        \paragraph{Dynamics implication}
        The increased contrast in east-west variability in mode 1 indicates a stronger variability of the EICC, which could affect coastal upwelling and nutrient distribution \citep{vinayachandran2009impact}. PC1 of the UNet-corrected DSL shows the intensity of the winter characteristics in the BoB, particularly after 2060, which could lead to improved winter mixing of freshwater from higher latitudes, modified stratification patterns \citep{shetye1996hydrography}, and altering the nutrient cycle that affects marine productivity \citep{narvekar2006seasonal, sarma2016effects}. The distinct spatial patterns in UNet projections better capture mesoscale features and boundary currents, suggesting more pronounced effects on coastal sea levels, sediment transport, and ecosystem dynamics. The increased variability in Far projections (2080-2100) indicates increased vulnerability to winter extremes, necessitating improved coastal management strategies and ecosystem adaptation measures.

\subsubsection{Pre-monsoon Season}

        The mean pre-monsoon DSL (2023 to 2100) exhibits intensified patterns with values ranging from 0.4-0.6~m. The raw CNRM-CM6 projections show higher DSL values in the central and northern bays, with a notable maximum in the northwestern region, as shown in Fig.\ref{fig: dsl_m1_ssp2}(2). The UNet-corrected projections improve these characteristics, showing more pronounced spatial gradients and stronger DSL values (0.5-0.6~m) in the central bay (Fig.\ref{fig: dsl_m1_ssp2}(6)). This pattern reflects the development of the Spring eddy and associated circulation characteristics in the Bay of Bengal prior to the onset of the monsoon. The enhanced spatial features in the UNet-corrected DSL suggest a better representation of the pre-monsoon oceanographic conditions that play a critical role in setting up the bay for the subsequent monsoon season.

        The first EOF mode spatial patterns reveal substantial differences between the raw and UNet-corrected DSL projections. In raw CNRM-CM6, mode 1 explains 31.2\% of the total variance, with spatial patterns showing maximum variability in the central and eastern bays, reflecting the development of the circular pattern and associated circulation characteristics (Fig.~\ref{fig: dsl_m1_ssp2}(10)). After UNet correction, the explained variance increases substantially to 87.3\%, with enhanced characteristics showing stronger variability and clear central bay maxima, better representing the development of the circulation patterns (Fig.~\ref{fig: dsl_m1_ssp2}(14)). This significant increase in explained variance indicates that UNet corrections capture more coherent spatial patterns of DSL variability during the pre-monsoon season.

        Principal components demonstrate notable differences in temporal evolution between the raw and UNet-corrected DSL. Although the raw CNRM-CM6 PC exhibits moderate variability (Fig.~\ref{fig: dsl_m1_ssp2}(17)), the UNet-corrected projections show stronger variability with a clear increasing trend after 2060 (Fig.~\ref{fig: dsl_m1_ssp2}(18)). This temporal difference suggests that UNet corrections capture more pronounced long-term changes in pre-monsoon circulation patterns, potentially indicating an intensification of the Spring eddy and associated circulation features in the future climate. These changes could have important implications for pre-monsoon oceanographic conditions and their influence on subsequent monsoon development.

        Throughout the three future periods, the raw CNRM-CM6 projections consistently project maximum DSL in central-eastern bay. The UNet corrections reveal modified DSL patterns across all periods, better representing regional features. 
        
        In raw CNRM-CM6 projections, the near-future period shows high DSL values (0.4-0.5~m) in the central and northern bays, with maximum values (0.5-0.6~m) concentrated in the central-eastern region, reflecting the development of the circulation pattern during spring in the Bay of Bengal. In UNet-corrected projections of the near-future, the spatial structure shows more distinct gradients and enhanced representation of the spring warm eddy, with clearer coastal features.
        
        The raw mid-future projections maintain a similar spatial pattern as the near-future but with slightly enhanced values, particularly in the central bay, where the warm eddy influence is strongest. However, the UNet-corrected mid-future demonstrates intensified DSL patterns, particularly in the central bay, suggesting stronger warm circulation development and associated circulation features.
        
        The raw far-future projections show further intensification of the patterns in near- and mid-futures, with expanded areas of high DSL values in the central and northern bay, indicating strengthening pre-monsoon conditions. In contrast, the UNet-corrected far-future exhibits the most significant changes, with markedly higher DSL values (0.5-0.6~m) in the central and eastern bay, showing an enhanced representation of both the eddies and coastal dynamics. 
        
        \paragraph{Dynamics implication}
        The UNet corrections consistently improve the depiction of pre-monsoon oceanographic features, particularly the warm eddy structure and associated eddy activities. These enhanced features have important implications for understanding future changes in the pre-monsoon oceanographic conditions that influence not only the regional circulation but also the subsequent development of the monsoon season.
        
        Intensified DSL patterns in the western bay (0.55-0.6~m) suggest enhanced spring anticyclonic circulation pattern development, affecting pre-monsoon convection and rainfall \citep{roy2011regional, choudhury2019impact}. The westward shift of maximum DSL values (centered around 85$^o$E) indicates modified eddy activity and circulation patterns, which influence MLD and stratification \citep{lin2024evolution}. Enhanced DSL gradients in the southern region influence the development of the BoB warm circulation, which affects monsoon onset conditions \citep{yanliang2013mixed, li2016possible, lin2024evolution}. The increased DSL values along the western boundary (0.5-0.6~m) suggest modified Indian coastal current systems, affecting the intensity of upwelling and the distribution of nutrients \citep{everett2014relative, jana2018sensitivity, madhu2006lack}. These changes affect marine ecosystem productivity and fisheries during the critical pre-monsoon period. Intensified variability of the BoB warm circulation and associated circulation patterns could lead to more intense pre-monsoon cyclones \citep{mishra2019pre, jena2024interdecadal}. The projections indicate increased vulnerability to pre-monsoon extreme events, modified marine productivity patterns, and potential changes in monsoon onset characteristics, suggesting the need for early-season preparedness strategies and adaptive fisheries management.

\subsubsection{Monsoon Season}

        The monsoon season exhibits the highest DSL values (0.5-0.7~m) with distinct spatial patterns. The raw CNRM-CM6 projections show elevated DSL in the northern and eastern regions (Fig.~\ref{fig: dsl_m1_ssp2}(3)), while UNet-corrected projections intensify these features (Fig.~\ref{fig: dsl_m1_ssp2}(7)). Raw and UNet-corrected DSL capture the monsoon-driven circulation, with maximum values along the eastern boundary. The southern bay maintains moderate DSL values, influenced by monsoon circulation. In particular, the UNet-corrected DSL captures the Sri Lankan Dome, a major monsoon feature in the Bay of Bengal. Crucially, the raw CNRM-CM6 DSL projection does not capture this important dynamical feature. These spatial patterns reflect strong seasonal circulation and enhanced coastal dynamics that characterize the monsoon season in the Bay of Bengal.

        In raw CNRM-CM6, Mode 1 explains 27.7\% of the total variance (Fig.~\ref{fig: dsl_m1_ssp2}(11)), with spatial patterns showing a monsoon-driven structure with high DSL values in the eastern bay, central and northern regions. After UNet correction, the explained variance increases dramatically to 90.0\% (Fig.~\ref{fig: dsl_m1_ssp2}(15)), with the spatial mode showing low variability in the western and northern bay before 2060 and higher variability afterward, significantly impacting Western Boundary Current (WBC) and eddy activities. This remarkable increase in explained variance indicates that UNet corrections capture much more coherent spatial patterns of DSL variability during the monsoon season.

        Principal components demonstrate significant differences in temporal evolution between the raw DSL and the UNet-corrected DSL. Although the raw CNRM-CM6 PC shows moderate variability without a clear trend (Fig.~\ref{fig: dsl_m1_ssp2}(17)), the UNet-corrected projections show stronger variability with a clear increasing trend after 2060 (Fig.~\ref{fig: dsl_m1_ssp2}(18)), suggesting an intensification of the monsoon circulation pattern in the far-future (2080-2100). This temporal difference highlights the potential impact of UNet corrections on capturing long-term changes in monsoon circulation patterns, with implications for regional climate dynamics and coastal processes in the Bay of Bengal region during the critical monsoon season.
        
        In raw CNRM-CM6 projections, the near-future shows DSL values of 0.5-0.55~m in the northern Bay of Bengal (16$^o$N-20$^o$N), with high DSL regions (0.58-0.6~m) near the Andaman Islands. The central bay (8$^o$N-16$^o$N) shows values between 0.48-0.52~m with a prominent western boundary current along the Indian coast, while the southern region (4N-8N) maintains lower values (0.42-0.46~m) due to remote equatorial forcing. The mid-future shows a similar spatial distribution but with slightly higher values, particularly in the central bay. The far-future period demonstrates enhanced monsoon patterns with intensified DSL values in the northern and eastern regions, indicating stronger monsoon circulation features. In contrast, UNet-corrected DSL projections show higher DSL values in the northern region (0.5-0.6~m), particularly along the eastern boundary and the Andaman Islands, reflecting stronger monsoon circulation. The Western Boundary Current signature intensifies (0.54-0.56~m) along the western boundary with associated eddy formations, while the southern bay shows stronger gradients (0.4-0.46~m). The mid-future and far-future projections show progressive intensification, with the far period showing maximum DSL values (0.58-0.6~m) in the northern bay and eastern boundary. 

        \paragraph{Dynamics implication}
        The UNet corrected DSL projections of JJAS (June-September) indicate substantial monsoon-related impacts in the Bay of Bengal. Intensified DSL patterns (0.56-0.6~m in northern regions) suggest enhanced monsoon circulation and air-sea interactions \citep{schott2001monsoon,li2017bay}, which could affect precipitation patterns and extreme rainfall events during the summer monsoon \citep{turner2012climate, goswami2022role}. The increased freshwater influx from monsoon rainfall and river discharge, reflected in higher DSL values near river mouths (0.58-0.6~m), strengthens stratification and modifies upper ocean mixing \citep{lucas2016adrift, fousiya2016interannual}. Consequently, the intraseasonal oscillation of the Indian summer monsoon is influenced \citep{vecchi2002monsoon}. In addition, the intensification of the western boundary currents and their eddy activity along the coast affects coastal upwelling and primary productivity \citep{nuncio2012life, shee2024three}. This leads to repercussions on regional fisheries, as migration patterns and spawning grounds are altered \citep{bakun2015anticipated, asch2019climate}. These predictions indicate an increased risk for coastal communities with respect to flooding, erosion, altered fishing areas, and potential changes in monsoon rainfall, necessitating effective adaptation strategies for coastal and disaster management.

\subsubsection{Post-monsoon Season}
       
        The post-monsoon season shows transitional DSL patterns with values ranging from 0.4-0.6~m, marking the shift from monsoon to winter conditions. The raw CNRM-CM6 projections show a relatively uniform distribution with moderate DSL values throughout the basin (Fig.~\ref{fig: dsl_m1_ssp2}(4)). In contrast, the UNet-corrected projections reveal high DSL values along the eastern boundary and low DSL values near the coast of India (Fig.~\ref{fig: dsl_m1_ssp2}(8)). This enhanced east-west gradient in the UNet-corrected projections better captures the transitional nature of the post-monsoon season, reflecting the developing East India Coastal Current (EICC) and the remnant features of the monsoon circulation that persist during this period.

        The first EOF mode spatial patterns demonstrate significant differences between the raw DSL projections and UNet-corrected DSL projections. In raw CNRM-CM6, mode 1 explains 35.6\% of the total variance (Fig.~\ref{fig: dsl_m1_ssp2}(12)), with spatial patterns showing a post-monsoon structure containing remnant monsoon features that reflect the transition to winter circulation and the development of EICC. After UNet correction, the explained variance increases significantly to 85.5\% (Fig.~\ref{fig: dsl_m1_ssp2}(16)), with spatial patterns showing enhanced features, stronger gradients, and clear coastal maxima that better represent the transitioning EICC, post-monsoon circulation, and developing eddy activities. This substantial increase in explained variance indicates that UNet corrections capture more coherent spatial patterns of DSL variability during this transitional season.

        Principal components show notable differences in temporal evolution between the raw DSL and the UNet-corrected DSL. The raw CNRM-CM6 PC1 shows moderate variability without a clear trend (Fig.~\ref{fig: dsl_m1_ssp2}(17)). In contrast, the UNet-corrected DSL projections demonstrate stronger variability with a clear increasing trend after 2060 (Fig.~\ref{fig: dsl_m1_ssp2}(18)), suggesting intensification of post-monsoon circulation patterns in the Far period (2080-2100). This temporal difference highlights the potential impact of UNet corrections on capturing long-term changes in post-monsoon circulation patterns, with implications for regional oceanographic processes during this transitional season.

        In raw CNRM-CM6 projections, the near-future period shows moderate DSL values (0.45-0.5~m) in the northern Bay of Bengal (16$^o$N-20$^o$N), with localized high values (0.5-0.55~m) near the Andaman Islands. The central bay (8$^o$N-16$^o$N) shows values of 0.45-0.5~m with distinct eddy features in the central-western region, while the southern region (4$^o$N-8$^o$N) maintains lower values (0.35-0.4m) influenced by remote equatorial forcing and transitioning EICC (Fig.\ref{fig: dsl_proj}(4)). The mid-future maintains similar spatial distributions but with slightly enhanced values, particularly in the central-eastern region (Fig.\ref{fig: dsl_proj}(12)). The far-future demonstrates intensified transitional patterns with higher DSL values in the northern and eastern regions (Fig.\ref{fig: dsl_proj}(20)), indicating stronger post-monsoon circulation features. However, the UNet-corrected DSL projections show more pronounced features across all future periods. In the near-future period, the northern region exhibits higher DSL values (0.55-0.65~m), particularly along the eastern boundary, with an intensified EICC signature (0.5- 0.55~m) along the western boundary. The southern bay shows stronger gradients (0.35- 0.45~m) (Fig.\ref{fig: dsl_proj}(8)). The mid-future period projections show values 0.02- 0.03~m higher than near-period levels (Fig.\ref{fig: dsl_proj}(16)), while far-future period projections reach maximum DSL values (0.6- 0.7~m) in the northern bay and eastern boundary (Fig.\ref{fig: dsl_proj}(24)). These enhanced features in the UNet-corrected projections suggest a significant intensification of the post-monsoon oceanographic processes in future climate scenarios, with potential implications for coastal circulation, marine ecosystems, and coastal communities around the Bay of Bengal during this transitional season.

        \paragraph{Dynamics implication}
        In raw CNRM-CM6 projections, moderate intensification of the EICC affects coastal upwelling \citep{mukherjee2014observed}, while increased eddy activity in the central bay influences the distribution of heat and salt \citep{trott2023eddy}. UNet-corrected projections indicate more pronounced impacts, particularly in coastal processes and ecosystem responses. These DSL changes could affect regional rainfall during the northeast monsoon \citep{yadav2013emerging}. The post-monsoon season is the peak cyclone period in the Bay of Bengal. The UNet corrected DSL shows elevated values (0.45-0.55~m in Near to 0.5-0.65~m in far-future projections), which combined with enhanced cyclonic activity could amplify storm surges along the northern and eastern coasts, particularly affecting the eastern Indian and Bangladesh coasts \citep{jisan2018ensemble, vinayachandran2003phytoplankton}. The more pronounced changes in the UNet projections suggest the need for stronger adaptation measures for coastal communities and marine resource management compared to the raw CNRM-CM6 projections.

\section{Summary and Future Directions}  \label{sec: conclusion}

\begin{table}[]
\centering
\caption{Key findings from the analysis of the corrected CNRM-CM6 projections} \label{tab: features}
\begin{tabular}{|p{0.48in}|p{1.05in}|p{1.3in}|p{1.2in}|}
\hline
\textbf{Season} & \multicolumn{1}{c|}{\textbf{SST}}                                                                        & \multicolumn{1}{c|}{\textbf{DSL}}                                                      & \multicolumn{1}{c|}{\textbf{Features Affected}}                                                 \\ \hline
\textbf{DJF}    & \begin{tabular}[c]{@{}l@{}}Weak north-south \\ temperature gradient\\ 27$^\circ$C isotherm\\ northward migrates\end{tabular} & Stronger along coast                                                                   & \begin{tabular}[c]{@{}l@{}}Monsoon dynamics\\ Coastal upwelling\end{tabular}               \\ \hline
\textbf{MAM}    & \begin{tabular}[c]{@{}l@{}}Intensified warming\\ in central bay\end{tabular}                             & \begin{tabular}[c]{@{}l@{}}Enhanced warm eddy\\ High at Indian coast\end{tabular} & \begin{tabular}[c]{@{}l@{}}Cyclogenesis\\ Monsoon onset\\ Marine productivity\end{tabular} \\ \hline
\textbf{JJAS}   & \begin{tabular}[c]{@{}l@{}}Stronger coastal\\ warm front \\ Central \& southern\\ bay warming\end{tabular}    & High near river mouths                                                                 & \begin{tabular}[c]{@{}l@{}}ISMR \\ Extreme rainfall\\ SMC propagation\end{tabular}         \\ \hline
\textbf{ON}     & Warming bay                                                                                              & Elevated DSL values                                                                    & \begin{tabular}[c]{@{}l@{}}EICC\\ Cyclogenesis\\ Storm surges\end{tabular}                 \\ \hline
\end{tabular}
\end{table}

 We developed and applied a deep neural network for bias correction of CNRM-CM6 projections of SST and DSL in the Bay of Bengal. This model was trained using climatology-removed CNRM-CM6 as input and the ORAS5 reanalysis product as output for the period for which both fields are available. The trained model is used to correct the CNRM-CM6 projections for 2024-2100. For the test period (2021-2023), the RMSE when using our correction method is 0.5 for SST and 0.65 for DSL, which is 15\% lower than the RMSE of the corrected projections using the EDCDF method. Using corrected future projections, monthly, seasonal (EOF analysis), near-, mid-, and far-future projections are analyzed. 
 
 Our analysis reveals major dynamical implications of the changing climate in the Bay of Bengal for each season. Table~\ref{tab: features} summarizes the key findings along with the features that are affected. Overall, the corrected projections reveal that the dynamics of the northeast and southwest monsoons will be altered and the EICC will show stronger variability, leading to changes in coastal upwelling, changes in nutrient distribution, and biological productivity. We observed an intensified coastal warm front and alterations in thermal structure in the central and southern bays, potentially affecting Indian summer monsoon rainfall and SMC propagation during the summer monsoon. Enhanced DSL patterns might influence precipitation and extreme rainfall events in the summer monsoon. The most concerning aspect is that the corrected projections indicate more favorable conditions for cyclogenesis and intensification of cyclones in the Bay of Bengal, with amplified storm surges. 

 In the future, a similar analysis can be completed for salinity. More research is needed to explore whether the correction neural network model can be used for other members of the CNRM-CM6 suite, other than the CNRM model used in the present study. Beyond the ocean, there are avenues for bias correction of atmospheric variables in the future.

 \section*{Acknowledgment}
 We thank the Ministry of Earth Sciences, Government of India, for the grant MoES/36/OOIS/Extra/84/2022 that allowed the purchase of the compute resources used in model training. The authors thank the members of the QUEST Lab at IISc for insightful discussion.


\newpage


\section*{Supplementary material}

# Supporting Information for paper titled "Data Driven Deep Learning for Correcting Global Climate Model Projections of Sea Surface Temperature and Dynamic Sea Level in the Bay of Bengal"

Abhishek Pasula, Deepak N. Subramani

This document presents additional analyses supporting the Data-Driven Deep Neural models for correcting CMIP6 projections of the Bay of Bengal.

## S1 UNet Training

The training and validation loss of the UNet model for SST and DSL are shown in Fig. S1.

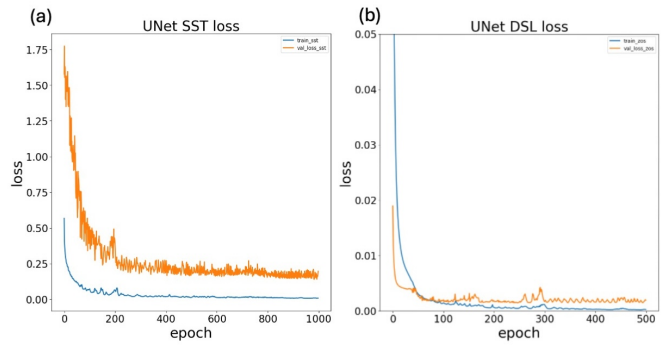

Figure S1: The training and validation loss for the UNet model of (a) SST, and (b) DSL.

## S2 Comparison of ORAS5, CMIP6 and UNet Corrected CMIP6 SST and DSL in 2022

### S2.1 Sea Surface Temperature projections

Bay of Bengal monthly SST for 2022 from ORAS5, CMIP6 and UNet corrected SST of SSP1-2.6 is shown in Figure S2, SSP2-4.5 is shown in Figure S3, SSP3-7.0 is shown in Figure S4 and SSP5-8.5 is shown in Figure S5

**Winter** During winter months, all SSP scenarios demonstrate a characteristic north-south temperature gradient in the Bay of Bengal. In SSP1 (Fig. S2), the ORAS5 reanalysis shows temperatures of 26-27°C in the northern bay, gradually increasing to 28-29° C in the southern regions, with CNRM-CM6 showing a moderate cold bias that UNet corrections effectively mitigate, closely matching observational patterns. SSP2-4.5 (Fig. S3) exhibits a similar gradient structure, though with slightly more pronounced cold bias in CNRM-CM6 (1-2°C lower than reanalysis), particularly in the northern and central regions where temperatures fall below 26°C; UNet corrections substantially improve these patterns, accurately representing both the spatial distribution and thermal boundaries. SSP3-7.0 (Fig. S4) demonstrates a more severe cold bias in raw model output, with temperatures underestimated by up to 2.5°C and the spatial extent of waters below 26°C significantly exaggerated across most of the northern and central bay; despite these challenges, UNet corrections successfully restore temperature patterns close to ORAS5 reanalysis, though some residual cold bias remains in the northwestern bay during January. SSP5-8.5 (Fig. S5) presents the most extreme case, with CNRM-CM6 exhibiting

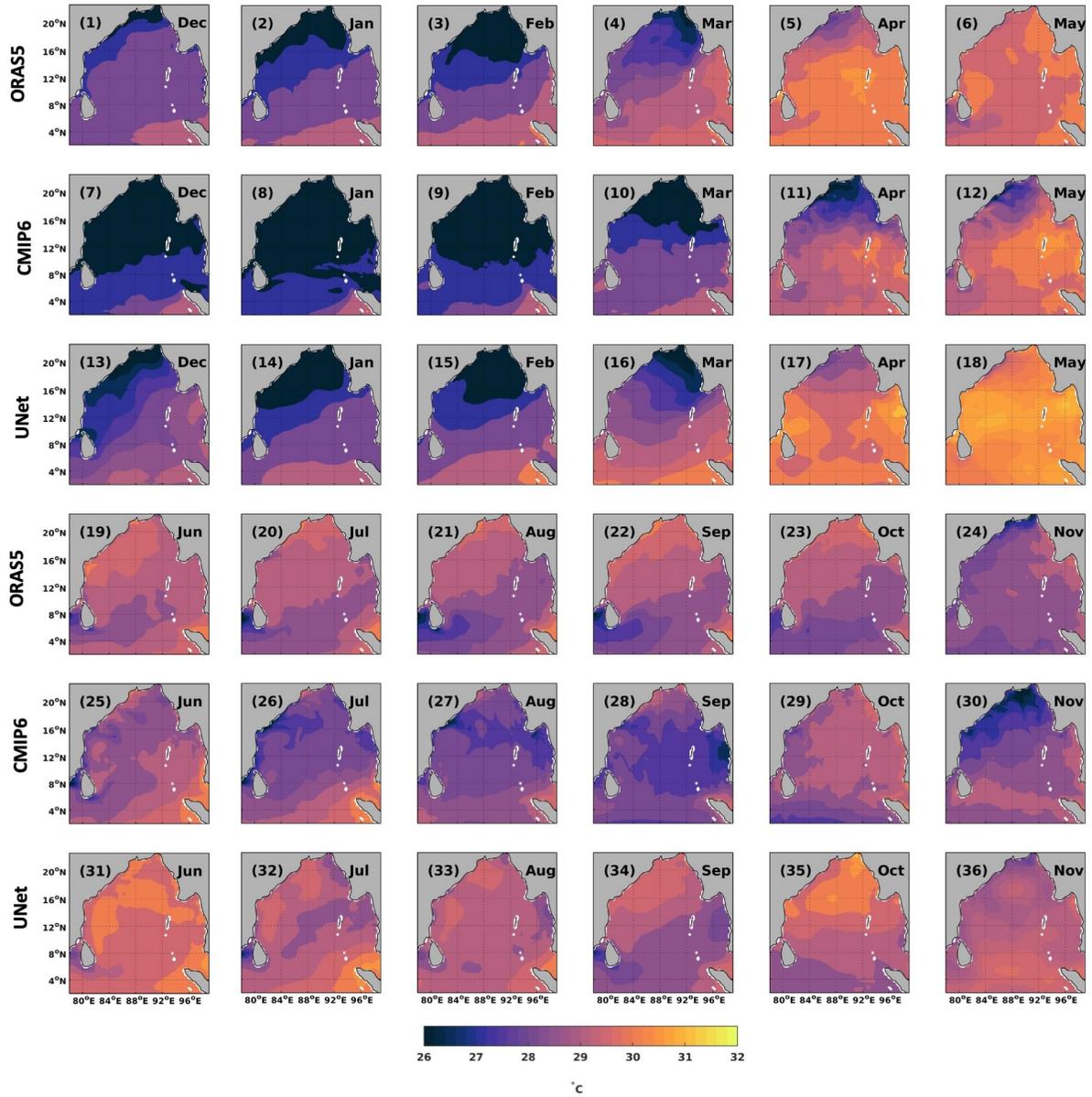

Figure S2: Monthly sea surface temperature (SST) in the BoB during 2022, ORAS5 reanalysis data comparing with raw CMIP6 SSP1-2.6 (CMIP6), and UNet-corrected USSP1-2.6 SST (UNet).

catastrophic cold bias where temperatures are underestimated by 2-3°C across much of the basin, with waters below 26°C extending through nearly two-thirds of the bay in January; remarkably, UNet corrections still substantially reduce this severe bias, restoring temperature patterns that reasonably approximate ORAS5 reanalysis, though with slightly greater residual errors than in lower emissions scenarios.

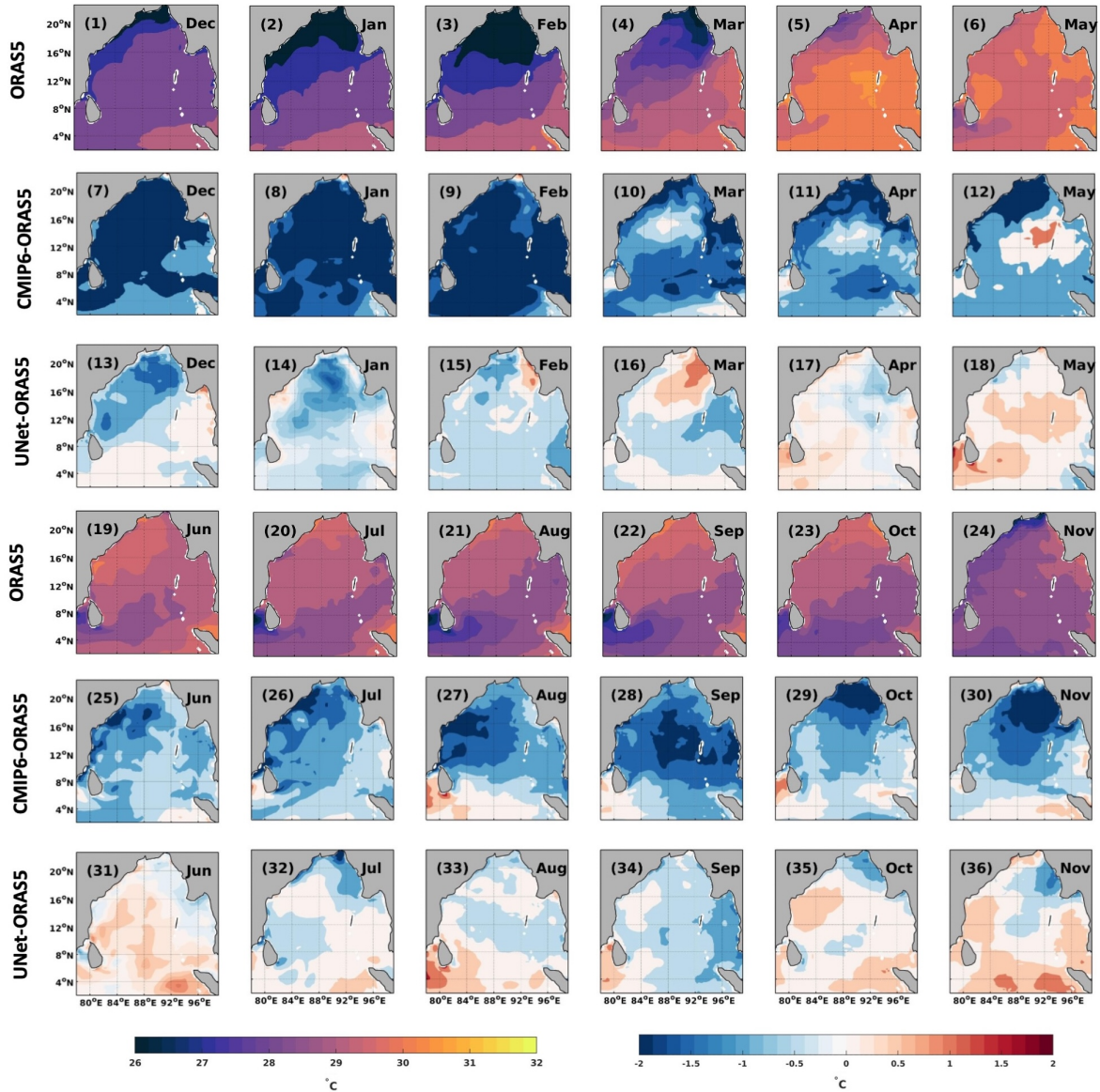

Figure S3: Monthly sea surface temperature (SST) in the BoB during 2022, ORAS5 reanalysis, comparing with raw CMIP6 SSP2-4.5 (CMIP6-ORAS5), UNet-corrected USSP2-4.5 SST (UNet-ORAS5).

**Pre-monsoon** The pre-monsoon season demonstrates progressive warming across all SSP scenarios, with important differences in intensity and spatial patterns. In SSP1 (Fig. S2), March initiates the warming phase with temperatures beginning to increase, particularly in the eastern bay, and by April temperatures reach approximately 30°C in the central and eastern regions, while May shows moderate warm pool development with temperatures reaching 31°C in parts of the central and northern bay; UNet corrections effectively capture these patterns, particularly the formation of warm pools. SSP2-4.5 shows slightly higher warming rates (Fig. S3), with April temperatures reaching 30°C in the central

3

and eastern regions and May exhibiting temperatures exceeding 31°C in more extensive areas of the central and northern bay; CNRM-CM6 consistently underestimates temperatures by 1-2°C and fails to properly capture warm pool formation, while UNet corrections significantly improve representations of both the spatial extent and intensity of warming patterns. SSP3-7.0 exhibits intensified warming with more rapid progression compared to lower emission scenarios (Fig. S4), with April temperatures in ORAS5 reaching 30-31°C in the central and eastern regions and May showing temperatures exceeding 31.5° C in most of the central and northern bay; CNRM-CM6 shows a persistent cold bias that underestimates temperatures by 1.5-2.5°C, while UNet corrections successfully reproduce the accelerated warming trajectory and expanded warm pool development. SSP5-8.5 demonstrates the most dramatic warming progression with highest temperature extremes (Fig. S5), where April shows temperatures exceeding 31°C in extensive areas of the central and eastern bay and May reaches up to 32°C in parts of the northern and central bay; despite CNRM-CM6's severe cold bias (2-3°C across much of the basin), UNet corrections demonstrate exceptional performance in reproducing both the accelerated warming trajectory and expanded warm pool, though with slight warm biases in some regions during April-May.

**Monsoon** The monsoon period exhibits distinctive SST patterns associated with monsoon circulation dynamics in all SSP scenarios, with progressive intensification from the lower to higher emission pathways. In SSP1 (Fig. S2), June marks the onset of monsoons with temperatures of 29-30°C in most of the bay and slightly cooler waters in regions affected by monsoon circulation, while July and August show established monsoon patterns with the Summer Monsoon Current (SMC) evident south of Sri Lanka and the Western Boundary Current flowing northward along the Indian coast; UNet corrections closely match ORAS5's depiction of both basin-wide patterns and localized features. SSP2-4.5 demonstrates similar patterns but with slightly elevated temperatures, showing 29-30°C across most of the bay in June and distinct SMC thermal signatures in July-August; CNRM-CM6 exhibits a cold bias throughout this period and fails to accurately represent the SMC's thermal signature, while UNet corrections substantially improve representations of both basin-wide patterns and localized features. SSP3-7.0 shows altered temperature patterns associated with intensified monsoon circulation (Fig. S4), with June temperatures of 29-30.5°C across most of the bay but more pronounced cooling in regions directly affected by strengthened monsoon winds and currents; the SMC shows a stronger thermal signature south of Sri Lanka and enhanced upwelling along the western boundary in July-August, patterns that UNet successfully captures despite CNRM-CM6's augmented cold bias. SSP5-8.5 (Fig. S5) exhibits the most dramatically altered temperature patterns reflecting an intensification of monsoon dynamics, with June showing earlier and more intense onset conditions, and July-August demonstrating significantly strengthened monsoon patterns with enhanced thermal signatures in the SMC and more extreme upwelling along the western boundary; despite CNRM-CM6's most significant biases (2-3°C across much of the basin), UNet corrections show remarkable skill in reproducing both the elevated baseline temperatures and enhanced cooling signatures associated with intensified circulation.

**Post-monsoon** The post-monsoon transition shows a progression from warmer to cooler conditions in all SSP scenarios, with significant differences in cooling rates and gradient intensities. In SSP1 (Fig. S2), October maintains relatively warm temperatures around 29-30°C throughout much of the bay, while November displays the beginning of winter cooling with decreasing temperatures in northern regions establishing a moderate north-south gradient; the East India Coastal Current (EICC) begins its seasonal reversal to southward flow, creating distinct temperature patterns along the western boundary that UNet corrections accurately capture. SSP2-4.5 shows similar patterns but with slightly enhanced gradients (Fig. S3), with October displaying temperatures around 29-30°C and November exhibiting more pronounced cooling in the northern bay; CNRM-CM6 shows its most significant cold bias during this transition period, particularly in November where temperatures are underestimated by more than 2°C, while UNet corrections demonstrate remarkable skill in reproducing both the spatial distribution and the onset of cooling. SSP3-7.0 displays more rapid cooling and enhanced temperature gradients (Fig. S4), with October maintaining elevated temperatures around 29.5-30.5°C

4

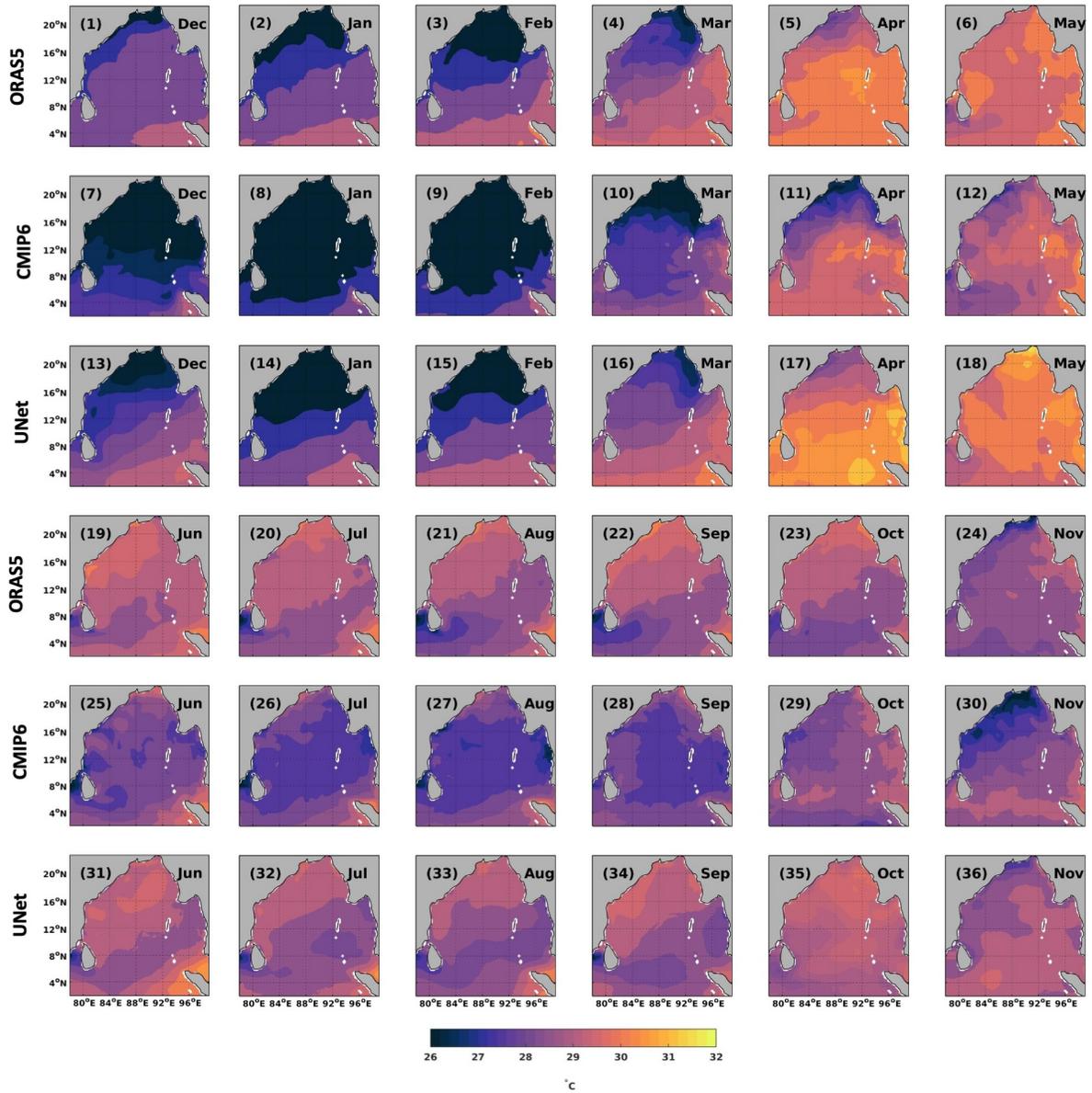

Figure S4: Monthly sea surface temperature (SST) in the BoB during 2022, comparing raw CMIP6 SSP3-7.0 output (CMIP6), UNet-corrected USSP3-7.0 SST (UNet), and ORAS5 reanalysis data.

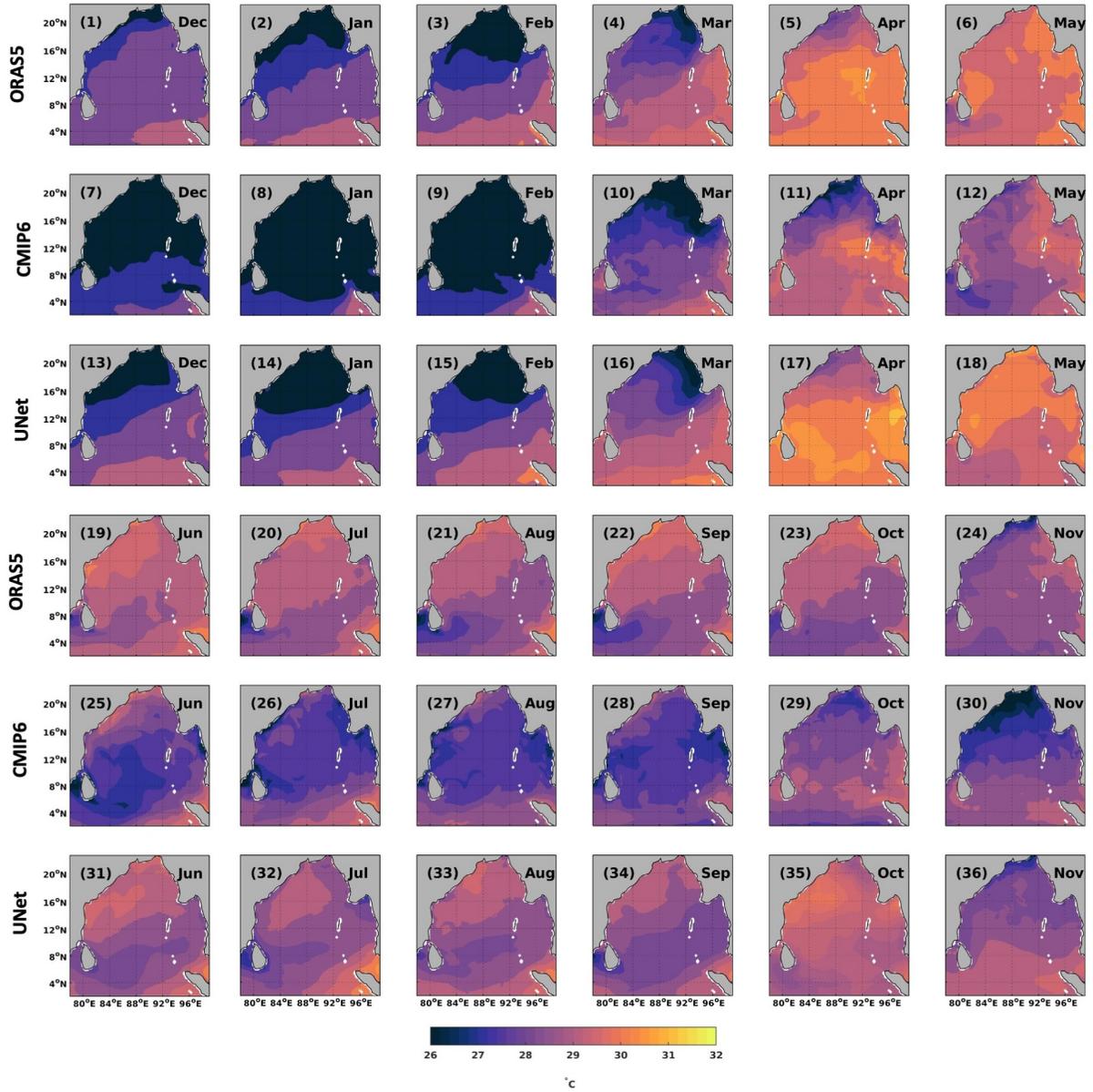

Figure S5: Monthly sea surface temperature (SST) in the BoB during 2022, comparing raw CMIP6 SSP5-8.5 (CMIP6), UNet-corrected USSP5-8.5 SST (UNet), and ORAS5 reanalysis data.

but November showing accelerated cooling with a more pronounced north-south gradient that is established earlier than in lower emissions scenarios; the EICC shows a stronger signal during its seasonal reversal, creating more distinct temperature patterns along the western boundary that UNet successfully reproduces despite CNRM-CM6's significant cold bias (more than 3°C in the northern bay). SSP5-8.5 exhibits the most extreme cooling trajectory and gradient intensification (Fig. S5), with October maintaining significantly elevated temperatures exceeding 30-31°C across much of the bay, but November showing dramatically accelerated cooling with an enhanced north-south gradient; the EICC displays its strongest signal during reversal, creating highly distinctive temperature patterns along the western boundary that UNet remarkably captures despite CNRM-CM6's catastrophic biases (3-4°C underestimation in the northern bay) that extend across nearly the entire basin.

### S2.2 Dynamic Sea Level projections

Bay of Bengal monthly DSL for 2022 from ORAS5, CMIP6 and UNet corrected SST of SSP1-2.6 is shown in Figure S6, SSP2-4.5 is shown in Figure S7, SSP3-7.0 is shown in Figure S8 and SSP5-8.5 is shown in Figure S9.

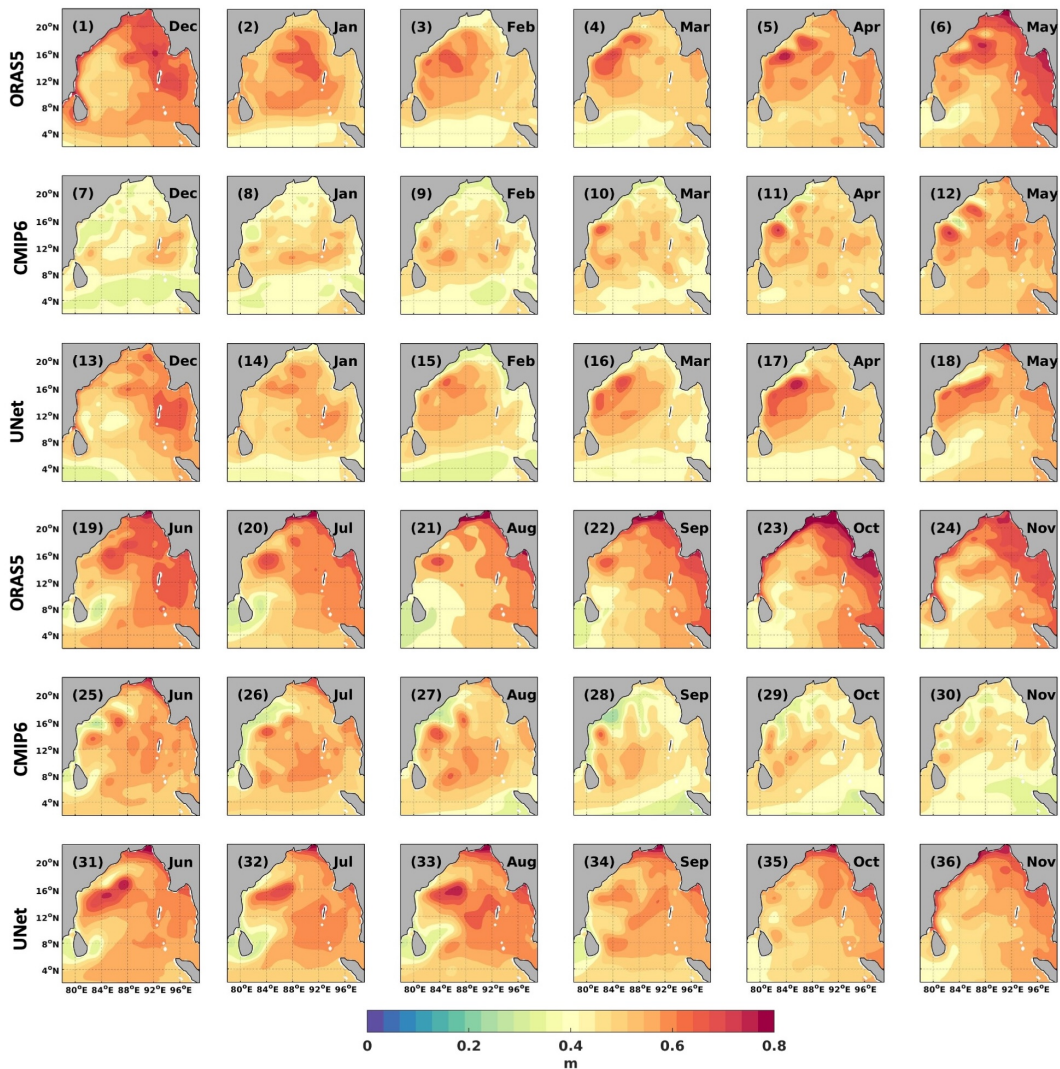

Figure S6: Monthly Dynamic Sea Level (DSL) in the BoB during 2022, comparing raw CMIP6 SSP1-2.6 (CMIP6), UNet-corrected USSP1-2.6 (UNet), and ORAS5 reanalysis data.

**Winter** During the winter months, all SSP scenarios demonstrate characteristic DSL patterns in the Bay of Bengal with important differences in bias and correction effectiveness. In SSP1-2.6 (Fig. S6), the ORAS5 reanalysis shows DSL values of 0.4-0.6m concentrated in the central bay with a distinct maximum (0.5-0.6m) centered around 15-18°N, 85-90°E and a secondary peak along the eastern boundary near the Andaman Sea; CNRM-CM6 underestimates these features, while UNet corrections successfully reproduce the dual maximum structure. SSP2-4.5 exhibits similar observed patterns (Fig. S7), though CNRM-CM6 shows a more significant underestimation of DSL values (0.3-0.5m compared to 0.5-0.6m in ORAS5); UNet corrections effectively restore both the magnitude and spatial distribution of these characteristics, closely matching ORAS5 reanalysis. SSP3-7.0 demonstrates more severe bias in raw model output (Fig. S8), with DSL values substantially underestimated across most of the bay and the dual maximum structure almost entirely absent; despite these challenges, UNet corrections successfully restore the primary features, although with slightly less accuracy in the northernmost regions. SSP5-8.5 presents the most extreme case (Fig. S9), with CNRM-CM6 showing not only significant underestimation, but also spatial misplacement of DSL features; remarkably, UNet corrections still substantially improve both the magnitude and spatial distribution of winter DSL patterns, although with some residual errors in representing the finer details of the dual maximum structure.

**Pre-monsoon** During winter months, all SSP scenarios demonstrate characteristic DSL patterns in the Bay of Bengal with important differences in bias and correction effectiveness. In SSP1-2.6 (Fig. S6), the ORAS5 reanalysis shows DSL values of 0.4-0.6m concentrated in the central bay with a distinct maximum (0.5-0.6m) centered around 15-18°N, 85-90°E and a secondary peak along the eastern boundary near the Andaman Sea; CNRM-CM6 underestimates these characteristics, while UNet corrections successfully reproduce the dual maximum structure. SSP2-4.5 exhibits similar observed patterns (Fig. S7), though CNRM-CM6 showed a more significant underestimation of DSL values (0.3-0.5m compared to 0.5-0.6m in the reanalysis); UNet corrections effectively restore both the magnitude and spatial distribution of these features, closely matching the ORAS5 reanalysis. SSP3-7.0 demonstrates more severe bias in raw model output (Fig. S8), with DSL values substantially underestimated across most of the bay and the dual maximum structure almost entirely absent; despite these challenges, UNet corrections successfully restore the primary features, although with slightly less accuracy in the northernmost regions. SSP5-8.5 presents the most extreme case (Fig. S9), with CNRM-CM6 showing not only significant underestimation, but also spatial misplacement of DSL features; remarkably, UNet corrections still substantially improve both the magnitude and spatial distribution of winter DSL patterns, although with some residual errors in representing the finer details of the dual maximum structure.

**Monsoon** The monsoon period exhibits distinctive DSL patterns associated with monsoon circulation dynamics in all SSP scenarios, with progressive differences in the representation of key oceanographic features. In SSP1-2.6 (Fig. S6), the reanalysis of ORAS5 shows a well-defined signature of the Summer Monsoon Current (SMC) (Sri Lankan Dome) near Sri Lanka and the Western Boundary Current (WBC) along the Indian coast; CNRM-CM6 underestimates these features, while UNet corrections successfully capture the characteristic circulation patterns. SSP2-4.5 (Fig. S7) shows similar observed patterns with the western boundary exhibiting increased variability with alternating bands of higher (0.5-0.6m) and lower (0.3-0.4m) DSL values reflecting enhanced eddy activity during the monsoon; CNRM-CM6 fails to represent these complex patterns, while UNet corrections substantially improve the representation of both the SMC signature and boundary current influences. SSP3-7.0 demonstrates more significant challenges for the raw model output (Fig. S8), with CNRM-CM6 showing a severe underestimate of monsoon circulation features, particularly in July and August; UNet corrections manage to restore most of these features, though with slightly reduced accuracy in capturing the most intricate details of eddy-current interactions. SSP5-8.5 presents the most complex case (Fig. S9), with CNRM-CM6 showing not only underestimation but also significant spatial distortion of monsoon features, including exaggerated high values along the northeastern coast in June that are not present in ORAS5; despite these complex biases, UNet corrections demonstrate remarkable effectiveness in restoring the major circulation features, though some residual errors remain in the

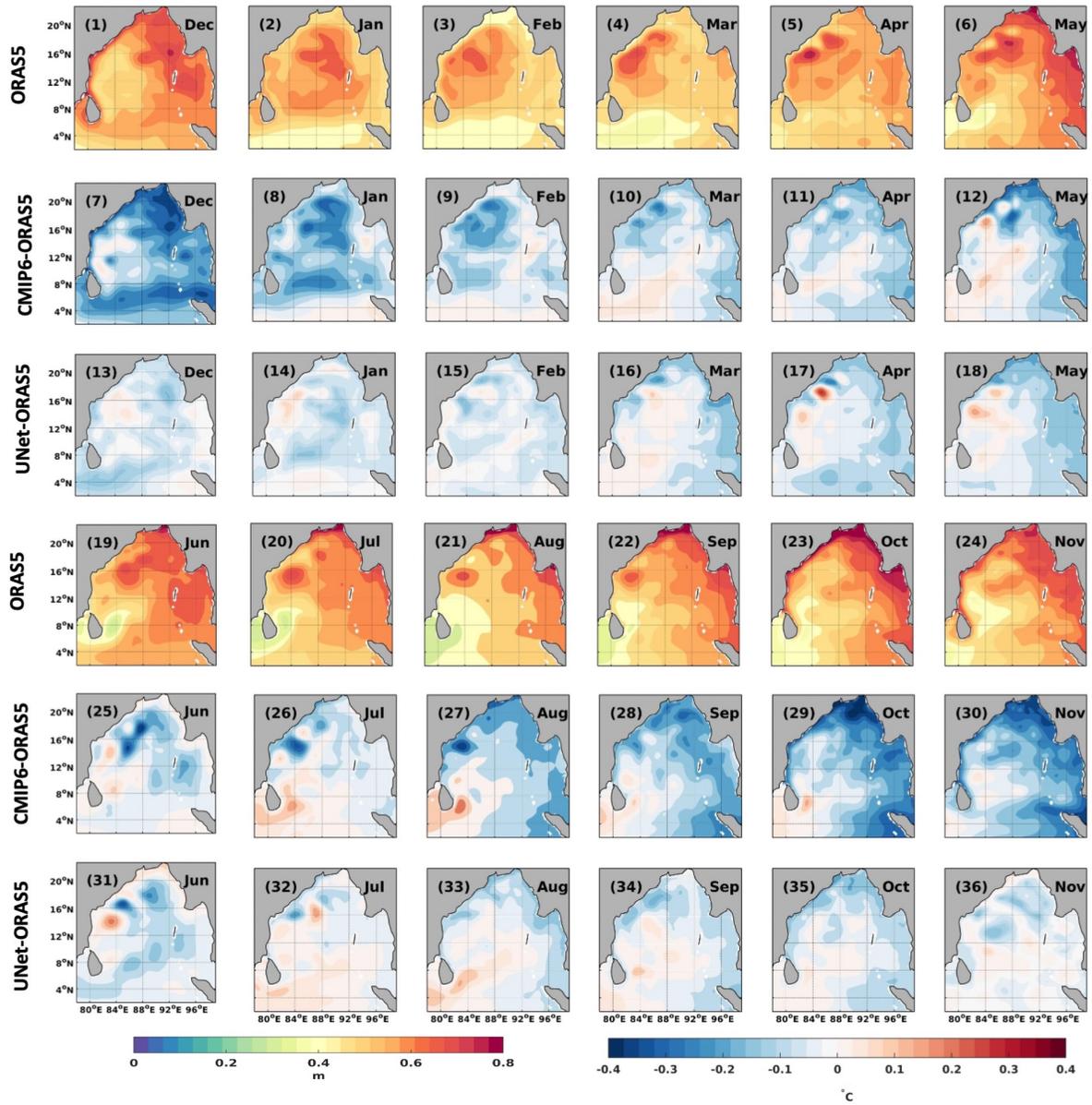

Figure S7: Monthly sea surface temperature (DSL) in the BoB during 2022, ORAS5 reanalysis, comparing with raw CMIP6 SSP2-4.5 (CMIP6-ORAS5), UNet-corrected USSP2-4.5 DSL (UNet-ORAS5).

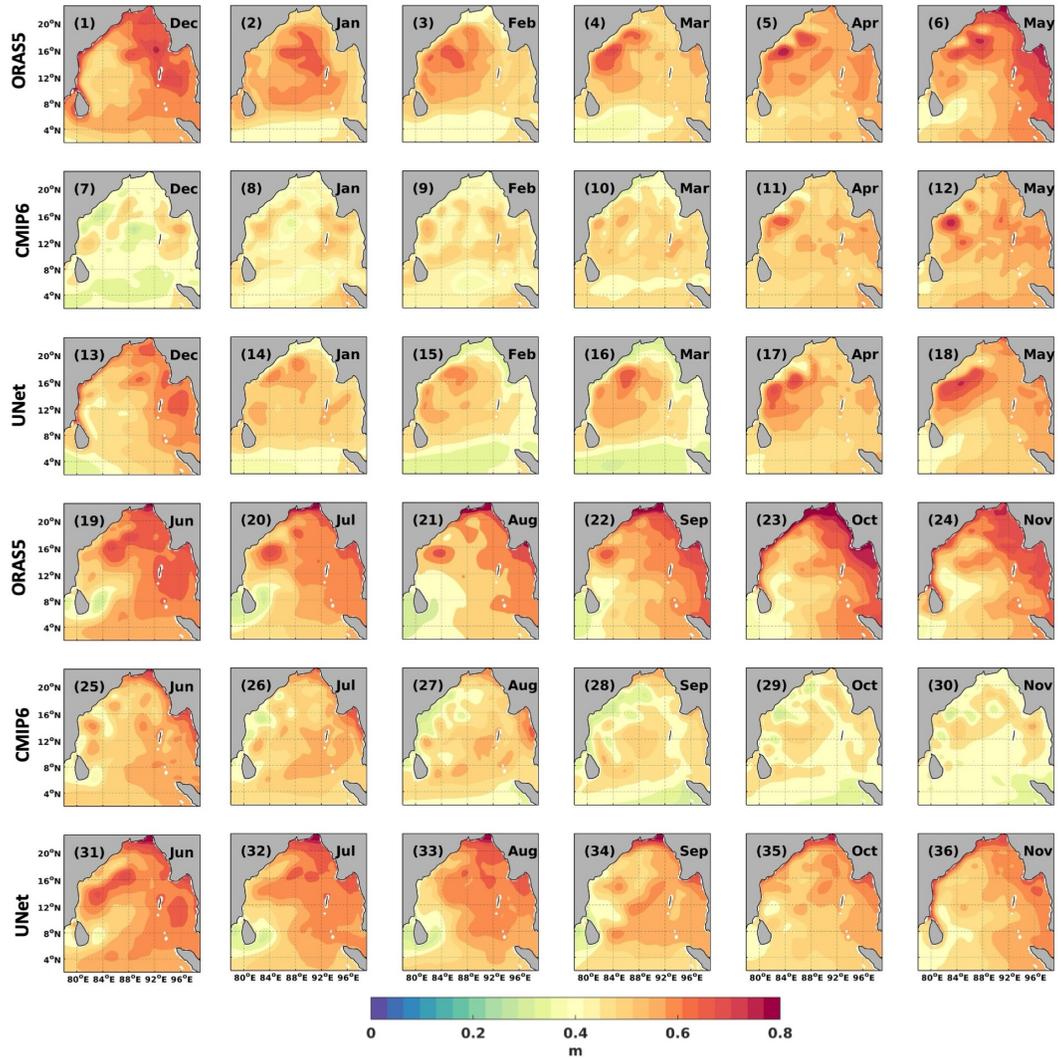

Figure S8: Monthly Dynamic Sea Level (DSL) in the BoB during 2022, comparing raw CMIP6 SSP3-7.0 output (CMIP6), UNet-corrected USSP3-7.0 (UNet), and ORAS5 reanalysis data.

representation of the finest-scale dynamics.

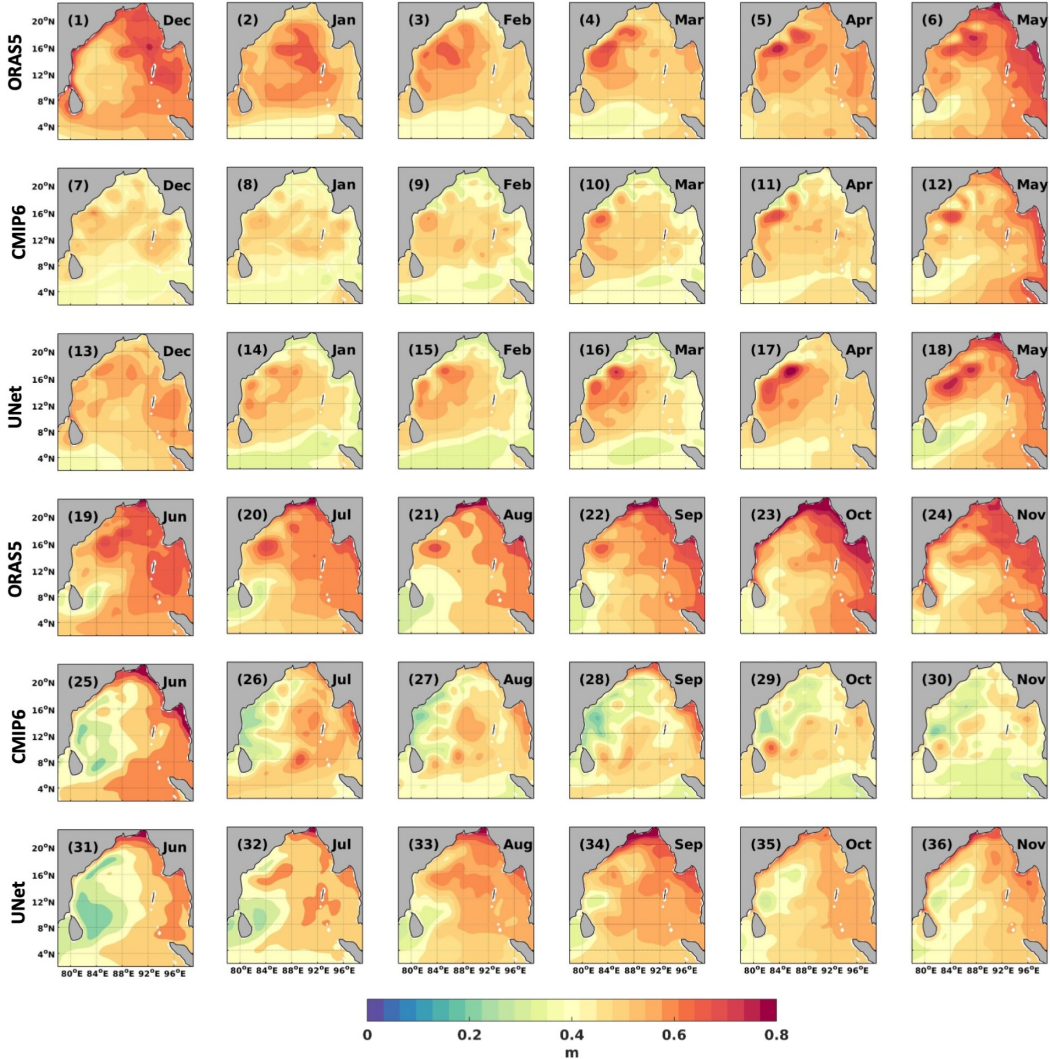

Figure S9: Monthly Dynamic Sea Level (DSL) in the BoB during 2022, comparing raw CMIP6 SSP5-8.5 output (CMIP6), UNet-corrected USSP5-8.5 (UNet), and ORAS5 reanalysis data.

**Post-monsoon** The post-monsoon transition shows a characteristic progression toward winter conditions in all SSP scenarios, with important differences in coastal features and correction performance. In SSP1-2.6 (Fig. S6), ORAS5 reanalysis shows that the reversal of the East India Coastal Current (EICC) begins and is evident in the higher DSL patterns along the coast of India; CNRM-CM6 fails to capture this feature, while UNet corrections successfully reproduce the transitional patterns. SSP2-4.5 shows similar observed patterns (Fig. S7) with the presence of eddies indicated by zones of varying DSL intensities in the western and central regions; CNRM-CM6 shows systematic errors in representing both the EICC reversal and eddy structures, while UNet corrections accurately capture these transitional features. SSP3-7.0 demonstrates more severe bias in the raw CNRM-CM6 projections during this period (Fig. S8), with almost no representation of the EICC reversal patterns in November; UNet corrections substantially improve the representation of these features, though with slightly reduced accuracy in some localized regions. SSP5-8.5 exhibits the most extreme biases (Fig. S9), with CNRM-CM6 showing severe underestimation and complete failure to capture the EICC reversal pattern; remarkably, UNet corrections still manage to restore the major features of the post-monsoon transition, including a recognizable EICC signature, although with some remaining inaccuracies in the

11

finer details of the coastal current structure. Across all scenarios, the east-west asymmetry in DSL patterns remains persistent throughout the year, with the eastern boundary consistently maintaining higher values compared to the western boundary, a feature that CNRM-CM6 fails to capture but is successfully reproduced after UNet-based correction.

## S3   Interannual Variability of Seasonal SST fro SSP3-7.0 and SSP5-8.5: An EOF Analysis

For analyzing the spatiotemporal patterns of seasonal SST (winter, pre-monsoon, monsoon, and post-monsoon) from 2023-2100, we examine the EOF for each season using raw CMIP6 and UNet-corrected CMIP6 projections for SSP3-7.0 and SSP5-8.5 scenarios. Figures S10, S12 show the mean, Mode 1, and principal component 1 of raw and corrected CMIP6 projections. Figures S11, S13 show the mean of raw and corrected CMIP6 projections in the near, mid, and far futures.

### S3.1   Winter

SSP3-7.0 demonstrates a slightly modified pattern with temperatures higher than SSP2-4.5 but a similar temperature pattern to SSP2-4.5, and its corresponding prominent Mode 1 explains 93.7% of the variance in the region, as shown in Figure S10 (1, 9). USSP3-7.0 shows a more pronounced warming, with temperatures $\approx 1.5°C$ higher than SSP3-7.0 in most of the bay. Mode 1 of the USSP3-7.0 projections explains 91.5% of the variance as shown in Figure S10 (5,13). The SSP3-7.0 Mode 1 indicates stronger variability extending into the central bay (Fig.S10 (9)), while the USSP3-7.0 Mode 1 shows intensified patterns along the Indian coastal boundary (Fig.S10 (13)).

SSP5-8.5 shows the most intense warming among raw projections; its Mode 1 explains 96.8% of the variance (Fig.S10 (1,9)), suggesting potential modifications in the features of the winter monsoon in the region. USSP5-8.5shows higher winter warming, with temperatures consistently in the entire basin higher than SSP5-8.5, its Mode 1 explains 95.0% of the variance as shown in Figure S12 (5,13). The SSP5-8.5Mode 1 demonstrates the most extensive variability throughout the bay, while the USSP5-8.5Mode 1 reveals intense patterns in the northern and eastern regions (Fig.S12 (9,13)), indicating potential changes in regional circulation dynamics. PC1 of SSP3-7.0 shows a stronger warming trend (Fig.S10 17), while PC1 of SSP5-8.5 shows the most pronounced upward trend (Fig.S12 17), suggesting an accelerated winter warming in the BoB.

SSP3-7.0 demonstrates slightly higher temperatures, approximately $\approx 0.5°C$ above SSP2-4.5, indicating reduced winter cooling while maintaining the basic north-south temperature structure [1]. USSP3-7.0 shows a more pronounced warming with temperatures 1.5°C above the SSP3-7.0 values, showing stronger north-south gradients (Fig.S11 (1,5)). The mid USSP3-7.0 shows warming 2.2°C higher than near SSP3. The far USSP3-7.0 indicates warming $\approx 3°C$ higher than near SSP3-7.0 (Fig.S11).

SSP5-8.5 shows the weakest winter cooling among the Near projections, with temperatures $1 - 1.5°C$ higher than SSP2-4.5 SST. USSP5-8.5exhibits maximum warming with temperatures 2°C higher than SSP5-8.5(Fig.S13 (1,5)), showing intense warming throughout the basin and the weakest winter cooling effects. Mid USSP5-8.5exhibits temperatures $\approx 2.5°C$ above Near SSP5. The far USSP5-8.5 shows temperatures $\approx 3.5°$ C above near SSP5-8.5 (Fig. S13). Spatial patterns show significantly modified gradient structures and almost complete erosion of the typical winter cooling pattern, particularly in Far USSP5.

### S3.2   Pre-monsoon

SSP3-7.0 demonstrates enhanced warming patterns with temperatures approximately $\approx 0.5 - 1°C$ higher than SSP2-4.5, its Mode 1 explains 94.8% of the variance (Fig.S10 (10)), indicating intensified pre-monsoon heating while maintaining the spatial SST distribution. USSP3-7.0 demonstrates more pronounced warming, with temperatures $\approx 1.5°C$ higher than SSP3-7.0 in many regions, its Mode 1 explains 87.7% of the variance (Fig.S10 (14)). SSP3-7.0 Mode 1 pattern shows low variability in the central bay, while USSP3's EOF shows intensified patterns along the eastern boundary (Fig.S10

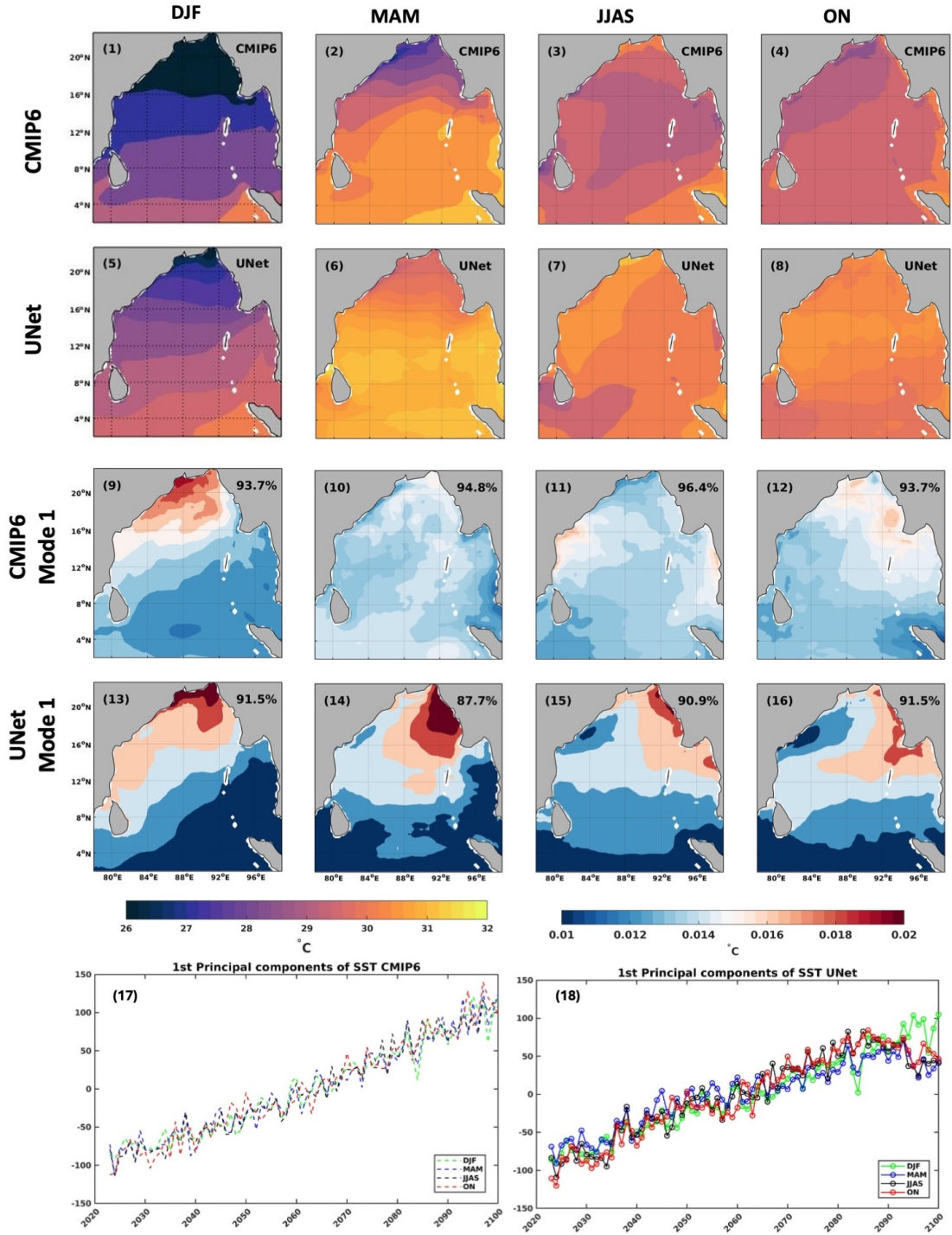

Figure S10: Seasonal SST temporal evolution from 2023 to 2100: Seasonal mean SST over the Bay of Bengal comparing raw CMIP6 SSP3-7.0 (1-4) and UNet-corrected (5-8) projections across DJF (winter), MAM (pre-monsoon), JJAS (monsoon), and ON (post-monsoon) seasons. First EOF modes of CMIP6 (9-12) with variances (DJF: 88.2%, MAM: 91.2%, JJAS: 93.2%, ON: 88.6%) and UNet-corrected projections (13-16; DJF: 84.5%, MAM: 85.1%, JJAS: 88.8%, ON: 89.7%). First Principal Component time series for CMIP6 (17) and UNet corrected (18) SST.

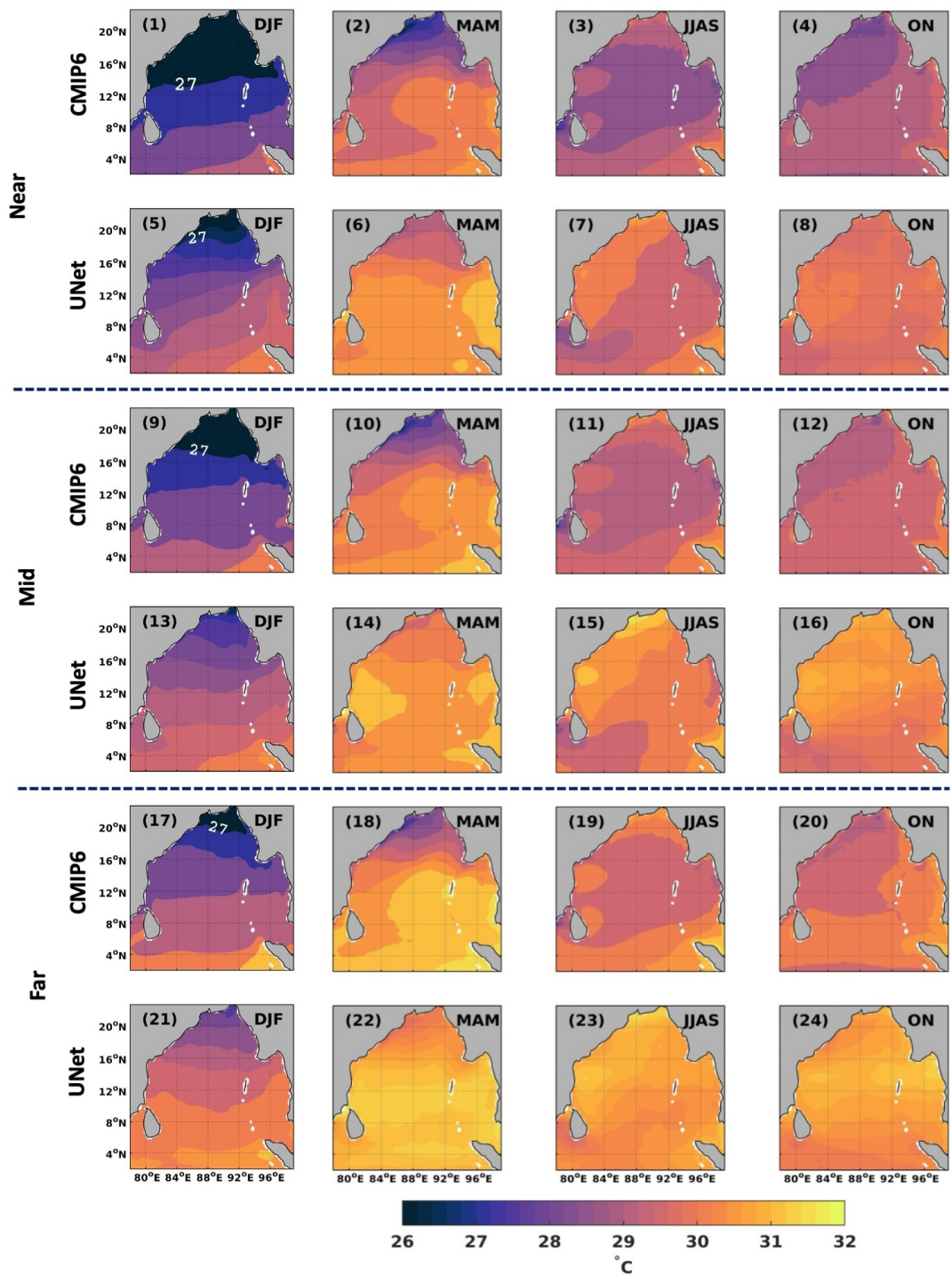

Figure S11: Seasonal SST patterns over the Bay of Bengal comparing raw CMIP6 SSP3-7.0 and UNet-corrected projections for Near future of CMIP6 (1-4) and UNet (5-8), Mid future of CMIP6 (9-12) and UNet (13-16), and Far future of CMIP6 (17-20) and UNet (21-24). Each row shows seasonal means for DJF (winter), MAM (pre-monsoon), JJAS (monsoon), and ON (post-monsoon) seasons.

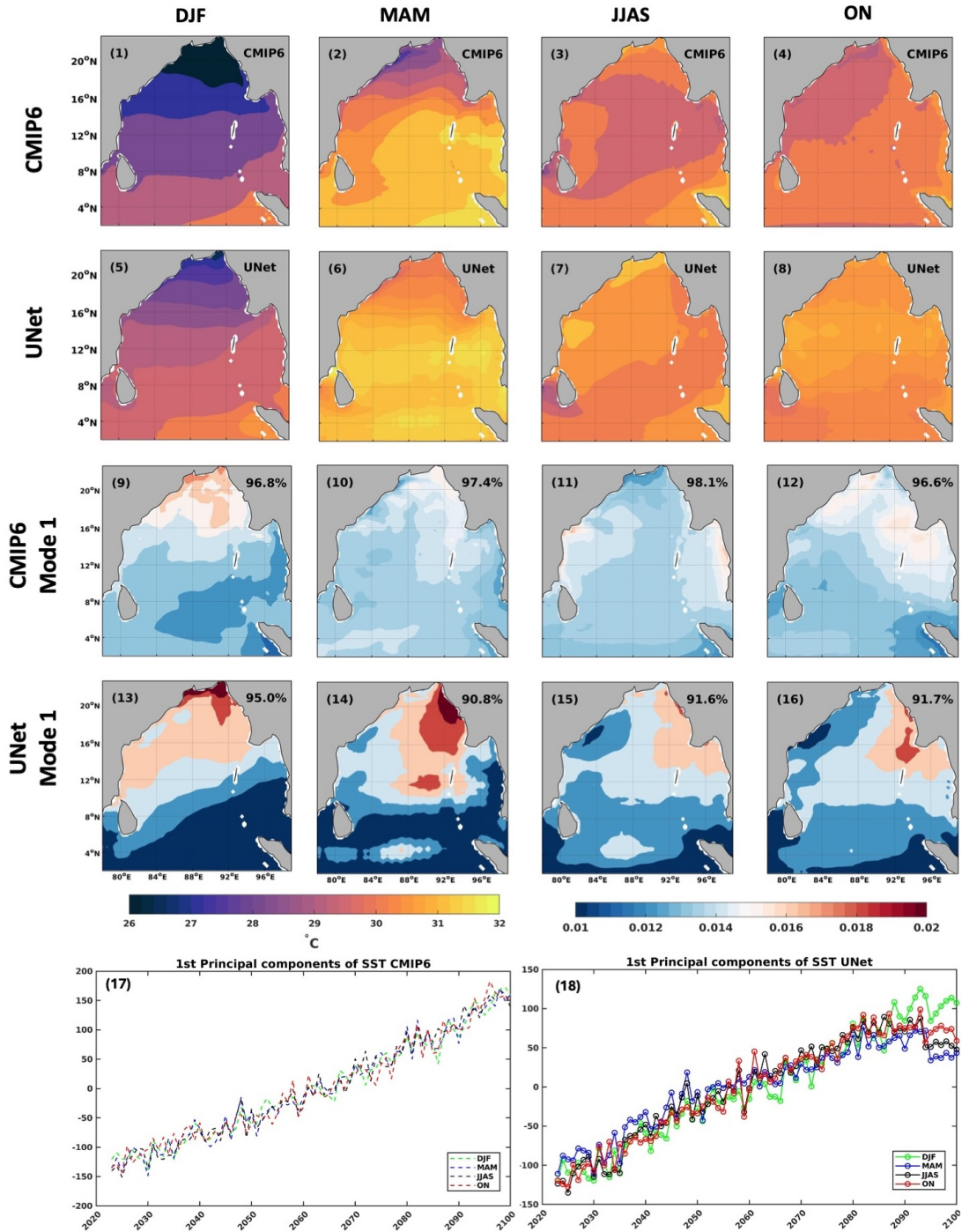

Figure S12: Seasonal SST fo SSP5-8.5 temporal evolution from 2023 to 2100: Seasonal mean SST over the Bay of Bengal comparing raw CMIP6 (1-4) and UNet-corrected (5-8) projections across DJF (winter), MAM (pre-monsoon), JJAS (monsoon), and ON (post-monsoon) seasons. First EOF modes of raw CMIP6 (9-12) with variances (DJF: 96.8%, MAM: 97.4%, JJAS: 98.1%, ON: 96.8%) and UNet-corrected projections (13-16; DJF: 95.0%, MAM: 90.8%, JJAS: 91.6%, ON: 91.7%). First Principal Component time series for CMIP6 (17) and UNet corrected (18) SST .

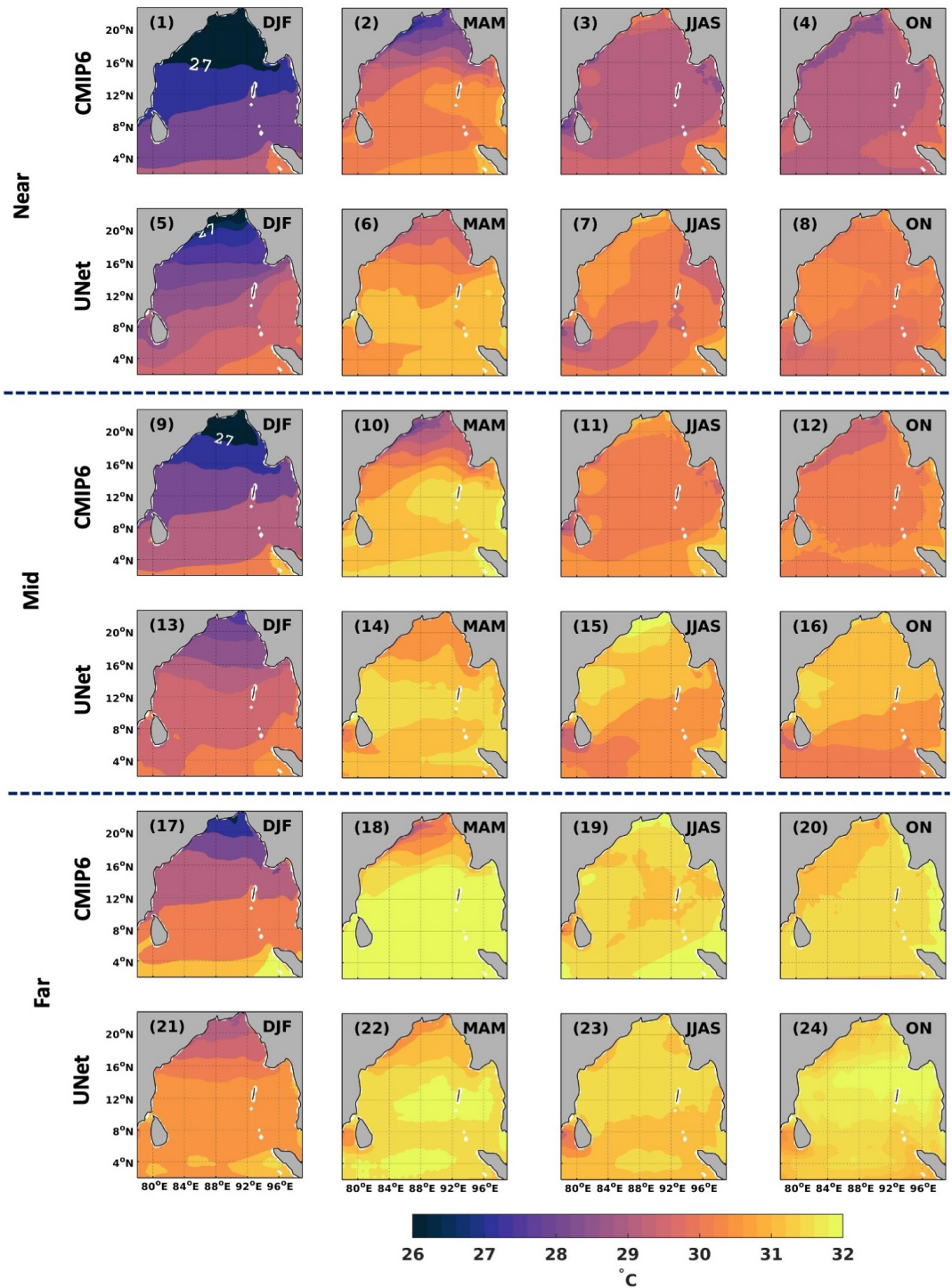

Figure S13: Seasonal SST patterns over the Bay of Bengal comparing raw CMIP6 SSP5-8.5 and UNet-corrected projections for Near future of raw CMIP6 (1-4) and UNet (5-8), Mid future of CMIP6 (9-12) and UNet (13-16), and Far future of CMIP6 (17-20) and UNet (21-24). Each row shows seasonal means for DJF (winter), MAM (pre-monsoon), JJAS (monsoon), and ON (post-monsoon) seasons.

(10,14)). The spatial pattern reveals stronger east-west gradients, indicating potential changes in pre-monsoon heating patterns. Near SSP3-7.0 demonstrates elevated temperatures approximately $\approx 0.5 - 0.8°C$ above Near SSP2-4.5 (Fig.S11), indicating enhanced warming while maintaining the basic spatial structure. Mid SSP3-7.0 exhibits temperatures $\approx 1.5°C$ above Near SSP3-7.0 (Fig.S11 (6,22)), Far SSP3-7.0 exhibits temperatures $2°C$ more than Near SSP3-7.0 (Fig.S11 (22)). USSP3-7.0 shows a more significant warming with temperatures $1.5°C$ higher than the SSP3-7.0 scenario, displaying stronger east-west gradients (Fig.S11 (6)). Mid USSP3-7.0 shows warming $2.5°C$ higher than Near SSP3-7.0 (Fig.S11 (14)), Far USSP3-7.0 indicates warming $3.2°C$ higher than near SSP3-7.0 (Fig.S11 (22)).

SSP5-8.5shows the most intense warming among raw projections, with temperatures $\approx 1 - 1.5°C$ higher than SSP2-4.5, explaining 97.4% of the variance (Fig.S12 (10)), suggesting potential modifications in the characteristics of the pre-monsoon. The SSP5-8.5EOF demonstrates the most extensive variability throughout the bay, while the USSP5-8.5EOF reveals intense patterns in the eastern and central regions (Fig.S12 (10,14)), indicating potential changes in regional circulation dynamics. USSP5-8.5shows the most significant pre-monsoon warming, with temperatures consistently higher throughout the basin than SSP5, explaining 90.8% of the variance (Fig.S10 (14)). SSP5-8.5shows the most intense warming among the Near projections, with temperatures 1-1.5°C above SSP2-4.5 levels (Fig.S13) and Mid SSP5-8.5demonstrates temperatures $\approx 2°C$ higher than Near SSP5-8.5(Fig.S13 (6,22)). The far SSP5-8.5 shows extreme warming with temperatures $\approx 2.5°C$ higher than the near SSP5-8.5 (Fig.S13 (6,22)). The SST patterns show significantly modified temperature distributions, with the most intense warming observed in the central and eastern bay. USSP5-8.5exhibits maximum warming with temperatures $\approx 2.5°C$ higher than SSP5-8.5(Fig.S13 6), showing intense warming throughout the basin. and Mid USSP5-8.5exhibits temperatures $\approx 3°C$ higher than Near SSP5-8.5(Fig.S13 (14)). Far USSP5-8.5 exhibits temperatures $\approx 3.5°$ C higher than near SSP5-8.5(Fig.S13 (22)). Spatial patterns show significantly modified gradient structures and enhanced warming throughout the basin.

### S3.3 Monsoon

SSP3-7.0 demonstrates slightly higher temperatures $\approx 0.5°C$ above SSP2-4.5, explaining 96.4% of the variance (Fig.S10 (3,11)), indicating reduced monsoon cooling while maintaining the basic spatial structure. SSP5-8.5shows the weakest monsoon cooling among the raw projections, with temperatures $\approx 1 - 1.5°C$ higher than SSP2-4.5, which explains 98.1% of the variance (Fig.S12 (3,11)), suggesting potential modifications in the effect of the monsoon in the region. USSP3-7.0 demonstrates a more pronounced warming, with temperatures $\approx 1.5°C$ higher than SSP3-7.0 in many regions, which explained 90.9% of the variance (Fig.S10 (7,15)). The spatial structure reveals stronger east-west gradients, indicating potential changes in monsoon circulation patterns. The SSP3-7.0 EOF pattern indicates stronger variability that extends into the northern bay, while the USSP3-7.0 Mode 1 shows intensified patterns along both coastal boundaries. SSP3-7.0 demonstrates reduced upwelling signatures along the western coast with temperatures reaching $\approx 30°C$, while USSP3-7.0 shows further warming (30.5°C) and significantly modified coastal-offshore temperature gradients, suggesting potential changes in upwelling intensity (Fig.S10 (3,7)).

SSP3-7.0 demonstrates increased temperatures $\approx 0.5 - 0.8°C$ more than SSP2-4.5, indicating a weakened monsoon cooling while maintaining the basic spatial structure (less temperature gradient between the north and south regions). Mid SSP3-7.0 exhibits temperatures $\approx 1.5°C$ more than Near SSP3-7.0 (Fig.S11 3,11), Far SSP3-7.0 exhibits temperatures 2.5°C higher than near SSP3-7.0 (Fig. S11 3,19).

SSP5-8.5shows the most reduced monsoon cooling among the Near projections, with temperatures $\approx 1 - 1.5°C$ above raw CMIP6 SSP2-4.5. and Mid SSP5-8.5shows temperatures $\approx 2°$ C higher than Near SSP5-8.5(Fig.S13 3,11). The far SSP5-8.5 shows extreme warming with temperatures 3°C higher than the near SSP5-8.5(Fig. S13 3,19).

USSP5-8.5shows the most significant reduction in monsoon cooling, with temperatures consistently higher than SSP5-8.5throughout the basin, explaining 93.6% of the variance (Fig.S12 (7,15)). The SSP5-8.5mode 1 demonstrates the most extensive variability throughout the bay, while the USSP5-

8.5mode 1 reveals intense patterns in the northern and eastern regions (Fig.S12 (12,16)).

PC1 of SSP3-7.0 demonstrates a stronger warming trend (Fig.S10 (17)), while SSP5-8.5PC1 shows the most articulated upward trend (Fig.S12 (17)), suggesting a significant reduction in the effectiveness of monsoon cooling.

### S3.4 Post-monsoon

SSP3-7.0 demonstrates slightly higher temperatures, $\approx 0.5°C$ above SSP2-4.5, explaining 93.7% of the variance (Fig.S10 4,12), indicating increased post-monsoon warming while maintaining the basic spatial structure. USSP3-7.0 shows a more pronounced warming, with temperatures $1.5°C$ higher than SSP3-7.0 in many regions, which explained 91. 5% of the variance (Fig.S10 8,16). The spatial structure reveals stronger east-west gradients, indicating potential changes in post-monsoon circulation patterns. The SSP3-7.0 EOF pattern indicates stronger variability that extends into the northern bay, while the USSP3-7.0 Mode 1 shows intensified patterns along the eastern boundary (Fig.S10 12,16).

Near SSP3-7.0 shows higher temperatures $\approx 0.5-0.8°C$ than Near SSP2-4.5 (Fig.S11 4), indicating increased postmonsoon warming. USSP3-7.0 shows significant warming with temperatures $2°C$ above SSP3-7.0 levels (Fig.S11 4,8), displaying stronger cross-basin gradients. Mid SSP3-7.0 shows warming $1.8°C$ above Near SSP3. Mid USSP3-7.0 showing warming $\approx 2.5°C$ higher than Near SSP3-7.0 (Fig.S11 4,16). Far SSP3-7.0 exhibits temperatures $2.5°C$ above Near SSP3-7.0 (Fig.S11 4,12). Far USSP3-7.0 indicates warming $3.5°C$ than Near SSP3-7.0 (Fig.S11 4,24).

SSP5-8.5shows the most intense post-monsoon warming among the raw projections, with temperatures $1-1.5°C$ higher than SSP2-4.5, explaining 96.6% of the variance (Fig.S12 4,12), suggesting possible modifications in seasonal transition characteristics of the BoB. USSP5-8.5shows the most significant warming, with temperatures consistently higher than SSP5-8.5 across the basin, explaining 91.7% of the variance (Fig.S12 8,16). SSP5 Mode 1 demonstrates the most extensive variability in the bay, while USSP5 Mode 1 reveals intense patterns in the eastern and northern regions (Fig.S12 12,16), indicating potential changes in post-monsoon circulation dynamics.

Near SSP5-8.5shows the most intense winter warming among the Near projections, with temperatures $1-1.5°C$ above SSP2-4.5 levels (Fig.S13 4), and Mid SSP5-8.5indicates temperatures $2°C$ higher than Near SSP5. Far SSP5-8.5 shows extreme warming with temperatures $3°C$ higher than Near SSP5-8.5(Fig.S13 4,12).

Near USSP5-8.5 shows maximum warming with temperatures $\approx 2.5°C$ higher than near SSP5-8.5 (Fig.S13 (4,8)), showing intense warming throughout the basin in the extreme scenario. Mid USSP5-8.5demonstrating temperatures $\approx 3.2°C$ over Near SSP5-8.5(Fig.S13 (4,16)). Far USSP5-8.5exhibits temperatures $4°C$ higher than Near SSP5-8.5(Fig.S13 (4,24)).

PC1 of SSP3-7.0 demonstrates a stronger warming trend (Fig.S10 (17)), while SSP5-8.5PC1 shows the most pronounced upward trend (Fig.S12 (17)), suggesting an accelerated post-monsoon warming in the basin.

## S4 Interannual Variability of Seasonal DSL fro SSP3-7.0 and SSP5-8.5: An EOF Analysis

### S4.1 Winter

The first mode explains 32.3% and 52.5% of total variance in SSP3-7.0 (Fig.S14 9) and SSP5-8.5 (Fig.S16 9), respectively, in CMIP6 projections. After UNet correction, the explained variance increases to 79.7% for SSP3-7.0 and 80.4% for SSP5-8.5. Both scenarios show higher DSL values compared to SSP2-4.5, with SSP3-7.0 showing values 0.1-0.2m above (Fig.S14 13) and SSP5-8.5 showing values 0.2-0.3m above the levels of SSP2-4.5 (Fig.S14 13). SSP5-8.5 notably shows an eddy-shaped structure in the central-western bay and higher DSL values around the Andaman Islands. In their UNet-corrected versions (USSP3-7.0 and USSP5-8.5), the eddy-shaped structure shifts to the northwest bay, influenced by the EICC. Both scenarios show progressive intensification from near to far periods, with enhanced DSL values consistently present around the Andaman Islands.

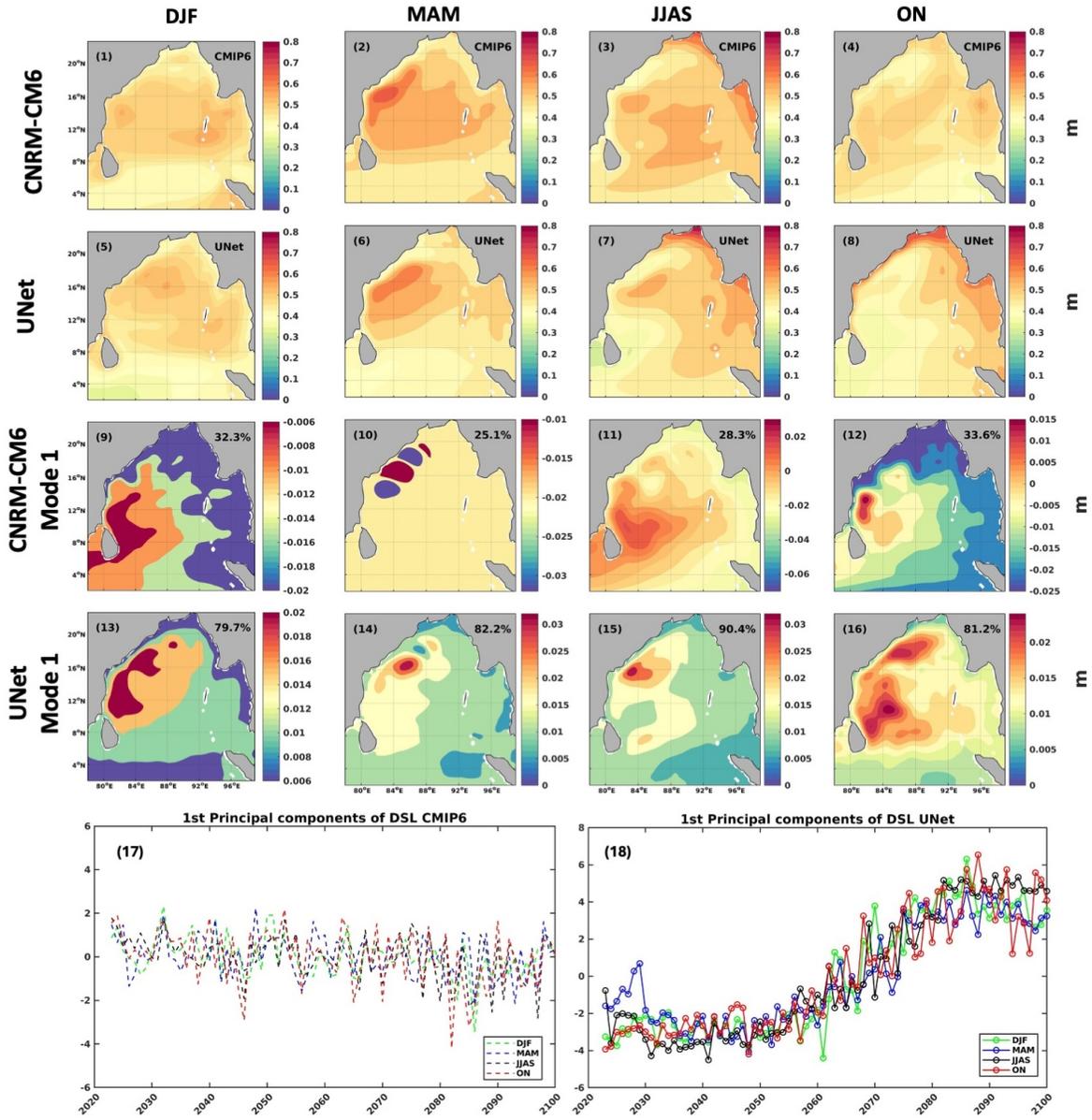

Figure S14: Seasonal DSL temporal evolution of SSP3-7.0 from 2023 to 2100: Seasonal mean DSL over the Bay of Bengal comparing raw CMIP6 (1-4) and UNet-corrected (5-8) projections across DJF (winter), MAM (pre-monsoon), JJAS (monsoon), and ON (post-monsoon) seasons. First, EOF modes of CMIP6 (9-12) show higher variance explained (DJF: 32.3%, MAM: 25.1%, JJAS: 28.3%, ON: 33.6%) compared to UNet-corrected projections (13-16; DJF: 79.7%, MAM: 82.2%, JJAS: 90.4%, ON: 81.2%). First Principal Component time series for CMIP6 (17) and UNet corrected (18) DSL.

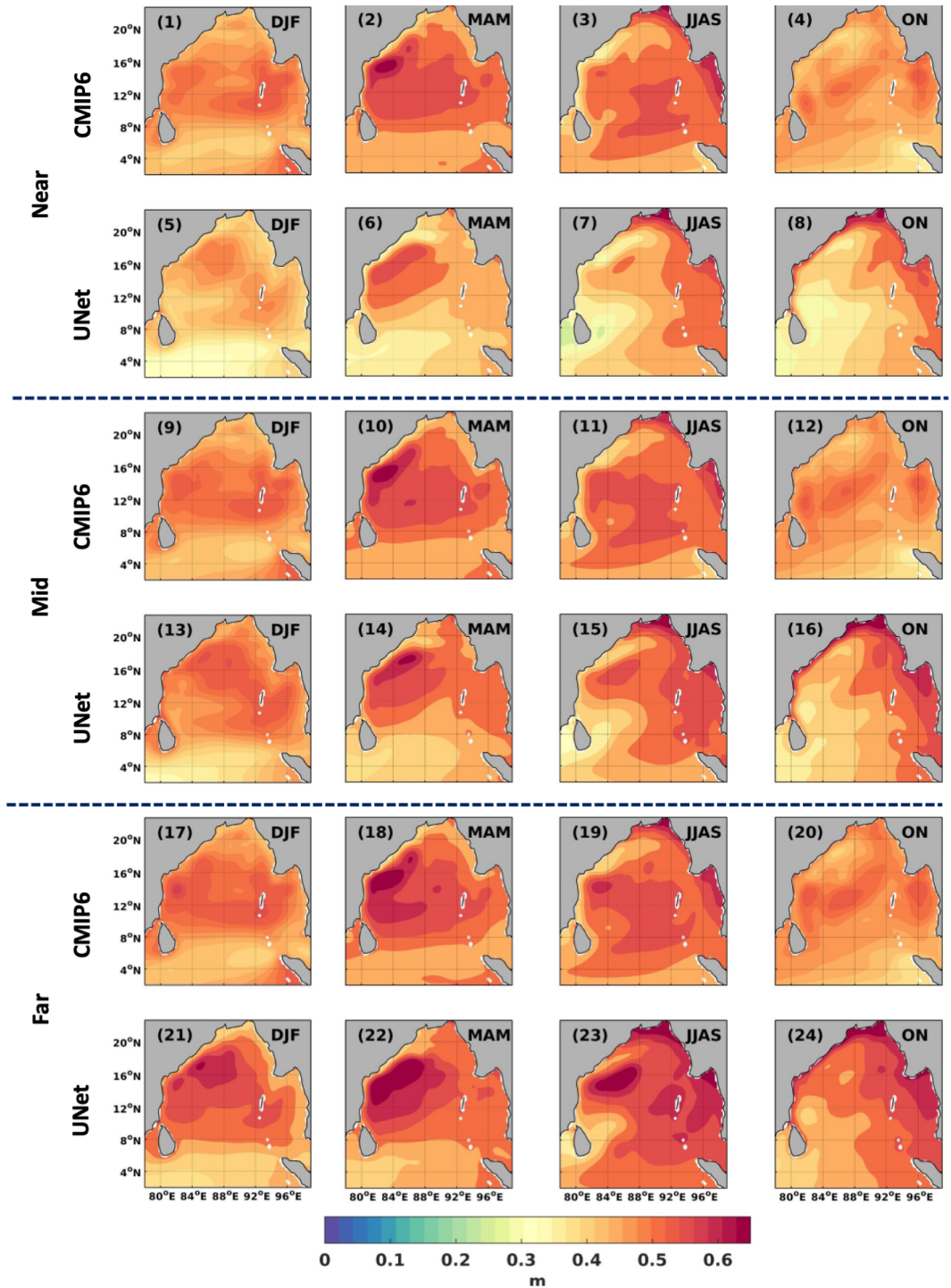

Figure S15: Seasonal DSL patterns over the Bay of Bengal comparing raw CMIP6 SSP3-7.0 and UNet-corrected projections for Near future of CMIP6 (1-4) and UNet (5-8), Mid future of CMIP6 (9-12) and UNet (13-16), and Far future of CMIP6 (17-20) and UNet (21-24). Each row shows seasonal means for DJF (winter), MAM (pre-monsoon), JJAS (monsoon), and ON (post-monsoon) seasons.

20

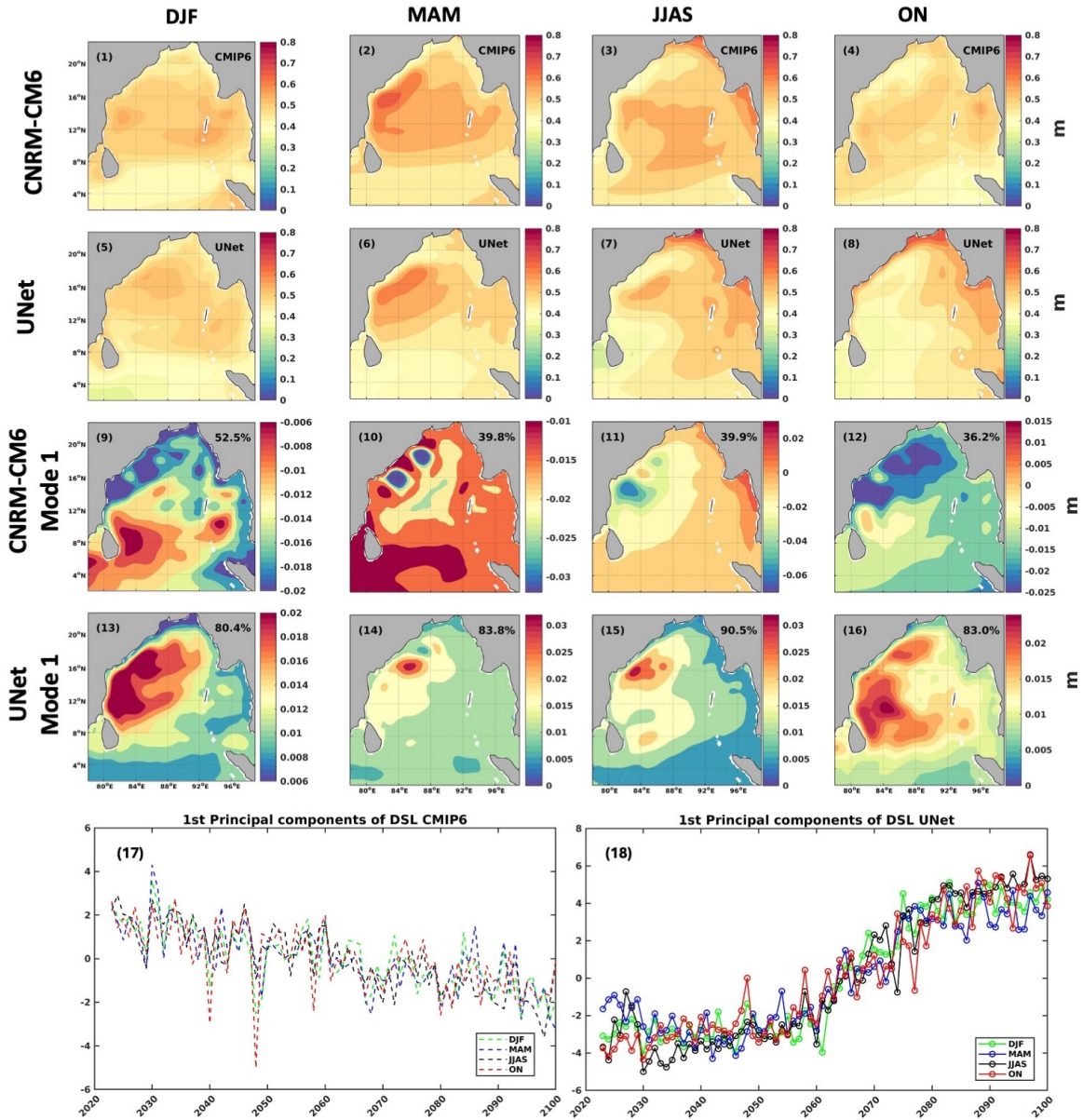

Figure S16: Seasonal DSL temporal evolution of SSP5-8.5 from 2023 to 2100: Seasonal mean DSL over the Bay of Bengal comparing raw CMIP6 (1-4) and UNet-corrected (5-8) projections across DJF (winter), MAM (pre-monsoon), JJAS (monsoon), and ON (post-monsoon) seasons. First, EOF modes of CMIP6 (9-12) show higher variance explained (DJF: 52.5%, MAM: 39.8%, JJAS: 39.9%, ON: 36.2%) compared to UNet-corrected projections (13-16; DJF: 80.4%, MAM: 83.8%, JJAS: 90.5%, ON: 83.0%). First Principal Component time series for CMIP6 (17) and UNet corrected (18) DSL.

21

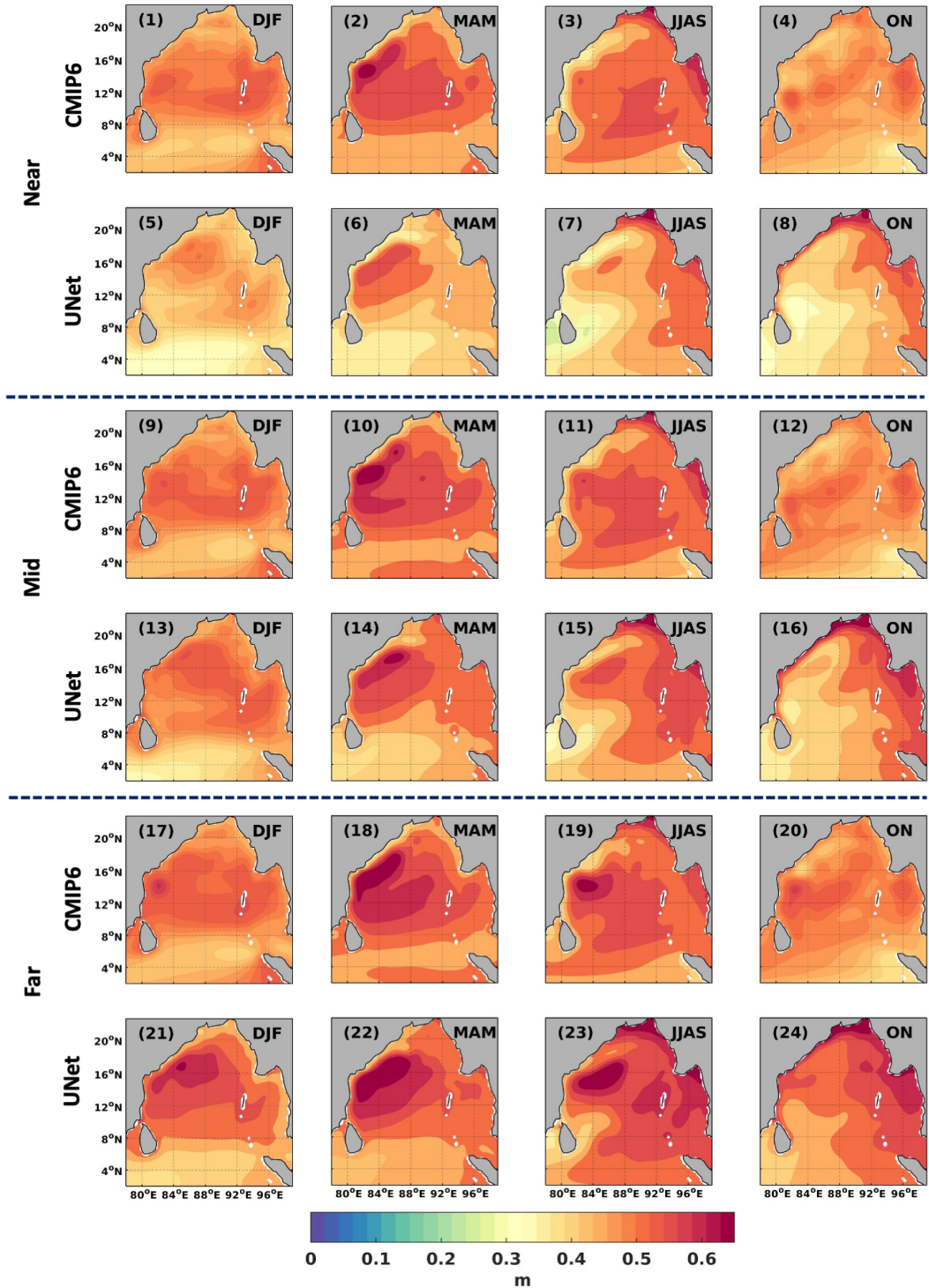

Figure S17: Seasonal DSL patterns over the Bay of Bengal comparing raw CMIP6 SSP5-8.5 and UNet-corrected projections for Near future of CMIP6 (1-4) and UNet (5-8), Mid future of CMIP6 (9-12) and UNet (13-16), and Far future of CMIP6 (17-20) and UNet (21-24). Each row shows seasonal means for DJF (winter), MAM (pre-monsoon), JJAS (monsoon), and ON (post-monsoon) seasons.

## S4.2 Pre-monsoon

The primary mode explains 33.5% and 38.7% of total variance in SSP3-7.0 (Fig.S14 10) and SSP5-8.5 (Fig.S16 10) respectively in CMIP6 projections. After UNet correction, the explained variance increases significantly to 88.1% for SSP3-7.0 (Fig.S14 (14)) and 88.7% for SSP5-8.5 (Fig.S16 (14)). Both scenarios maintain similar spatial patterns to SSP2-4.5, with maximum variability in the central and eastern bay. Their UNet-corrected versions show enhanced features with stronger variability and clear central bay maxima, better capturing the Spring Warm Pool development and associated eddy activities along the coast. Although CMIP6 PC exhibits moderate variability (Fig.S14 (17)), the UNet projections demonstrate stronger variability with a clear increasing trend after 2060 (Fig.S14 (18)).

## S4.3 Monsoon

The first mode explains 28.3% and 39.9% of total variance in SSP3-7.0 (Fig. S14 (11)) and SSP5-8.5 (Fig. S16 (11)) respectively in the CMIP6 projections. After UNet correction, the explained variance increases substantially to 90.4% for SSP3-7.0 (Fig. S14 (15)) and 90.5% for SSP5-8.5 (Fig. S16 (15)). Both scenarios maintain similar monsoon-driven spatial patterns with high DSL values in the western, eastern bay, and northern regions. Their UNet-corrected versions show similar characteristics to SSP2-4.5, with low variability in the western and northern bay before 2060 and higher variability afterward, significantly impacting WBC and eddy activities. The temporal evolution suggests an intensification of the monsoon circulation pattern, particularly in the Far period.

In raw projections, SSP3-7.0 shows DSL values 0.02-0.03m above SSP2-4.5 (Fig.S15 (3)), while SSP5-8.5 shows the highest values ( 0.04-0.05m above SSP2-4.5) with increased eddy activity in the central bay (Fig.S17 (3)). Their UNet-corrected versions show patterns similar to USSP2-4.5 but with greater intensification (Fig.S15 (7)). Both scenarios demonstrate intensified eddy-driven features, stronger WBC signatures, and enhanced freshwater-induced stratification effects compared to their raw projections. The enhanced DSL values around the Andaman Islands and strong WBC signatures persist in both scenarios, indicating robust monsoon-driven oceanographic features. The mid projections show accelerated changes with DSL values 0.02-0.04m higher than Near levels, particularly in regions of strong eddy activity and boundary currents.

## S4.4 Post-monsoon

The first mode explains 36.2% and 41.3% of total variance in SSP3-7.0 (Fig.S14 (12)) and SSP5-8.5 (Fig.S16 (12)) respectively in the CMIP6 projections. After UNet correction, the explained variance increases substantially to 86.2% for SSP3-7.0 (Fig.S14 (16)) and 86.7% for SSP5-8.5 (Fig.S16 (16)). Both scenarios maintain similar post-monsoon spatial patterns with remnant monsoon features. Their UNet-corrected versions show enhanced features with stronger gradients and clear coastal maxima, better capturing the transitioning EICC, post-monsoon circulation, and developing eddy activities. The PC's suggest an intensification of post-monsoon circulation patterns, particularly in the Far period, with stronger variability and increasing trends after 2060 (Fig.S14 (18)).

The raw projections show SSP3-7.0 with DSL values 0.02m above SSP2-4.5 (Fig.S15 4), while SSP5-8.5 shows the highest values ( 0.03-0.04m above SSP2-4.5) (Fig.S17 4). Their UNet-corrected versions show patterns similar to USSP2-4.5 but with greater intensification (Fig.S15 8). Both scenarios feature enhanced EICC flow, eddy activities, and more pronounced remote forcing effects. The mid-period shows accelerated changes, particularly in regions of eddy activity and along the EICC pathway (Fig.S15 16), while far projections demonstrate the most extreme conditions with maximum DSL values in the northern bay and along the eastern boundary (Fig.S15 20). The USSP projections better capture post-monsoon circulation features compared to their raw counterparts (Fig.S15 24).



\end{document}